\documentclass[11pt,a4paper,english,twoside]{tesisFC}

\usepackage[left=3cm,right=3cm,bottom=3.5cm,top=3.5cm]{geometry}
\usepackage[english]{babel}
\usepackage{amsmath, amscd, amssymb, amsthm}
\usepackage{latexsym, enumerate, xspace, dsfont, url, makeidx, graphicx, tikz, hyperref}
\usepackage[draft]{fixme}
\usepackage{lipsum}

\definecolor{darkgreen}{rgb}{0,0.5,0}

\hypersetup{
    bookmarks=true,         
    unicode=false,          
    pdftoolbar=true,        
    pdfmenubar=true,        
    pdffitwindow=false,     
    pdfstartview={FitH},    
    pdftitle={Modal first-order fragments},    
    pdfauthor={Facundo Carreiro},     
    pdfsubject={Modal Logic},   
    pdfkeywords={modal logic, correspondence theory, bisimulation, simulation}, 
    pdfnewwindow=true,      
    colorlinks=true,       
    linkcolor=blue,          
    citecolor=darkgreen,        
    filecolor=magenta,      
    urlcolor=cyan           
}

\usetikzlibrary{arrows,chains,matrix}
\tikzstyle{point}=[circle,draw]
\tikzstyle{edge}=[style=-latex]
\tikzstyle{every loop}=[]

\theoremstyle{definition}
\newtheorem{thm}{Theorem}[section]
\newtheorem{lem}[thm]{Lemma}
\newtheorem{propo}[thm]{Proposition}
\newtheorem{cor}[thm]{Corollary}

\newtheorem*{thm*}{Theorem} 

\theoremstyle{definition}
\newtheorem{defn}[thm]{Definition}
\newtheorem*{defn*}{Definition}
\newtheorem{nota}[thm]{Notation}
\newtheorem{exam}[thm]{Example}

\def\modelfont{\mathcal}
\def\langfont{\mathfrak}
\def\signaturefont{\mathcal}

\newcommand{\tiff}{\mbox{\xspace iff \xspace}}
\newcommand{\liff}{\leftrightarrow}

\newcommand{\efeq}{\risingdotseq}
\newcommand{\noitemsep}{\setlength{\itemsep}{0pt}}

\newcommand{\model}{\ensuremath{\modelfont{M}}\xspace}
\newcommand{\nodel}{\ensuremath{\modelfont{N}}\xspace}
\newcommand{\adel}{\ensuremath{\modelfont{A}}\xspace}
\newcommand{\tmodel}{\ensuremath{\modelfont{M}^{t}}\xspace}
\newcommand{\fmodel}{\ensuremath{\modelfont{M}^{f}}\xspace}
\newcommand{\fnodel}{\ensuremath{\modelfont{N}^{f}}\xspace}

\newcommand{\Th}{\ensuremath{\mathsf{Th}}\xspace}
\newcommand{\NTh}{\ensuremath{\mathsf{NTh}}\xspace}
\newcommand{\Trf}{\ensuremath{\mathsf{Tf}_{x}}\xspace}
\newcommand{\Trm}{\ensuremath{\mathsf{T}}\xspace}
\newcommand{\Tlm}{\ensuremath{\mathsf{T}^{\text{\raisebox{0.35em}{\tiny -1}}}}\xspace}

\newcommand{\Trmk}[1]{\ensuremath{\mathsf{T}_{\!#1}}\xspace}

\newcommand{\cA}{\ensuremath{\langfont{A}}\xspace}
\newcommand{\cP}{\ensuremath{\langfont{P}}\xspace}
\newcommand{\cL}{\ensuremath{\langfont{L}}\xspace}
\newcommand{\cFO}{\ensuremath{\langfont{F}}\xspace}

\newcommand{\cFOu}{\ensuremath{\cFO^{1}}\xspace}
\newcommand{\bml}{\textsc{\small BML}\xspace}
\newcommand{\bmlsn}{\ensuremath{\mbox{\bml}^{\mbox{\!-}}}\xspace}

\newcommand{\sig}{\ssig}
\newcommand{\ssig}{\ensuremath{\signaturefont{S}}\xspace}
\newcommand{\tsig}{\ensuremath{\signaturefont{T}}\xspace}
\newcommand{\fsig}{\ensuremath{\signaturefont{F}}\xspace}

\newcommand{\fprop}{\textsc{fprop}\xspace}
\newcommand{\fconst}{\textsc{fconst}\xspace}
\newcommand{\frel}{\textsc{frel}\xspace}
\newcommand{\fvar}{\textsc{fvar}\xspace}
\newcommand{\ffunc}{\textsc{ffunc}\xspace}

\newcommand{\rels}[1]{(R_r)_{r \in #1}}

\newcommand{\trels}[1]{(R^{t}_r)_{r \in #1}}
\newcommand{\tprops}[1]{(P^{t}_i)_{i \in #1}}
\newcommand{\tconsts}[1]{(c^{t}_i)_{i \in #1}}

\newcommand{\form}[1]{\textrm{FORM}(#1)}
\newcommand{\mods}[1]{\textrm{MODS}(#1)}
\newcommand{\pmods}[1]{\textrm{PMODS}(#1)}

\newcommand{\slc}[1]{\textrm{SLC}(#1)}

\newcommand{\theq}[1]{\equiv_{#1}}
\newcommand{\thsub}[1]{\sqsubseteq_{#1}}

\newcommand{\classK}{\ensuremath{\mathsf{K}}\xspace}
\newcommand{\classM}{\ensuremath{\mathsf{M}}\xspace}
\newcommand{\classC}{\ensuremath{\mathsf{C}}\xspace}
\newcommand{\CclassK}{\ensuremath{\overline{\mathsf{K}}}\xspace}
\newcommand{\CclassM}{\ensuremath{\overline{\mathsf{M}}}\xspace}

\newcommand{\wsat}{\ensuremath{\omega}-saturated\xspace}
\newcommand{\elementary}{\ensuremath{EC_{\Delta}}\xspace}
\newcommand{\belementary}{\ensuremath{EC}\xspace}
\newcommand{\setcomp}[2]{\{{#1} : {#2}\}}

\newcommand{\finv}[1]{#1^{\text{$\scriptstyle-\!1$}}}

\newcommand{\pw}[1]{\mathcal{P}\left(#1\right)}
\newcommand{\nat}{\ensuremath{\mathds{N}}\xspace}

\newcommand{\tup}[1]{\langle #1 \rangle}
\newcommand{\bisim}{\mathrel{\,\raisebox{.3ex}{$\underline{\makebox[.7em]{$\leftrightarrow$}}$}}}

\newcommand{\rsim}{\mathrel{\,\raisebox{.3ex}{$\underline{\makebox[.8em]{\hspace{0.05em}$\rightarrow$}}$}}}
\newcommand{\pisom}{\cong}

\newcommand{\cMLRK}{\ensuremath{\mathcal{ML}(\diam{r})}\xspace}
\newcommand{\cMLRKM}{\ensuremath{\mathcal{ML}(\ddiam{r})}\xspace}
\newcommand{\cMLRKF}{\ensuremath{\mathcal{ML}(\diam{r}, \forget)}}
\newcommand{\cMLRKMF}{\ensuremath{\mathcal{ML}(\ddiam{r},  \forget)}}
\newcommand{\cMLRKE}{\ensuremath{\mathcal{ML}(\diam{r},  \erase)}}
\newcommand{\cMLRKME}{\ensuremath{\mathcal{ML}(\ddiam{r},  \erase)}}
\newcommand{\cMLRKA}{\ensuremath{\mathcal{ML}(\diam{r}, \forget, \erase)}}

\newcommand{\diam}[1]{\langle #1 \rangle}
\newcommand{\ddiam}[1]{\ttup{#1}}
\newcommand{\ttup}[1]{\langle\!\langle #1 \rangle\!\rangle}
\newcommand{\bbox}[1]{[\![ #1 ]\!]}

\newlength{\ksize}
\newcommand{\known}{{\bigcirc\hspace*{-1.45\ksize}\textnormal{\tt k}}\,}
\newlength{\sksize}

\newlength{\rsize}
\newcommand{\remember}{{\bigcirc\hspace*{-1.4\rsize} \textnormal{\tt r}}\,}
\newlength{\srsize}

\newlength{\fsize}
\newcommand{\forget}{{\bigcirc\hspace*{-1.45\fsize} \textnormal{\tt f}}\,}
\newlength{\sfsize}

\newlength{\esize}
\newcommand{\erase}{{\bigcirc\hspace*{-1.45\esize} \textnormal{\tt e}}\,}
\newlength{\sesize}

\newcommand{\prop}{\textsc{prop}}
\newcommand{\rel}{\textsc{rel}}

\newcommand{\hl}{\ensuremath{\mathcal{HL}}\xspace}
\newcommand{\hlA}{\ensuremath{\mathcal{HL}($@$)}\xspace}

\newcommand{\nom}{\textsc{nom}}

\definecolor{orange}{rgb}{1,0.8,0}
\definecolor{green}{rgb}{0,0.8,0}

\tikzstyle{notestyleOR} = [
	draw = black,
	fill = orange!60,
	text width = 2.4cm,
	font=\scriptsize, 
	rounded corners = 2pt
]
\tikzstyle{notestyleGR} = [
	draw = black,
	fill = green!50,
	text width = 2.4cm,
	font=\scriptsize, 
	rounded corners = 2pt
]
\tikzstyle{notestyleRD} = [
	draw = black,
	fill = red!80,
	text width = 2.4cm,
	font=\scriptsize, 
	rounded corners = 2pt
]

\tikzstyle{connectstyleOR} = [draw = orange, thick]
\tikzstyle{connectstyleGR} = [draw = green, thick]
\tikzstyle{connectstyleRD} = [draw = red, thick]

\let\origdoublepage\cleardoublepage
\newcommand{\clearemptydoublepage}{%
\clearpage
{\pagestyle{empty}\origdoublepage}%
}
\let\cleardoublepage\clearemptydoublepage

\begin{document}

\newcommand{\HRule}{\rule{\linewidth}{0.2mm}}
\thispagestyle{empty}

\begin{center}\leavevmode

\vspace{-2cm}

\begin{tabular}{l}
\includegraphics[width=2.6cm]{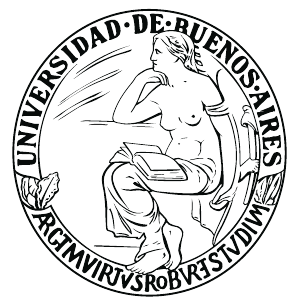}
\end{tabular}

{\large \sc Universidad de Buenos Aires

Facultad de Ciencias Exactas y Naturales

Departamento de Computaci\'on

}

\vspace{6.0cm}


{\huge\bf Caracterizaci\'on y Definibilidad en Fragmentos Modales de \mbox{Primer Orden}}
\HRule\\
\medskip
{\Large Characterization and Definability in Modal First-Order Fragments}

\vspace{2cm}

{\large Tesis presentada para optar al t\'{\i}tulo de\\
Licenciado en Ciencias de la Computaci\'on}

\vspace{1cm}

{\Large Facundo Mat\'ias Carreiro}

\end{center}

\vfill

{\large

{Director: Santiago Daniel Figueira}

\vspace{.2cm}

{Codirector: Carlos Eduardo Areces}

\vspace{.2cm}

Buenos Aires, 5 de Marzo de 2010
}

\newpage\thispagestyle{empty}


\thispagestyle{empty}
\thispagestyle{plain}


\frontmatter

\tableofcontents

\abstract
Model theoretic results such as Characterization and Definability give important information about different logics. It is well known that the proofs of those results for several modal logics have, somehow, the same `taste'. A general proof for most modal logics below first order is still too ambitious. In this thesis we plan to isolate sufficient conditions for the characterization and definability theorems to hold in a wide range of logics. Along with these conditions we will prove that, whichever logic that meets them, satisfies both theorems. Therefore, one could give an unifying proof for logics with already known results. Moreover, one will be able to prove characterization and definability results for logics that have not yet been investigated. In both cases, it is only needed to check that a logic meets the requirements to automatically derive the desired results. 


\addcontentsline{toc}{chapter}{Abstract}

\chapter*{Agradecimientos}\addcontentsline{toc}{chapter}{Agradecimientos}
\label{chap:ack}
Una gran cantidad de personas invirti\'o tiempo, esfuerzo y buena voluntad para que pudiera llegar a este momento. La siguiente lista de agradecimientos no planea ser extensiva ni mucho menos. Probablemente olvide nombres pero todo el que sienta que haya hecho un aporte puede sentirse incluido.

Primero que nada quiero agradecer a Santiago Figueira, por confiar en mi desde el principio y abrirme un mundo de oportunidades y personas nuevas. Gracias por incentivarme a la investigaci\'on y poner todo tu empe\~no para crear un ambiente de trabajo genial. Tambi\'en por ser una persona abierta, dispuesta a compartir mucho m\'as que algunas reuniones por semana.

Le debo mucho tambi\'en, a Carlos Areces, por orientarme en mi tesis y fuera de ella. Por darme la oportunidad de conocer lugares nuevos y personas muy importantes. Por encargarse de todos los arreglos necesarios, siempre!
Quiero agradecer a ambos, Carlos y Luciana Benotti, por darme el mejor desayuno de mi vida, recibirme en su casa como si fuera la mia, ayudarme y compartir un mont\'on de cosas. Son personas realmente bondadosas y les estoy muy agradecido.

A todo el Grupo de L\'ogica y Computabilidad, por haberme recibido con los brazos abiertos sin pedir nada a cambio. Gracias a Sergio Mera y Daniel Gor\'in por haberme ense\~nado innumerables cosas sobre l\'ogica y sobre c\'omo ser un buen docente, con teor\'ia y dando el ejemplo. A Ricardo Rodriguez, por mostrar la mejor disposici\'on al recibirme en sus proyectos y por actuar como jurado de mi tesis. Estoy en deuda con todos ellos por haber revisado detalladamente mi trabajo y haber aportado comentarios (o discutido) al respecto. Tambi\'en quisiera agradecer a Ver\'onica Becher por mostrar un contagioso gran entusiasmo por su trabajo y responder mis preguntas sobre `que hace un investigador'.

Un muy merecido agradecimiento a la `Gente de la Facu', por ser mis compa\~neros dentro y fuera de las aulas en los \'ultimos seis a\~nos: Mat\'ias Blanco, Fernando Bugni, Luis Brassara, Bruno Cuervo Parrino, Diego Freijo, Maxi Giusto, Pablo Laciana, Sergio \mbox{Medina}, Ariel Neisen, Santiago Palladino, Leandro Radusky, Leo Rodriguez, Nati Rodriguez, Viviana Siless, Javier Silveira, Leo Spett, Andr\'es Taraciuk, Mart\'in Verzilli, Pablo Zaidenvoren, Eddy Zoppi. El apoyo del grupo fue muy importante para poder concretar varios proyectos.

Al Departamento de Computaci\'on de la Universidad de Buenos Aires, por superar todas mis expectativas. Encontr\'e all\'i, adem\'as de un excelente nivel acad\'emico, una calidad y calidez humana que ser\'a dificil de superar.

A mi familia, menci\'on especial para mis padres por intentar siempre resolverme la mayor cantidad de problemas para que yo pueda continuar tranquilo con mis cosas. Tambi\'en quiero destacar a Jos\'e Cuba y Mar\'ia In\'es Cuba por demostar real inter\'es en mi inentendible trabajo.

A mis amigos Gabriel, Murray, Nico, Martina, Gon y el resto del grupo por ayudarme a despejar mi cabeza y acompa\~narme en el mundo real.

Finalmente, agradezco profundamente a Lucila Schmidt, por su apoyo constante de mil maneras distintas y soportar mis mejores y peores momentos. Por ayudarme a ser una mejor persona.

\chapter*{Abstract extendido}\addcontentsline{toc}{chapter}{Abstract extendido}
\label{chap:resumen}
Para una amplia variedad de aplicaciones que usan la l\'ogica como
herramienta, la l\'ogica de primer orden es suficiente para resolver
sus problemas de manera \emph{te\'orica}. Sin embargo, cuando se
considera el comportamiento \emph{pr\'actico} de la l\'gica de
primer orden uno se encuentra con varias complicaciones. Primero que
nada, la l\'ogica de primer orden es \emph{indecidible}. Esto quiere
decir que no existe un algoritmo general para decidir si una
f\'ormula arbitraria es satisfacible. Segundo, en general la
mayor\'ia de las aplicaciones que usan esta l\'ogica no aprovechan
al m\'aximo su poder. Por lo tanto, incluso cuando se trabaja con
fragmentos decidibles de primer orden, se puede estar pagando un
precio excesivo por cualidades que no ser\'an utilizadas.

Las l\'ogicas modales proposicionales ofrecen una alternativa a los
lenguajes tradicionales. Pueden ser pensadas como un conjunto de
herramientas que permiten dise~nar l\'ogicas espec\'ificamente
construidas para una tarea en particular, posibilitando un control
fino en su expresividad. M\'as a\'un, las l\'ogicas modales
resultaron tener un buen comportamiento computacional que prob\'o
ser bastante robusto frente a extensiones. Estas caracter\'isticas,
entre otras, ubicaron a las l\'ogicas modales como una alternativa
atractiva con respecto a los lenguajes cl\'asicos como por ejemplo
la l\'ogica de primer orden.

En esta tesis trabajaremos con l\'ogicas modales que son \emph{a lo sumo} tan
expresivas como la l\'ogica de primer orden. Informalmente, esto quiere decir que
si uno puede expresar una propiedad con una formula de dicha l\'ogica modal entonces
existe una manera de expresar \emph{la misma} propiedad en primer orden.
En otras palabras, uno puede decir que si una formula modal $\varphi$ denota
una propiedad dada entonces existe alg\'un tipo de \emph{traducci\'on}
cuyo resultado es una f\'ormula de primer orden $\varphi^{t}$ que denota
la misma propiedad.

\bigskip
En ciencias de la computaci\'on, una bisimulaci\'on es, a grandes
rasgos, una relaci\'on binaria entre modelos que asocia aquellos que
se comportan de la misma manera. As\'i, dos modelos son bisimilares
cuando no pueden ser distinguidos mutuamente por un observador. La noci\'on de
bisimulaci\'on es ampliamente empleada en varias \'areas como la l\'ogica modal,
la teor\'ia de concurrencia, la teor\'ia de conjuntos, la verificaci\'on formal,
etc.

La noci\'on de bisimulaci\'on fue  descubierta de manera independiente y
relativamente simult\'anea por van~Benthem, en el contexto de teor\'ia
de correspondencia modal; Milner y Park, en teor\'ia de la
concurrencia; y Forti y Honsell en teor\'ia de conjuntos sin axioma de
buena fundaci\'on. Estos \'ultimos utilizan bisimulaciones  para mostrar
la equivalencia de objetos con estructura infinita no-inductiva y
garantizar as\'i extensionalidad de los modelos de su
teor\'ia~\cite{FH83}. Van~Benthem~\cite{vB76} obtiene la idea de
bisimulaci\'on como una generalizaci\'on del concepto de $p$-morfismo
entre modelos; con ella caracteriza a la l\'ogica modal b\'asica como el
fragmento de primer orden invariante bajo bisimulaciones (lo que se
conoce como \emph{Teorema de Caracterizaci\'on de van~Benthem}).
Milner y Park fueron los que acu\~naron el t\'ermino
\emph{bisimulaci\'on}, t\'ecnica que utilizaron como herramienta para
probar la equivalencia de procesos concurrentes~\cite{M80,P81}.
En~\cite{S08} se da un interesante panorama hist\'orico del \'area.

La {\em bisimulaci\'on} es una herramienta crucial en el proceso de
estudiar estructuras relacionales y abre el camino para poder
analizar formalmente caracterizaciones de la expresividad de los
lenguajes modales. Intuitivamente, fijada una l\'ogica $\mathcal{L}$,
la noci\'on de bisimulaci\'on define cu\'ando dos modelos son
indistinguibles para $\mathcal{L}$ (es decir, no existe una f\'ormula
de ${\cal L}$ que sea verdadera en un modelo y falsa en otro).
Existe una gran variedad de \'areas en donde la bisimulaci\'on juega
ahora un rol central: l\'ogica modal~\cite{vB76}, teor\'ia de
concurrencia~\cite{P81}, teor\'ia de conjuntos~\cite{FH83},
verificaci\'on formal~\cite{CGP00}, generaci\'on de lenguaje natural
\cite{AKS08}, etc.

El teorema de caracterizaci\'on de van~Benthem para la l\'ogica modal b\'asica
caracteriza el fragmento de primer orden invariante bajo la definici\'on
de bisimulaci\'on. Informalmente, puede ser enunciado de la siguiente manera.

\medskip

\noindent {\bf Teorema.} Una formula de primer orden $\alpha$ es
equivalente a la traducci\'on de una f\'ormula de la l\'ogica modal
b\'asica si y solo si $\alpha$ es invariante bajo bisimulaciones.

\medskip

Ahora bien, desde un punto de vista l\'ogico, no existe una \'unica
noci\'on de \emph{bisimulaci\'on}. A cada lenguaje modal le corresponde
una noci\'on de bisimulaci\'on distinta (o, en el caso de l\'ogicas
sub-booleanas, una \emph{simulaci\'on}~\cite{KR97,KR99}).

En general, cada combinaci\'on de l\'ogica y bisimulaci\'on tiene su
demostraci\'on de un teorema equivalente a la caracterizaci\'on de
van~Benthem. Un problema esencial es que no parece haber una
demostraci\'on general y cada caso necesita una nueva prueba usando
herramientas ad-hoc.

\bigskip

El nacimiento del concepto de bisimulacion y la teor\'ia de
correspondencia ayud\'o a responder nuevas preguntas desde una
perspectiva puramente de teor\'ia de modelos. Un ejemplo es la
caracterizaci\'on de \emph{definibilidad} en l\'ogica modal.
Informalmente decimos que una clase de modelos es {\em definible}
por un conjunto de formulas $\Gamma$ si est\'a compuesta por
exactamente todos los modelos donde $\Gamma$ es v\'alida. Una clase
se dice definible por una f\'ormula modal si es definible por un
conjunto singleton.

Ser\'ia interesante saber qu\'e propiedades deber\'ia cumplir una
clase de modelos para ser definible ya ser por un conjunto de
f\'ormulas o por \'unica una f\'ormula modal. Esta pregunta ya se ha
enunciado y respondido para la l\'ogica de primer orden. Para ese
caso, la respuesta est\'a formulada en t\'erminos de
\emph{isomorfismos potenciales}. En cambio, en el caso de las
l\'ogicas modales la noci\'on de bisimulaci\'on juega un rol
esencial. Para dar un ejemplo citamos el siguiente resultado para la
l\'ogica modal b\'asica~\cite{MLBOOK}.

\medskip

\noindent {\bf Teorema.} Una clase de modelos \classK es definible
por una formula modal si y solo si \classK y \CclassK est\'an
cerrados por bisimulaciones y ultraproductos.

\medskip

Por el momento, no es necesario preocuparse por la definici\'on formal de
`ultraproductos'. S\'olo es necesario saber que los ultraproductos
son una construcci\'on de modelos (con or\'igenes algebraicos) muy \'utiles.
Inicialmente, dicha construcci\'on llam\'o la atenci\'on a los l\'ogicos porque
pod\'ia ser usada para dar una demostraci\'on \emph{puramente algebraica}
del Teorema de Compacidad para primer orden. Para un desarrollo detallado
sobre ultraproductos recomendamos la lectura de~\cite{KULT08}.

Como con el teorema de caracterizaci\'on, resultados de definibilidad
similares al aqu\'i presentado valen para una amplia variedad de
l\'ogicas modales. De la misma manera, cada l\'ogica tiene
su propia demostraci\'on especialmente dise\~nada para ese
caso en particular.

\bigskip
Claramente, resultados como los de Caracterizaci\'on y Definibilidad nos
sirven para entender mejor una l\'ogica. Es m\'as, estos resultados tambi\'en
tienen un gran impacto en aplicaciones pr\'acticas de Ciencias de la Computaci\'on.

Consideremos el siguiente problema: Supongamos que estamos interesados en realizar \emph{model checking}, esto quiere decir, dado el modelo de un sistema, verificar autom\'aticamente si el modelo cumple una cierta especificaci\'on. Supongamos tambi\'en que la especificaci\'on puede ser escrita como una f\'ormula de primer orden $\varphi$.

Siempre podemos usar herramientas de primer orden para verificar si
el modelo satisface $\varphi$ pero eso puede resultar, como ya hemos
mencionado, en un alto costo en cuanto a complejidad computacional.
Uno podr\'ia tratar de encontrar l\'ogicas m\'as `baratas' que
puedan resolver el problema. Si la misma propiedad pudiera ser
expresada en una l\'ogica modal entonces probablemente podr\'iamos
mejorar la performance del proceso dr\'asticamente.

Discutamos un ejemplo concreto: Supongamos que los puntos del dominio de nuestro modelo son diferentes estados en la ejecuci\'on de un programa. De esta manera, hay una transici\'on desde un punto a otro si es posible ejecutar una transformaci\'on del programa que lo lleve del estado $a$ al estado $b$. Pensando en el modelo de esta manera se puede ver que los estados sin sucesores representan estados donde el programa ha finalizado.

Una propiedad deseable del modelo podr\'ia ser que ``en cada estado del programa se debe poder `escapar' del flujo de ejecuci\'on''. Esto quiere decir que todo punto debe poder ver directamente a un estado sin sucesores. Esta propiedad puede ser verificada probando que la formula de primer orden
$$\varphi(x) = \exists y. R(x,y) \to (\exists z. R(x,z) \;\land\; \forall w. \lnot R(z,w))$$
sea v\'alida en el modelo pero tambi\'en puede ser verificada
probando que la f\'ormula de la l\'ogica modal b\'asica ${\psi =
\Diamond\top \to \Diamond\Box\bot}$ sea v\'alida en el modelo. Como
estas dos f\'ormulas representan la misma propiedad (son
equivalentes) podemos usar \emph{model checkers} que acepten
f\'ormulas de la l\'ogica modal b\'asica como entrada para poder
resolver nuestro problema.

Aparte de ser mucho m\'as `amigable', la simple existencia de la f\'ormula $\psi$ nos dice que la propiedad en cuesti\'on es invariante bajo bisimulaciones. Esta informaci\'on nos brinda un beneficio extra. Supongamos que el modelo es automaticamente generado a partir de una porci\'on de c\'odigo. Si, por ejemplo, alimentamos al generador con el c\'odigo de un sistema operativo entero, el modelo resultante ser\'a \emph{muy} grande.

No es el objetivo de esta tesis meterse en estos temas pero existen algoritmos eficientes para minimizar el modelo autom\'aticamente. Estos algoritmos preservan la verdad de las f\'ormulas invariantes por bisimulaci\'on~\cite{H71, G73}. Por lo tanto, al tener una f\'ormula modal que representa nuestra propiedad, uno podr\'ia primero minimizar el modelo y hacer \emph{model checking} sobre el modelo resultante que ser\'a, muy probablemente, mucho m\'as chico que el original.

Por otra parte, supongamos que se quiere verificar si el modelo es `irreflexivo'. Esto quiere decir, que ningun elemento est\'a relacionado consigo mismo. Si interpretamos esta propiedad en el escenario descripto anteriormente, la propiedad dir\'ia que ningun estado debe poder quedarse `colgado' en si mismo.

Para este caso, incluso cuando la propiedad puede ser puesta a prueba verificando la validez de la f\'ormula de primer orden $\lnot R(x,x)$, no existe ninguna f\'ormula de la l\'ogica modal b\'asica que sea equivalente. Esto puede ser demostrado facilmente ya que la `irreflexividad' no es invariante bajo bisimulaciones. Es m\'as, la l\'ogica modal b\'asica tiene una propiedad llamada la `propiedad de modelo de \'arbol' o \emph{tree model property}. Esto quiere decir que cualquier f\'ormula satisfacible es tambi\'en satisfacible en un modelo que tiene forma de \'arbol. Como corolario surge que no existe ninguna f\'ormula de la l\'ogica modal b\'asica que caracterice irreflexividad, antisimetr\'ia ni intransitividad.

Es este el fin de la l\'ogica modal? Estamos condenados a usar la l\'ogica de primer orden en este caso? Afortunadamente, la respuesta es \emph{no}. Aunque la l\'ogica modal b\'asica no pueda expresar esas propiedades, existen l\'ogicas modales m\'as ricas (que a\'un se mantienen por debajo del poder expresivo de primer orden) que pueden hacerlo. Como ejemplo, pueden usarse las l\'ogicas h\'ibridas~\cite{MLBOOK} y las menos conocidas Memory Logics~\cite{AFFM08} que ser\'an oportunamente introducidas.

\bigskip
En s\'intesis, resultados de teor\'ia de modelos como los de Caracterizaci\'on y Definibilidad dan informaci\'on importante sobre las distintas l\'ogicas. Se sabe que las demostraciones de estos resultados para las diferentes l\'ogicas modales tienen, de alg\'un modo, el mismo `sabor'. Una prueba general que cubra todas las l\'ogicas modales por debajo de primer orden es, por el momento, un plan demasiado ambicioso.

En esta tesis damos condiciones muy generales pero suficientes para que las propiedades de Caracterizaci\'on y Definibilidad valgan en una amplia gama de l\'ogicas modales: cualquier l\'ogica modal que cumpla nuestras condiciones verificar\'a las propiedades de Caracterizaci\'on y Definibilidad. El resultado se puede aplicar tanto a l\'ogicas para las que se saben ciertas las propiedades en cuesti\'on, como a l\'ogicas para las que se desconoc\'ia si estas propiedades val\'ian o no. En el primer caso, obtenemos demostraciones nuevas de resultados ya conocidos (en particular, aplicamos nuestro esquema a las l\'ogicas h\'ibridas con el operador @ y nominales). En el segundo caso, obtenemos resultados novedosos, aplicando nuestras herramientas a las memory logics, una familia de l\'ogicas modales con comportamiento din\'amico introducidas recientemente en el \'area.

\mainmatter

\chapter{Introduction}\label{chap:intro}
\section{A bit of history}
The first traces of modal logic go back to 1918 with the work of C. Lewis~\cite{lewis18}. In this publication he enriched the propositional calculus with two operators to try to solve some problems with material implication. In a modern notation these operators would be $\Box$ and $\Diamond$. Given a formula $\varphi$, then, $\Box\varphi$ was meant to be interpreted as ``it is necessary that $\varphi$'' and $\Diamond\varphi$ as ``it is possible that $\varphi$''. At this point (called the `syntactic era') all the work on Modal Logic was strictly \emph{syntactical}, there was no model theory for it.

Later, during the end of 1950s and early 1960s (sometimes called the `classical era') the first ideas on modal logic semantics were born. The seminal work of Prior~\cite{pri57} (with \emph{tense logic}) and J\'onson and Tarski with boolean algebras with operators~\cite{tarski51,tarski52} later gave birth to Kripke semantics for modal logics. Kripke's work~\cite{kripke63,kripke63b} proposes a relational semantic for modal logic, that is, a suitable model to evaluate a modal formula is just a set of \emph{worlds} (or points) and \emph{relations} among them.

With these semantics, many difficult problems (such as knowing whether two axiomatic systems are equivalent) had now turned a lot easier. The emergence of \emph{cannonical models} and \emph{completeness} results were predominant in this period which helped link the ancient `syntactic era' with the new semantics. Although the research made in the `classical era' was not syntactical, it was anyways \emph{syntactically driven}. That is, relational semantics, were used as a tool to analyze logics and prove syntactical results. Model theory, was not playing a big role by itself.

The so-called `modern era' goes from the 1970s to the present days. In this period, modal logic started to be used to \emph{describe} relational structures and not just as a mere tool. The germ of modal logic also started to spread to other fields, as an example, computer scientists started to use modal logics to reason about \emph{programs} represented as relational models. The first steps in this line of work were taken by Pratt~\cite{pratt79} with his work on \emph{propositional dynamic logic} (PDL). Computer scientists added new problems to the already growing pool of questions. Complexity of the satisfiability problem for modal logics started to be studied with the work of Ladner~\cite{ladner77} for normal logics and Ladner, Fischer and Pratt~\cite{fischer79, pratt79} for PDL.

The discovery of \emph{frame incompleteness} results showed that there are classes of models for which there is no possible axiomatization (Thomason~\cite{thomason72, thomason74} and Fine~\cite{fine74}). This shows that modal logics can't be analyzed from a purely syntactical perspective.

Modal logic is not isolated from the rest of the world. During this period, the \emph{expressive} power of modal logics was put into question. Which logic is the best to describe certain relational structures? Now that we know that different logics have different computational complexities, which is the `cheapest' logic that solves my problem? The power of these logics could be compared between each other and also with respect to classical logics such as first and second order logic.

The results brought to light by this period helped shift the view of modal logics as `intensional' formalisms that were only able to talk about `modes of truth' to a much broader panorama, which constitutes the current way of looking at modal logics.

\section{Basic modal logic}\label{sec:intro:bml}

It is now time to formally meet the modal logics we are going to work with and its relational semantics. We start by defining the \textit{basic modal language} $\bml$. Because we are interested in working with many modalities at the same time, the diamond ($\Diamond$) and box ($\Box$) operators are going to turn into the operators $\diam{r}$ and $[r]$, where $r$ indicates the modality we are working with. When we are in a case where there is a single modality, we are going to use $\Diamond$ and $\Box$ again.

\begin{defn}[Syntax]\label{def:ml-syntax}
Suppose we have a set of propositional symbols $\prop=\{p_1, p_2, \dots\}$ and a set of modality symbols
$\rel=\{r_1, r_2, \dots\}$. We assume that both sets are pairwise disjoint and countable infinite. A specific choice of $\prop$ and $\rel$ is called the \emph{signature} of the language. We define the set of formulas of the \textit{basic modal language} over the signature $\tup{\prop,\rel}$ as:
$$
\varphi ::= \top 
	   \mid \bot 
           \mid p 
           \mid \lnot \varphi
           \mid \varphi \land \psi
           \mid \varphi \lor \psi
           \mid \varphi \to \psi
           \mid \varphi \liff \psi
           \mid \diam{r} \varphi
           \mid [r] \varphi
$$
where $p \in \prop$, $r \in \rel$  and $\varphi, \psi$ are formulas.
\end{defn}

Of course this is not a minimal definition. One can fix an adequate set of primitive boolean connectors (like $\lnot$ and $\land$) and define all the other boolean connectors in terms of that primitive set. Also, as it will follow from the satisfaction definition we are going to present below, diamond and box are dual operators, and therefore for all $r \in \rel$, $\diam{r}\varphi$ can be defined as $\lnot [r] \lnot \varphi$, and conversely, $[r]\varphi$ is equivalent to $\lnot\diam{r}\lnot\varphi$. We are not going to bother yet to pick a set of primitives operators, since it is not really important at this point. When we do that, we will only have to worry about choosing a convenient set that allows us to generate the whole language.\\

Now we formally define the models for the basic modal language. As we mention before, Kripke semantics define models as graphs, and in fact, as \textit{directed} graphs with \textit{decorations}.

\begin{defn}[Kripke models]\label{def:ml-models}
Let $\sig=\tup{\prop,\rel}$ be a signature. A \emph{Kripke model} $\model$ for $\sig$
is a tuple $\tup{W, \rels{\rel}, V}$ satisfying the following
conditions:
\begin{enumerate}[(i)]
\item $W$, the \textit{domain}, is a nonempty set whose elements are called points, but also, depending on the context, states, worlds, times, etc.
\item Each $R_r$, an \textit{accessibility relation}, is a binary relation on $W$.
\item $V: \prop \to \pw{W}$, the \textit{valuation}, is a labeling function that assigns to each propositional symbol $p \in \prop$ a subset of $W$. We can think of $V(p)$ as the set of points in $\model$ where $p$ holds.
\end{enumerate}
Given a model $\model$ and $w \in |\model|$, we call $\tup{\model,w}$ a \emph{pointed
model}.
\end{defn}

Before moving on, let us see an example of a Kripke model, in order to clarify the concept. In the following model we will give a graphical representation of the domain and the relations of the model. A node represents an element in the domain and an edge from $w$ to $w'$ labeled as $R$ means that $wRw'$.\\

\begin{exam}
Consider the following model $\model=\tup{W,\rels{\rel}, V}$:

\begin{scriptsize}
\begin{center}
\begin{tikzpicture}
  \node (n1) at (-2,1) [point] {$w_1$} edge [edge, in=180, out=130, loop] node [left] {$R_{2}$} ();
  \node (n2) at (1,0) [point] {$w_2$} ;
  \node (n3) at (3,2) [point] {$w_3$} ;
  \node (n4) at (3,0) [point] {$w_4$} ;

  \node [below] at (n1.south) {$p$};
  \node [above] at (n3.north) {$p,q$};
  \node [below] at (n2.south) {$q$};

  \draw (n1) edge [->, edge, bend right] node [above] {$R_{1}$} (n2);
  \draw (n2) edge [->, edge, bend left] node [above=2pt] {$R_{1}$} (n3);
  \draw (n3) edge [->, edge, bend left=40] node [right] {$R_{2}$} (n4);

  \draw (n1) edge [->, edge, bend left] node [above] {$R_{2}$} (n3);
  \draw (n2) edge [->, edge, bend right] node [above] {$R_{1}$} (n4);
\end{tikzpicture}
\end{center}
\end{scriptsize}

This model has a domain of four points, $W=\{w_1, w_2, w_3, w_4\}$. The signature in which it is based on is $\tup{\prop=\{p,q\}, \rel=\{r_{1},r_{2}\}}$, that is, it has two modalities, $r_{1}$ and $r_{2}$, and two propositional symbols, $p$ and $q$. We explicitly indicate in the picture the places where the propositional symbols hold. Translated to the valuation function $V$, that means that $V(p)=\{w_1, w_3\}$ and $V(q)=\{w_2, w_3\}$. Observe that at $w_4$ no propositional symbol holds.
\end{exam}

Now we are ready to define the semantics for the basic modal language, since we already have both the syntax and the structures the language is going to talk about. Recall that modal logics describe Kripke structures \textit{from an internal perspective}. This means that, in contrast with first order logic in which formulas see models from some kind of omniscient lookout point, modal formulas are evaluated \emph{at some particular point} of the model. 

\begin{defn}\label{def:ml-semantics}
Given the model $\model = \tup{W, \rels{\rel}, V}$ and $w \in W$, we inductively define the notion of a formula $\varphi$ being satisfied (or \textit{true}) in $\model$ at the point $w$ as follows:
$$
\begin{array}{rcl}
\model, w \models \top & &\mbox{always} \\
\model, w \models \bot & &\mbox{never} \\
\model, w \models p & \tiff & w \in V(p) \quad p \in \prop\\
\model, w \models \neg \varphi & \tiff & \model, w \not\models \varphi\\
\model, w \models \varphi \land \psi & \tiff & \model, w \models \varphi \mbox{ and } \model, w \models \psi \\
\model, w \models \varphi \lor \psi & \tiff & \model, w \models \varphi \mbox{ or } \model, w \models \psi \\
\model, w \models \varphi \to \psi & \tiff & \model, w \not\models \varphi \mbox{ or } \model, w \models \psi \\
\model, w \models \varphi \liff \psi & \tiff & \model, w \models \varphi \mbox{ if and only if } \model, w \models \psi \\
\model, w \models \diam{r}\varphi & \tiff & \mbox{there is a $w'$ such that $wR_rw'$} \mbox{ and } \model, w' \models \varphi\\
\model, w \models [r] \varphi & \tiff & \mbox{for all $w'$ such that $wR_rw'$}, \model, w' \models \varphi\\
\end{array}
$$
\end{defn}

Given a model $\model$, we say that $\varphi$ is \emph{globally satisfied} (or \emph{globally true}) on $\model$, and
write $\model \models \varphi$, if for all points $w$ in the domain of $\model$ we have that $\model,w \models \varphi$. A formula $\varphi$ is \emph{universally valid} if it is globally satisfied in all models, and in that case we write $\models \varphi$. A formula $\varphi$ is \emph{satisfied in a model} $\model$ when there is a point in $\model$ where $\varphi$ is true, and $\varphi$ is \emph{satisfiable} if there is some point in some model at which it is satisfied. When working with sets of formulas, these definitions are lifted in the expected way.

\section{Model equivalence}
\label{sec:modelequiv}
Let $\model$ and $\model'$ be two models for a logic \cL, and $w$ and $w'$ be two points in $\model$ and $\model'$ respectively. We say that $w$ and $w'$ are \cL-equivalent (notation: $w \theq{\cL} w'$) if they make the same \cL-formulas true.\footnote{When the logic is clear from context we don't add the subscript \cL.} This means that, although the models may be different, if we look at $w$ and $w'$ ``through the glasses of the logic \cL'' they are indistinguishable. Consider now the following two models.

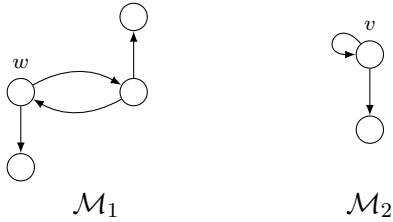
\begin{figure}[h]
\centering
\begin{tikzpicture}[>=latex]
  \node (n1) at (0,0) [point] {} ;
  \node (n2) at (1.5,0) [point] {} ;
  \node (n3) at (0,-1) [point] {} ;
  \node (n4) at (1.5,1) [point] {} ;

  \node [above] at  (n1.north) {\scriptsize $w$};

  \draw (n1) edge [->, bend left] (n2);
  \draw (n2) edge [->, bend left] (n1);
  \draw (n1) edge [->] (n3);
  \draw (n2) edge [->] (n4);
  
  \node at (1,-1.5) {$\model_1$};
\end{tikzpicture}
\hspace{2cm}
\begin{tikzpicture}[>=latex]
  \node (m1) at (0,0.5) [point] {} edge [edge, in=180, out=130, loop]  ();
  \node (m2) at (0,-0.5) [point] {} ;

  \node [above] at  (m1.north) {\scriptsize $v$};

  \draw (m1) edge [->] (m2);

  \node at (0,-1.5) {$\model_2$};
\end{tikzpicture}
\caption{The points $w$ and $v$ are \bml equivalent.}
\label{fig:mod2eq}
\end{figure}

Let us consider the basic modal language. Assuming that $V(p) = \emptyset$ in both models for all $p \in \prop$, is there a way to distinguish $w$ from $v$ in $\bml$? That is, is there a basic modal formula that is true at $w$ and false at $v$? It doesn't seem to be easy to find one. On the other hand, if we can use first order logic this is quite straightforward: the formula $\lnot R(x,x)$ is true if we assign $w$ to $x$, and false in the case of $v$.

\subsection*{Equivalence as a structural notion}

One could pick two pointed models $\tup{\model,w}$ and $\tup{\nodel,v}$ and ask wether they are \cL-equivalent for a given logic \cL without checking every possible formula. For example, in Figure~\ref{fig:mod2eq}, we would like to know if there is a \emph{structural} relationship between the models that makes them equivalent for \bml.

In classical first-order logic this relationship corresponds to \emph{potential isomorphisms}, which is defined as follows in~\cite{CK90}.\footnote{In the literature, such as~\cite{CK90}, potential isomorphism are sometimes called `partial isomorphism' because they are formed of sequences of isomorphism with restricted domain.}
\begin{defn}[Potential isomorphism]
Let $\fmodel$ and $\fnodel$ be first order models with domains $M$ and $N$ respectively. A \emph{potential isomorphism} between $\fmodel$ and $\fnodel$ is a relation $Z$ on the set of pairs of finite sequences $\tup{a_{1},\dots,a_{n}}$, $\tup{b_{1},\dots,b_{n}}$ of elements of $A$ and $B$ of the same length such that:
\begin{enumerate}[(i)]
\noitemsep
\item $\emptyset \;Z\; \emptyset$.
\item If $\tup{a_{1},\dots,a_{n}} \;Z\; \tup{b_{1},\dots,b_{n}}$ then $(\fmodel,a_{1},\dots,a_{n})$ and $(\fnodel,b_{1},\dots,b_{n})$ satisfy the same \emph{atomic} formulas.
\item If $\tup{a_{1},\dots,a_{n}} \;Z\; \tup{b_{1},\dots,b_{n}}$ then for all $c \in M$ there exists $d \in N$ such that $\tup{a_{1},\dots,a_{n},c} \;Z\; \tup{b_{1},\dots,b_{n},d}$ and vice versa.
\end{enumerate}
We use $\fmodel \pisom \fnodel$ to note that there exists a potential isomorphism between $\fmodel$ and $\fnodel$. Observe that potential isomorphism relations are symmetrical, that is, $\fmodel \pisom \fnodel$ if and only if $\fnodel \pisom \fmodel$.
\end{defn}

Given two first-order models $\fmodel,\fnodel$ if $\fmodel \pisom \fnodel$ then the models are indistinguishable by first-order logic~\cite[Proposition 2.4.4]{CK90} (they are also called \emph{elementary equivalent}).

In the modal domain, take the basic modal logic as an example, the notion of \emph{bisimulation} between models gives a structural notion which implies that the models are equivalent when looking through the glasses of basic modal logic~\cite[Section 2.2]{MLBOOK}. For a detailed historical insight on bisimulation we recommend~\cite{S08}.

Let's take a look at the definition of bisimulation for the basic modal logic. We will give the definition for the monomodal version of \bml because its simplicity is suitable for this introduction but all the definitions and results of this chapter also hold for the multimodal case.

\begin{defn}[Bisimulation for \bml]\label{def:bml-bisim}
A \emph{bisimulation} between two \bml models ${\model = \tup{W, R, V}}$ and ${\model'=\tup{W', R', V'}}$ is a non-empty binary relation $Z \subseteq W \times W'$ between their domains such that whenever $wZw'$ we have that: 

\smallskip

\textbf{Atomic harmony}:  $w$ and $w'$ satisfy the same propositional symbols. 

\textbf{Forth}: if $wRv$, then there exists a point $v'$ in $\model'$ such that $vZv'$ and $w'R'v'$.

\textbf{Back}: if $w'R'v'$, then there exists a point $v$ in $\model$ such that $vZv'$ and $wRv$. 

\medskip
\noindent If there is a bisimulation between two models $\model$ and $\model'$ we say that $\model$ and $\model'$ are bisimilar and we write $\model \bisim \model'$. Moreover, we say that two points $w \in \model$ and $w' \in \model'$ are bisimilar if they are related by some bisimulation, and we write $\model, w \bisim \model', w'$. We write $w \bisim w'$ when the models are clear from context.
\end{defn}

Returning to the models $\model_1$ and $\model_2$ we have just presented in Figure~\ref{fig:mod2eq}, it is easy to see that $\model_1, w \bisim \model_2, v$. The bisimulation would be as follows (the dotted line indicates the pairs in the bisimulation relationship):

\begin{figure}[h]
\centering
\begin{tikzpicture}[>=latex]
  \node (n1) at (0,0) [point] {} ;
  \node (n2) at (1.5,0) [point] {} ;
  \node (n3) at (0,-1) [point] {} ;
  \node (n4) at (1.5,1) [point] {} ;

  \node [above] at  (n1.north) {\scriptsize $w$};

  \draw (n1) edge [->, bend left] (n2);
  \draw (n2) edge [->, bend left] (n1);
  \draw (n1) edge [->] (n3);
  \draw (n2) edge [->] (n4);

  \node (m1) at (4,0) [point] {} edge [edge, in=0, out=50, loop]  ();
  \node (m2) at (4,-1) [point] {} ;
  \node [above] at  (m1.north) {\scriptsize $v$};

  \draw (m1) edge [->] (m2);

  \draw (n1) edge [bend right=40, dashed] (m1);
  \draw (n2) edge [bend right=20, dashed] (m1);
  \draw (n3) edge [bend right, dashed] (m2);
  \draw (n4) edge [bend right, dashed] (m2);
\end{tikzpicture}
\caption{Bisimilar models.}
\end{figure}
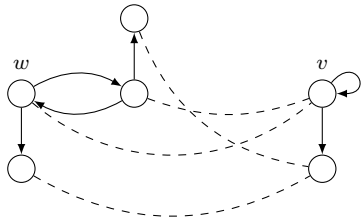

The definition of bisimulation we just gave is specifically designed for the basic modal logic, and thus the expected property is that satisfiability of formulas in the basic modal logic is invariant under bisimulations as proved in \cite{MLBOOK}.

\begin{thm}\label{thm:bisimpreserv}
Let $\model$ and $\model'$ be two Kripke models over the same signature. Then, for every $w \in \model$ and $w' \in \model'$, if $w \bisim w'$ then for every formula $\varphi$ of \bml, $\model,w \models \varphi$ if and only if $\model',w' \models \varphi$.
\end{thm}

The aforementioned logics, namely \bml and first order logic, have both negation and disjuction in their languages and both ``model equivalence'' notions (bisimulation and potential isomorphisms) are symmetrical. In many areas of computer science one finds logical formalisms that lack some of the 
standard Boolean connectives `and', `or' and `not'. In particular, negation-free logics are widely used in areas as diverse as semantics of programming and knowledge representation. In some applications, such as the generation of referring expressions~\cite{AKS08}, Boolean negation may be unnatural.

Take now the basic sub-boolean modal logic, \bmlsn, which is defined as \bml but \emph{doesn't} have negation nor the $\Box$ modality. As the language is weaker, the notion of model equivalence should change. Bisimulations, for example, are too strong for negation-free logics because they preserve negation. In~\cite[Definition 2.77]{MLBOOK} we can find the definition of the concept of \emph{simulation} for negation-free logics. If there is a simulation from $\model,w$ to $\nodel,v$ then every formula true at $\model,w$ is also true at $\nodel,v$.  The formal definition is as follows.

\begin{defn}[Simulation for \bmlsn]\label{def:bmlsn-sim}
A \emph{simulation} between two models $\model = \tup{W, R,$ $V}$ and $\model'=\tup{W', R', V'}$ is a non-empty binary relation $Z \subseteq W \times W'$ between their domains such that whenever $wZw'$ we have that: 

\medskip

\textbf{Atomic condition}: If $w \in V(p)$ then $w' \in V(p)$ for all $p \in \prop$.

\textbf{Forth}: if $wRv$, then there exists a point $v'$ in $\model'$ such that $vZv'$ and $w'R'v'$.


\medskip
\noindent If there is a simulation between two models $\model$ and $\model'$ we write $\model \rsim \model'$. Moreover, we say that two points $w \in \model$ and $w' \in \model'$ are similar if they are related by some simulation, and we write $\model, w \rsim \model', w'$. We write $w \rsim w'$ when the models are clear from context.
\end{defn}

Observe that, in this case, half of the ``Atomic harmony'' condition has been removed. Another point to be taken into account is that even though \bml's bisimulation is symmetrical, simulations need not to be. This notion is specially suited for \bmlsn and preserves every formula formed from $\land, \lor$ and $\Diamond$. The following theorem  states this formally.

\begin{thm}
Let $\model$ and $\model'$ be two Kripke models over the same signature. Then, for every $w \in \model$ and $w' \in \model'$, if $w \rsim w'$ then for every formula $\varphi$ of \bmlsn; $\model,w \models \varphi$ implies $\model',w' \models \varphi$.
\end{thm}

As the notion of simulation is less restrictive than the notion of bisimulation it should be no surprise to find models which are similar but not bisimilar. Take, for example, the following two models.

\begin{figure}[h]
\centering
\begin{tikzpicture}[>=latex]
  \node (n1) at (0,0) [point] {};
  \node (n2) at (1,0.5) [point] {};
  \node (n3) at (1,-0.5) [point] {};

  \node [left] at  (n1.west) {\scriptsize $w_0$};
  \node [below] at  (n2.south) {\scriptsize $p$};
  \node [below] at  (n3.south) {\scriptsize $p,q$};
  
  \draw[->] (n1) edge (n2);
  \draw[->] (n1) edge (n3);
    
  \node (m1) at (5,0) [point] {};
  \node (m2) at (4,1) [point] {};
  \node (m3) at (4,0) [point] {};
  \node (m4) at (4,-1) [point] {};

  \node [right] at  (m1.east) {\scriptsize $v_0$};
  \node [below] at  (m2.south) {\scriptsize $p$};
  \node [below] at  (m3.south) {\scriptsize $p,q$};
  \node [below] at  (m4.south) {\scriptsize $p,r$};
  
  \draw[->] (m1) edge (m2);
  \draw[->] (m1) edge (m3);
  \draw[->] (m1) edge (m4);

  \draw[->] (n1) edge [bend left=100, dashed] (m1);
  \draw[->] (n3) edge [dashed] (m3);
  \draw[->] (n2) edge [dashed] (m3);
  
  \node at (0.0,2.0) {$\model_{1}$};
  \node at (5.0,2.0) {$\model_{2}$};
\end{tikzpicture}
\caption{Similar but not bisimilar.}
\end{figure}
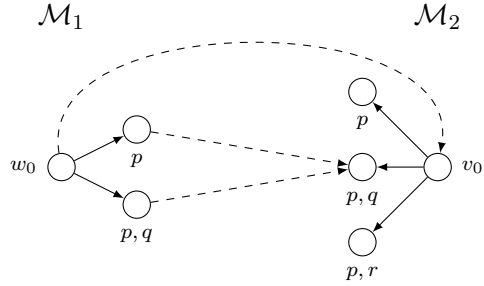

Again, the dashed lines indicate the pairs in the simulation relation. We can see, that $\model_{1}, w_{0} \rsim \model_{2},v_{0}$, on the other hand, there is no bisimulation linking them. To show this, it is enough to exhibit a formula $\varphi$ such that $\model_{2},v_{0} \models \varphi$ and $\model_{1}, w_{0} \not\models \varphi$. In this case, a possible formula is $\Diamond r$.

\subsection*{Equivalence as a game}

These notions of model equivalence can also be presented using a more dynamic perspective, closer to a form of process equivalence. For example, the task of determining whether two models are bisimilar can be recast in the form of an Ehrenfeucht-Fra\"iss\'e game~\cite{ebbi:math84}. 

Let $\tup{\model_1, w_1}$ and $\tup{\model_2, w_2}$ be two pointed models. An \emph{Ehrenfeucht-Fra\"iss\'e game for the basic modal logic} is defined as follows. There are two players called \emph{Spoiler} and
\emph{Duplicator}. Intuitively, Spoiler tries to devise a property true in one model and false in the other. Conversely, Duplicator tries to `copy' the property from one model to the other by imitating Spoiler's movements.

The two players compare successive pairs, starting from $(\model_1, w_1)$ and $(\model_2, w_2)$. Duplicator immediately loses if $w_1$ and $w_2$ do not coincide in the propositional symbols. Otherwise, the game starts, with the players moving alternatively. Spoiler always makes the first move of the game. In a turn of the game, Spoiler starts by choosing in which model he will make a move. After that, he chooses a point which is a successor of the current $w_1$ or $w_2$, and Duplicator responds with a matching successor in the other model. If the chosen points differ in the atomic propositions, Spoiler wins. If one player cannot move, the other wins. Duplicator wins on infinite runs.

Note that with this definition, exactly one of Spoiler or Duplicator wins
each game. A \emph{strategy for Duplicator} is a function that takes a valid state of the game (i.e. a pair $\tup{a,b}$ with $a \in |\model_{1}|$ and $b \in |\model_{2}|$) and returns a possible next move for Duplicator. A strategy for Spoiler is defined in the same way but note that the function should also return the model in which Spoiler should make the move. We say that a player is \emph{following a strategy $\sigma$} when all his moves in a game comply with the answer of $\sigma$ for every stage of the game. A strategy is \emph{winning} if the player following it necessarily wins the game, no matter what his opponent plays. Given two pointed models $\tup{\model_1,w_1}$ and $\tup{\model_2,w_2}$ we will write $\tup{\model_1,w_1} \efeq_{b} \tup{\model_2,w_2}$ when Duplicator has a winning strategy for the game.

Intuitively, this game captures exactly the zigzag behavior of bisimulations, and the atomic harmony condition. The two notions are equivalent, but depending on the context, one can be more natural than the other.

\begin{propo}\cite{goranko05}
Let $\tup{\model_1,w_1}$ and $\tup{\model_2,$ $w_2}$ be two \bml pointed models, then $\tup{\model_1,w_1} \efeq_{b} \tup{\model_2,w_2}$ if and only if $\tup{\model_1,w_1} \bisim \tup{\model_2,w_2}$.
\end{propo}

The perspective of model equivalence as a game is not restricted to the basic modal logic. With minor modifications to this notion of game we can create a notion that is suitable for \bmlsn, for instance. An \emph{Ehrenfeucht-Fra\"iss\'e game for \bmlsn} is the same as the game for the basic modal logic \emph{but} spoiler can't choose the model where he is playing. That is, Spoiler starts playing in $\model_{1}$ and he won't be able to change to $\model_{2}$.

Suppose that Spoiler and Duplicator start a game standing in $(\model_{1},w_{1})$ and $(\model_{2},w_{2})$ respectively. If Duplicator has a winning strategy then $\model_{1},w_{1} \rsim \model_{2},w_{2}$. On the other hand, if Spoiler has a winning strategy then $\model_{1},w_{1} \not\rsim \model_{2},w_{2}$. We will write $\tup{\model_1,w_1} \efeq_{s} \tup{\model_2,w_2}$ when Duplicator has a winning strategy for the game. The following proposition states the equivalence between \bmlsn simulation and the game definition.

\begin{propo}
Let $\tup{\model_1,w_1}$ and $\tup{\model_2,$ $w_2}$ be two \bmlsn pointed models, then $\tup{\model_1,w_1} \efeq_{s} \tup{\model_2,w_2}$ if and only if $\tup{\model_1,w_1} \rsim \tup{\model_2,w_2}$.
\end{propo}

Summing up, simulations and bisimulations are very powerful tools to measure the expressivity of a logic: they provide us with \emph{structural} conditions on the models that characterizes the appropriate structure preserving morphisms. Since simulations are directly linked to the expressivity of a given logic, there is not a \emph{unique} notion of simulation. Here we have just presented the notion of bisimulation for \bml and simulation for \bmlsn, but for \emph{every} logic we need to find a suitable definition, and this notion will reflect the logic's expressive power. In this sense, looking for the appropriate notion of model equivalence allows us to learn about the logic we are working with.

\section{Saturation}
We have discussed a lot about model equivalence notions, in particular about simulations and bisimulations, but we have been avoiding a fundamental question. Let's focus on bisimulations, we know that $\model,w \bisim \nodel,v$ implies $\model,w \theq{} \nodel,v$. Does, in general, the converse hold? That is, does $\model,w \theq{} \nodel,v$ imply $\model,w \bisim_{} \nodel,v$? The answer is \emph{no}. Consider the following two models:

\begin{center}
\includegraphics{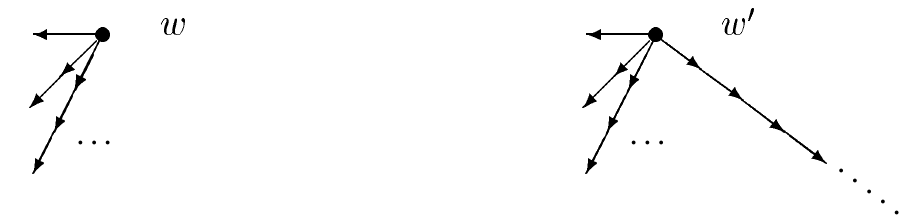}\\
\hspace{-1.5cm}$\model_{1}$\hspace{5cm}$\model_{2}$
\end{center}

It can be shown that, although $w$ and $w'$ satisfy the same \bml formulas, there is no possible bisimulation between them. Recall that in Section~\ref{sec:modelequiv} we presented an alternative interpretation of bisimulations as games. We will now use that notion to prove that these models are not bisimilar.

Set a game between Duplicator and Spoiler with them starting at $(\model_{1},w)$ and $(\model_{2},w')$ respectively. The first turn is for Spoiler. He chooses to stay in $\model_{2}$ and move to the successor of $w'$ that lays in the infinite branch of the model. Now it is Duplicator's turn, he must move to a matching world in $\model_{1}$. As the \emph{atomic harmony} condition is trivially satisfied by any two pairs of these models, the only problem could arise if Spoiler makes a move and Duplicator has no possible successors to move to. Duplicator has to choose a branch in $\model_{1}$ and move to it. Suppose, without loss of generality, that Duplicator chooses a branch with $k$ nodes. You can easily see that, as there is no going back, if Spoiler carries on moving in his infinite branch there will be a moment (after $k$ moves) when Duplicator hits the end of his branch in $\model_{1}$. In that moment Spoiler wins the match.

The strategy we've described guarantees that Spoiler will win no matter what Duplicator chooses. Therefore, as Spoiler has a winning strategy the models are not bisimilar. In fact, as Spoiler starts the game in $\model_{2}$ and never changes the model this argument proves that $\model_{2},w' \not\rsim \model_{1},w$ which is a stronger result.

Do not panic. There \emph{are} some classes of models where modal equivalence implies bisimilarity. A very useful one is the class of \wsat models. In order to present this class we need some previous definitions.

The following notions will be given in terms of first order models and not \bml models but this shouldn't carry any problem. Later, in Section~\ref{sec:charact}, we will see that there is a straightforward formalization that lets us think of a model as a \bml or first order model interchangeably.

\begin{nota}
We will use $\varphi(x)$ to note first order formulas with at most one free variable $x$ this notation extends to sets $\Gamma(x)$ as expected.
The notation $g[x \mapsto w]$ denotes a valuation $g'$ that is the same as $g$ on every parameter except on $x$ where $g'(x) = w$. Given a first order formula $\varphi(x)$ we will hereafter note $\fmodel \models \varphi(x)[w]$ to mean $\fmodel,g[x\mapsto w] \models \varphi(x)$. Observe that, as $\varphi$ has only one free variable $x$ the valuation will be irrelevant. Given a model \model we use $|\model|$ to denote the domain (or universe) of \model.
\end{nota}

\begin{defn} A set of first order formulas with (at most) one free variable is called a \emph{type}. Given a model $\fmodel$ we say that a type $\Gamma(x)$ has a \emph{witness} if there exists a state $w$ such that for every formula $\varphi(x) \in \Gamma(x)$ we have $\fmodel \models \varphi(x)[w]$. A type is \emph{finitely realizable} if every finite subset has a witness.
\end{defn}

\begin{defn}[Expansion] Let $\fmodel$ be a first order model with domain $W$. For $A \subseteq W$, the \emph{expansion} of $\cFO$ with $A$ (noted $\cFO[A]$) is obtained by extending $\cFO$ with new constants $\underline{a}$ for every element $a \in A$. The model $\fmodel_{A}$ is the same as $\fmodel$ but interprets the constants as expected.
\end{defn}

We are now ready to define $\omega$-saturation. Informally, it resembles some kind of `intra-model' compactness. That is, given a type $\Gamma(x)$ if every finite subset is satisfied in (possibly different) elements in $\fmodel$ then there is \emph{a single element} in $\fmodel$ which satisfies the whole set. Formally speaking the definition is as follows.

\begin{defn}[$\omega$-saturation]\label{def:omegasat} A first order model $\fmodel$ is called \emph{\wsat} if for every finite $A \subseteq |\fmodel|$ the expansion $\fmodel_{A}$ has a witness for every type $\Gamma(x)$ that is finitely realizable in $\fmodel_{A}$.
\end{defn}

In the beginning of this section we presented two models. One of them had branches of increasing length, the other one was an exact copy of the first but with an extra \emph{infinite} branch. We have already seen that, in some sense, the first model was `lacking' something that the second one had. The saturation that \wsat models have make them complete in this sense. The following theorem is a very important result which gives strength to the class of \wsat models.
\begin{thm}[\cite{MLBOOK}]\label{thm:bml:wsatbisim}
Let $\tup{\model,w}$ and $\tup{\nodel,v}$ be two \wsat models,
$$\text{If } \model,w \theq{} \nodel,v \text{ then } \model,w \bisim{\;} \nodel,v.$$
\end{thm}
Therefore, in the \wsat class, bisimulation and \bml equivalence coincide. This proof strongly uses the structural definition of bisimulations and thus we will not get into this kind of detail until we need it in Chapter~\ref{chap:app}.

A particularly interesting fragment of the \wsat class is the \emph{finitely branching} fragment, that is, every world has only finitely many successors. Another different (and more restrictive) example of \wsat class is the class of finite models.

To finish this section we want to say some final words about \wsat models. The use of \wsat models will be crucial to prove the results in this thesis. Because of their special properties one could think that these models are rather scarce but, fortunately, they abound. Moreover, there is a standard way of, given an \cFO-model $\fmodel$, construct an \wsat model $\fmodel_{*}$ such that $\fmodel \theq{\cFO} \fmodel_{*}$. This theorem is stated as Theorem~\ref{thm:satultra} and proved in the Appendix.

\section{What this thesis is about}\label{sec:aboutthesis}
For a wide spectrum of applications, which use logic as a tool, first order logic is enough to \emph{theoretically} solve their problems. However, complications arise when we consider the behavior of first order logic in \emph{practice}. First of all, first order logic is \emph{undecidable}, that is, there is no algorithm to decide whether an arbitrary formula is a satisfiable. Second, in general, most applications do not use the entire expressive power that first order gives. Therefore, even when working in decidable fragments of first order logic, they may be paying an excessive payload for things they will not be using.

Modal logics are very good at molding themselves to fit a particular purpose. If you know what you need it is most likely that you can end up with a modal logic which has exactly the required expressive power but with better properties than first order logic in terms of complexity and decidability. For example, \bml is decidable and has a \mbox{PSPACE-complete} satisfiability problem.

Along this thesis we will work with logics that are less (or equally) expressive than first order logic. Informally, this means that if one can express a property with a modal formula then there is always some way to express the same property in first order. In other words, one can say that if a modal formula $\varphi$ denotes some property then there exists some kind of \emph{translation} to a first order formula $\varphi^{t}$ which denotes the same property.

Johan van~Benthem studied the connection between modal and first order logic~\cite{benthem84}. One of his best known results in this area is the `Characterization Theorem' which identifies \bml as the bisimulation-invariant fragment of first order logic. Informally, one can state the theorem as follows.

\begin{thm*}
A first order formula $\alpha$ is equivalent to the translation of a \bml formula if and only if $\alpha$ is invariant under bisimulations.
\end{thm*}

Note that in this case the notion of bisimulation is that of \bml. As we have said before, every modal logic should have a potentially different notion of bisimulation. For example, we have already seen the notion of simulation for \bmlsn. Using this notion, Kurtonina and de Rijke proved that \bmlsn is the simulation-invariant fragment of first order logic.

\bml is just the tip of the iceberg, there exist plenty of extensions of \bml to suit particular needs. Many modal logics admit a translation to first order logic and a characterization of this kind has been given for some of them. One essential problem is that there seems to be no general proof and every case needs a new ad-hoc proof.

The birth of the concept of bisimulation and correspondence theory helped answer new questions from a purely model-theoretic perspective. One example is the characterization of modal \emph{definability}. Informally, we say that a class of models is definable by a set of formulas $\Gamma$ if it is composed of exactly all the models where $\Gamma$ is valid. A class is definable by a single formula if it is definable by a singleton set.

It would be interesting to know which properties should a class of models satisfy in order to be defined by a modal formula or by a set of modal formulas. This question had previously been stated and answered for classical first order logic~\cite{CK90}. Whereas the answer for first order logic is presented in terms of \emph{potential isomorphisms}, in the case of modal logics, the notion of bisimulation plays an essential role. To uncover the panorama we cite the following result for \bml which can be found in~\cite{MLBOOK}.

\begin{thm*}
A class of models \classK is definable by means of a single \bml formula if and only if both \classK and \CclassK are closed under bisimulations and ultraproducts.
\end{thm*}

Do not worry about what `ultraproducts' means right now. They will be introduced when needed. Just bear in mind that they are useful model construction tools (with algebraic roots) which first caught the attention of logicians because they could be used to give a purely algebraic proof of the Compactness Theorem for first order logic. For a detailed survey on ultraproducts we recommend~\cite{KULT08}.

As with the characterization theorem, definability results similar to the one presented here also hold for a vast number of modal logics. Similarly, every logic has a proof that is specially crafted for that case.

\bigskip
Clearly, characterization and definability results help us to better understand a logic. Interestingly, these results also have a great impact in practical computer science.

Consider the following problem: Suppose you are into model checking, that is, given a model of a system, test automatically whether this model meets a given specification. Suppose that the specification can be written as a first-order formula $\varphi$.

You could always use first-order tools to check if the model satisfies $\varphi$ but that can result in a high complexity cost as we have already mentioned. One could try to see if there are `cheaper' logics that can be used to solve the problem. If we can express the same property in some modal logic we may be able to drastically optimize the process.

Let's discuss a concrete example. Suppose that the points in our domain model the different states in the execution of a program and there is a transition from one point to another if there is a possible transformation that brings state $a$ into $b$. Thinking of the model in this way would imply that states without successors (also called endpoints) are states were the program has halted.

One possible property to be checked could be ``every point should be able to immediately `escape' from the flow of execution'', meaning that, every point should be able to directly see an endpoint. This property can be verified by checking that the first order formula $$\varphi(x) = \exists y. R(x,y) \to (\exists z. R(x,z) \;\land\; \forall w. \lnot R(z,w))$$ is valid in the model but it can also be checked by verifying that the \bml formula ${\psi = \Diamond\top \to \Diamond\Box\bot}$ is also valid in the model. As we have an equivalent \bml formula, we can now use model checkers that accept \bml formulas to solve our problem.

Apart from being more `user friendly', the sole existence of the formula $\psi$ tells us that the property is invariant under bisimulations and this information bears an extra benefit. Suppose that the model is automatically generated from a piece of code. If, for example, we feed the generator with the code of an entire operating system, the resulting model will be \emph{very} large.

It is not the purpose of this thesis to get into this topic but there are (efficient) algorithms to automatically minimize the model which preserve the truth of formulas invariant under bisimulations~\cite{H71, G73}. Therefore one could first minimize the model and then model check over the resulting model which will most likely be small with respect to the original one.

On the other hand, suppose now that we want to check whether the model is `irreflexive', that is, no element is related with itself. If we interpret this property in the setting described above, it would mean that no state has the possibility to `hang' in itself.

In this case, although the property can be verified checking the validity of the first order formula $\lnot R(x,x)$ in the model, there is no \bml formula which does the job. This can be shown easily because `irreflexivity' is \emph{not} invariant under bisimulations. Moreover, \bml has the so-called \emph{tree model property} which means that every formula satisfiable in a model is also satisfiable in a model which is a tree. As a corollary we get that there is no \bml-formula characterizing irreflexivity, antisymmetry nor intransitivity.

Is this the end of modal logic? Are we condemned to use first-order logic in this case? Fortunately, the answer is \emph{no}. Although \bml can't express those properties, there are richer logics (which still lay below first order) which can do the job. For example, Hybrid Logics~\cite{MLBOOK} and the less known Memory Logics~\cite{AFFM08} which will be introduced later in this thesis.

\bigskip
To summarize, model theoretic results such as Characterization and Definability give important information about different logics. It is well known that the proofs of those results for several modal logics have, somehow, the same `taste'. A general proof for most modal logics below first order is still too ambitious. In this thesis we plan to isolate sufficient conditions for the characterization and definability theorems to hold in a wide range of logics. Along with these conditions we will prove that, whichever logic that meets them, satisfies both theorems. Therefore, one could give an unifying proof for logics with already known results. Moreover, one will be able to prove characterization and definability results for logics that have not yet been investigated. In both cases, it is only needed to check that a logic meets the requirements to automatically derive the desired results.

\chapter{Known results for BML}\label{chap:knownbml}
If we want to generalize a result we'd better understand how it works in specific cases. This chapter is devoted to sketching the proof of some theorems for \bml. This will be helpful to identify the main ideas in their proofs and, with them in mind, get ready to undertake a generalization.

\section{Characterization}
We have talked about van~Benthem's characterization theorem. We know that \bml is strictly less expressive than first order logc, therefore, there are some `statements' that you can make in first order logic which can't be made in \bml. Informally, the Characterization Theorem identifies which first order formulas have an equivalent formula in the language of \bml. More formally it is stated as follows.
\begin{thm*}[van~Benthem]
A first order formula $\alpha(x)$ with at most one free variable is equivalent to the translation of a \bml formula if and only if $\alpha(x)$ is invariant under bisimulations.
\end{thm*}
Some work is needed for this wording to be precise. First of all, we are comparing modal formulas with first order formulas. Also, implicitly, when we talk about two formulas being `equivalent', we are evaluating them in some model. That's a problem because \bml formulas are evaluated in Kripke models and first order formulas aren't.

For us to be able to do such comparison between \bml and first order logic we need to define a formula translation and a way to interpret every \bml model as a first order model and vice-versa.

For this chapter we will set the signature for \bml to be $\ssig = \tup{\prop,\rel}$ with $\prop = \{p_{1},p_{2},\dots\}$ and $\rel = \{R\}$ therefore we will use a single diamond. This restriction to the unimodal case is only to make this introduction simpler. All these results also hold for the multimodal case.

\begin{defn}[Standard Translation] The \emph{Standard Translation} function $ST_{x}$ takes a \bml formula and returns a first order formula with at most one free variable $x$. It is defined as follows.
\begin{center}
$
\begin{array}{rcl}
ST_x(p_{i}) & = & P_{i}(x) \hfill \text{where $p_{i} \in \prop$}\\
ST_x(\lnot\varphi) & = & \lnot ST_x(\varphi)\\
ST_x(\varphi \land \psi) & = & ST_x(\varphi) \land ST_x(\psi)\\
ST_x(\Diamond\varphi) & = & \exists y (xRy \land ST_y(\varphi)) \qquad\hfill \text{where $y$ is a fresh variable}
\end{array}
$
\end{center}
We only define the translation for a basic (and adequate) connective set. It extends to the full set of connectives as expected.
\end{defn}

In this definition, we can already see that first order formulas include relations $P_{i}$ and $R$, this may give us a hint to define the first order signature. A first order signature is a tuple $\tup{\frel,\fconst,\ffunc}$ where \frel is the set of relation symbols, \fconst is the set of constant symbols and \ffunc is the set of function symbols. In our case we define the first-order signature to be $\fsig = \tup{\{R,P_{1},P_{2},\dots\}, \emptyset,\emptyset}$.

\begin{defn}[First order model]
A \emph{first order model} over the signature \fsig is a tuple $$\fmodel = \tup{A,(R^{I})_{R \in \frel},(f^{I})_{f \in \ffunc},(c^{I})_{c \in \fconst}}$$ where $A$ is the (non empty) domain, each $R^{I}$ is the interpretation of the relation symbol $R$, each $f^{I}$ is the interpretation for the function symbol $f$ and each $c^{I}$ is the interpretation for the constant symbol $c$.

In general we add a superscript or subscript $f$ to first order models so it is easier to distinguish them from modal models at first sight. We use $e,w,v,\dots$ to refer to elements of the domain of some model and $g,h,\dots$ to refer to first order valuations.
\end{defn}

The crucial point now is to see that there is a bijection between \bml models over the signature \ssig and first order models over the signature \fsig. Given a \bml model ${\model = \tup{W,R,V}}$ we can think of it as a first order model defined as $$\fmodel = \tup{W,\{R,P_{1},P_{2},\dots\},\emptyset,\emptyset}$$ where $P_{i} = V(p_{i})$. With this definition one can easily see that, given a \bml model $\model$ and $w \in W$; $\model,w \models p_{i}$ if and only if $\fmodel \models P_{i}(x)[w]$.

\bigskip
On the other hand, observe that \emph{any} first order model in this signature should be of the form $\fmodel = \tup{W,\{R,P_{1},P_{2},\dots\},\emptyset,\emptyset}$ and one can therefore build a \bml model analogously.

As it is usual in the literature, we will use, for this chapter only, the same model and think of it as a \bml or first order model as convenient. Now we can state the theorem that links \bml with first order.
\begin{thm}[Truth preservation]\label{thm:bml:sttruth}
Let \model be a \bml model, $w\in|\model|$ and $\varphi$ be a \bml formula, $$\model, w \models \varphi \text{ if and only if } \model \models ST_{x}(\varphi)[w].$$
\end{thm}
This theorem states that for every \bml formula there is a first order formula which is true in exactly the same worlds, thus, they are equivalent. Now that we have this theorem at hand it becomes clearer that we can compare formulas and models between \bml and first order logic.


\bigskip
The Characterization Theorem is stated in terms of `bisimulations' and uses notions we haven't yet defined. To begin with, we copy the definition of \bml bisimulation given in Section~\ref{sec:intro:bml}.
\begin{defn}[Bisimulation]
A \emph{bisimulation} between two models $\model = \tup{W, R, V}$ and $\model'=\tup{W', R', V'}$ is a non-empty binary relation $Z \subseteq W \times W'$ between their domains such that whenever $wZw'$ we have that: 

\smallskip

\textbf{Atomic harmony}:  $w$ and $w'$ satisfy the same propositional symbols. 

\textbf{Forth}: if $wRv$, then there exists a point $v'$ in $\model'$ such that $vZv'$ and $w'R'v'$.

\textbf{Back}: if $w'R'v'$, then there exists a point $v$ in $\model$ such that $vZv'$ and $wRv$. 
\end{defn}
In particular, the theorem talks about first order formulas being \emph{bisimulation-invariant}. Van~Benthem defines this concept as follows:
\begin{defn}[Bisimulation invariance]\label{thm:bml:bisiminv}
A first order formula $\alpha(x)$ is \emph{invariant for bimulations} if for all \bml models $\model, \nodel$ and $w \in |\model|, v \in |\nodel|$ such that $\model,w \bisim \nodel,v$ the following holds: 
$$\model \models \alpha(x)[w] \text{ iff } \model \models \alpha(x)[v].$$
\end{defn}
Observe that, so far, we only know that \bml formulas \emph{are} invariant for bisimulations and we don't have a result regarding first order formulas. The result for \bml formulas was stated in Theorem~\ref{thm:bisimpreserv}. On the other hand, when talking about first order formulas, some may be invariant for bisimulations and some others not. The set of formulas that are invariant for bisimulations is exactly the one identified by the characterization theorem.

\bigskip
As an example, take the following two \bml-bisimilar models. The first model is a single reflexive point and the second one is isomorphic to $\tup{\nat,<}$.\footnote{Therefore \bml can't distinguish between a single reflexive point and the naturals. It is surprising that, as weak as it is, \bml is still useful in practice.} The dashed lines represent the pairs in the bisimulation relation.
\begin{center}
\begin{tikzpicture}[>=latex]
  \tikzstyle{p} = [circle,fill=black,inner sep=0pt,minimum width=7pt]

  \node[p] (m1) at (0,2) {} edge [edge, loop]  ();
  \node[left] at (m1.west) {$w$};
  \node[left] at (7,2) {$\model_{1}$};

  \node[p] (n1) at (0,0) {};
  \node[p] (n2) at (1,0) {};
  \node[p] (n3) at (2,0) {};
  \node[p] (n4) at (3,0) {};
  \node[p] (n5) at (4,0) {};
  \node    (nf) at (5.5,0) {$\dots$};
  \node[left] at (n1.west) {$w'$};
  \node[left] at (7,0) {$\model_{2}$};
  
  \draw[->] (n1) edge (n2);
  \draw[->] (n2) edge (n3);
  \draw[->] (n3) edge (n4);
  \draw[->] (n4) edge (n5);
  \draw[->] (n5) edge (nf);
    
  \draw (n1) edge [dashed] (m1);
  \draw (n2) edge [bend right, dashed] (m1);
  \draw (n3) edge [bend right, dashed] (m1);
  \draw (n4) edge [bend right, dashed] (m1);
  \draw (n5) edge [bend right, dashed] (m1);
\end{tikzpicture}
\end{center}
Now take the first order formula $\varphi(x) = R(x,x)$. This formula holds at an element of the domain if and only if it is reflexive. It is clear that $\model_{1} \models \varphi(x)[w]$ and $\model_{2} \not\models \varphi(x)[w']$. As there is a bisimulation $w \bisim{} w'$, this two models serve as a proof that reflexivity is \emph{not} invariant under bisimulations.

This also means that there is no possible \bml formula equivalent to $\varphi(x)$. Suppose that there exists a \bml formula $\psi$ whose translation is equivalent to $\varphi$. By Theorem~\ref{thm:bml:sttruth} we have that $\model_{1},w \models \psi$ iff $\model_{1} \models \varphi(x)[w]$ and $\model_{2},w' \models \psi$ iff $\model_{2} \models \varphi(x)[w']$. We can conclude that $\model_{1},w \models \psi$ and $\model_{2},w' \not\models \psi$. This contradicts Theorem~\ref{thm:bml:bisiminv} because, as it is a \bml formula, $\psi$ should be invariant under \bml bisimulations.

\bigskip
We have proved that as $\varphi(x)$ is not invariant under bisimulations it is not equivalent to the translation of any \bml formula. What we have done for one particular case, the characterization theorem proves for an arbitrary first order formula. Moreover, it also proves the converse. Now that we understand what we are trying to prove we are ready to begin with the proof itself.

\subsection*{The proof of the Characterization Theorem}
In this section we will skim through the proof of van~Benthem's Characterization Theorem. It is not the goal of this section to give a detailed proof but to review the main ideas that support it. For a detailed proof refer to~\cite[Section 2.6]{MLBOOK}.
\begin{thm*}
A first order formula $\alpha(x)$ with at most one free variable is equivalent to the translation of a \bml formula if and only if $\alpha(x)$ is invariant under bisimulations.
\end{thm*}
\begin{proof}[Left to right]
This direction is easy, we argue by contradiction. Suppose that $\alpha(x)$ is equivalent to the translation of a \bml formula $\varphi$ and it is not invariant under bisimulations. That is, there exist $\tup{\model, w}$ and $\tup{\nodel,v}$ such that $\model, w \bisim \nodel,v$ but $\model \models \alpha(x)[w]$ and $\nodel \not\models \alpha(x)[v]$.

Using Theorem~\ref{thm:bml:sttruth} we get that $\model,w \models \varphi$ and $\nodel,v \not\models \varphi$. As we have a bisimulation linking those points and $\varphi$ is a \bml formula this drives us to a contradiction to Theorem~\ref{thm:bisimpreserv}. Absurd.
\end{proof}

\begin{proof}[Right to left]
All the magic is in the proof of this direction. Suppose that $\alpha(x)$ is invariant under bisimulations. Define the `modal consecuences of $\alpha$' as follows.
\begin{equation*}
MOC(\alpha) = \setcomp{ST_{x}(\varphi)}{\varphi \text{ is a \bml formula and } \alpha(x) \models ST_{x}(\varphi)}
\end{equation*}
It is trivial (by definition) that $\alpha(x) \models MOC(\alpha)$. As $MOC(\alpha)$ is formed by the translation of \bml formulas, if we can show that $MOC(\alpha) \models \alpha(x)$ then we are done. We first sketch the proof for this statement and then carry on.

Suppose that $MOC(\alpha) \models \alpha(x)$, by compactness of first order logic there exists a finite subset ${\Delta \subseteq MOC(\alpha)}$ such that $\Delta \models \alpha(x)$. We therefore have $\models \bigwedge\Delta \liff \alpha(x)$. As every formula in $\Delta$ is the translation of a \bml formula and $ST_{x}(\varphi \land \psi) = ST_{x}(\varphi) \land ST_{x}(\psi)$ we can conclude that $\bigwedge\Delta$ is also the translation of some \bml formula. Therefore we have proved that $\alpha(x)$ is equivalent to the translation of a \bml formula.

Hence, it all boils down to proving that $MOC(\alpha) \models \alpha(x)$. Assume that an arbitrary model satisfies $\model \models MOC(\alpha)[w]$; we need to show that $\model \models \alpha(x)[w]$. The proof goes as follows (we now focus on the ideas and then provide more detailed steps):
\begin{enumerate}
\item We first `create' a new model $\tup{\nodel,v}$ such that $\model,w \theq{} \nodel, v$ and $\nodel,v \models \alpha(x)$. We would like to transfer the validity of $\alpha(x)$ in $\nodel,v$ to $\model,w$.
\item Using standard model theoretic tools (that will be explained later) we construct, for $\model,w$ and $\nodel,v$, \wsat extensions $\model^{*},w^{*}$ and $\nodel^{*},v^{*}$ which are elementary equivalent to their originators. That is, they have the same first-order theory and they are \wsat. Observe that this implies $\model^{*},w^{*} \theq{} \nodel^{*},v^{*}$ and $\nodel^{*} \models \alpha(x)[v^{*}]$.
\item Using Theorem~\ref{thm:bml:wsatbisim} seen in Chapter~\ref{chap:intro}, as $\model^{*},w^{*} \theq{} \nodel^{*},v^{*}$ (and they are saturated) we have that $\model^{*},w^{*} \bisim \nodel^{*},v^{*}$.
\item Finally, as $\nodel^{*} \models \alpha(x)[v^{*}]$ and $\alpha(x)$ is invariant under bisimulations we get that $\model^{*} \models \alpha(x)[w^{*}]$. As $\model^{*}$ has the same first order theory that its originator we conclude that $\model \models \alpha(x)[w]$. \qedhere
\end{enumerate}
\end{proof}
That's it! Those are the main points to bear in mind. The whole idea is to make a `detour' through the class of first order \wsat models where bisimulation and equivalence do coincide. We can now proceed with the dissection of each point.

For the first point we do the following. Define the set $\Gamma$ as the translation of the \bml theory of $\model, w$, formally speaking
$$\Gamma = \setcomp{ST_{x}(\varphi)}{\varphi \text{ is a \bml formula and } \model \models ST_{x}(\varphi)[w]}.$$
We claim that $\Gamma \cup \{\alpha(x)\}$ is satisfiable. Suppose not, by first order compactness there is an unsatisfiable finite subset $\Gamma_{0} \subseteq \Gamma\cup\{\alpha(x)\}$. Observe that $\Gamma_{0} = \{\alpha(x),\gamma_{1},\dots,\gamma_{n}\}$ should include $\alpha(x)$. If it is unsatisfiable it means that $\models \alpha(x) \to \lnot\bigwedge\gamma_{i}$. Hence, $\lnot\bigwedge\gamma_{i} \in MOC(\alpha)$ because it is a modal consequence of $\alpha(x)$. Remember that one of our hypothesis was $\model \models MOC(\alpha)[w]$ therefore $\model \models \lnot\bigwedge\gamma_{i}$ but this is impossible since every formula in $\Gamma$ was true at $\model,w$ by definition. Absurd.

As $\Gamma \cup \{\alpha(x)\}$ is satisfiable we can say that there exists a model $\nodel$ and an element $v \in |\nodel|$ such that $\nodel \models \Gamma[v]$ and $\nodel \models \alpha(x)[v]$.

For the next three points the explanation given above should suffice. For further details we give the references where the theorems that we used are proved. For the second point we use Theorem~\ref{thm:satultra} of the Appendix on \model and \nodel and conclude exactly what we need. The theorems needed for the third point are already mentioned in the enumeration so there is nothing to add. The last point is the \emph{grande finale} where the validity of $\alpha(x)$ is transfered over the bisimulation to end up in $\model, w$.

Observe that this proof works for \bml which is a logic that has negation and disjunction. For negation-free logics the proof needs to be changed a little and for logics lacking disjunction the proof really changes a lot. In~\cite{KR97} you can find proofs for these languages.

In Chapter~\ref{chap:framework} we propose a framework to generalize this proof. Although the proof developed in this chapter looks pretty simple, in every step it makes use of a lot of suppositions that we may not be aware of. In a general scenario we will be working with an arbitrary translation, an arbitrary signature, an almost unknown model structure, etc. Because of the amount of uncertainty that we will have, we will need to do more complex detours to take the flux of the proof to some better known landscape. Be sure to remember this proof when reading Chapter~\ref{chap:framework} and \ref{chap:main}. Going back and forth may be useful to understand the motivation for some definitions.

\section{Definability}
In Section~\ref{sec:aboutthesis} we presented a piece of one of the Definability results for \bml. In this section we start by defining the concepts needed to state the full result.

\begin{nota}
Let \classK be a class of models we write \CclassK to denote the complement of \classK with respect to the class of all models. This notation will be used for both modal and first order models.
\end{nota}

\begin{defn} Let \classK be a class of pointed \bml models.
\begin{enumerate}[(i)]
\item \classK is \emph{definable by a set of formulas} $\Gamma$ if and only if for every $\tup{\model,w}$ it holds that $\model,w \models \Gamma$ if and only if $\tup{\model,w} \in \classK$.
\item \classK is \emph{definable by a single formula} if it is definable by a singleton set.
\end{enumerate}
\end{defn}

\begin{thm}\label{thm:bml:def}
Let \classK be a class of \bml models.
\begin{enumerate}[(i)]
\item \classK is definable by means of \emph{a set} of \bml formulas if and only if \classK is closed under ultraproducts and bisimulations and \CclassK is closed under ultrapowers.
\item \classK is definable by means of \emph{a single} \bml formula if and only if both \classK and \CclassK are closed under bisimulations and ultraproducts.
\end{enumerate}
\end{thm}
As before, we have to define what the closure condition mean for this theorem to make sense. We start by giving an informal introduction to ultraproducts. For this section it is enough to think of ultraproducts as follows: Given a family of first order models $(\model_{i},w_{i})_{i\in I}$ we can combine them and get a resulting model which is called the \emph{ultraproduct}.\footnote{Strictly speaking, there is also another ingredient which is called `ultrafilter'. Consult Appendix B for further information on ultrafilters and ultraproducts. We recommend its lecture.} When every model in the family is the same we call the resulting model an \emph{ultrapower}.

This new model satisfies some nice properties that will be useful for us. We take one of them from Appendix B to illustrate the idea.
\begin{thm*} Let $\fmodel,w$ be the ultraproduct of $(\model_{i},w_{i})_{i\in I}$ and let $\Gamma$ be a set of first order formulas.
\begin{itemize}
\item If every $\model_{i},w_{i} \models \Gamma$ then $\model,w \models \Gamma$.
\item In the particular case of an ultrapower this implies that $\model_{i},w_{i} \models \varphi$ if and only if $\model,w \models \varphi$.
\end{itemize}
\end{thm*}

We are ready to define the closure under ultraproducts and ultrapowers. These definitions should \emph{only} be used for the special case of \bml. In the next sections we will need to redefine these notions to have a broader reach.
\begin{defn} A class \classK of pointed \bml models is \emph{closed under ultraproducts} if and only if, for every family of \bml models $(\model_{i})_{i\in I}$ with $\model_{i} \in \classK$ the ultraproduct of those models also belongs to \classK. The closure under ultrapowers is defined as expected.
\end{defn}

With respect to the closure under bisimulations, it is a lot easier to imagine what it means. We define the notion of closure under bisimulations for the special case of \bml bisimulations.
\begin{defn} Let \classK be a class of \bml models, it is \emph{closed under bisimulations} if and only if the following holds: For every $\tup{\model,w} \in \classK$, if $\tup{\nodel,v}$ is such that $\model,w \bisim{} \nodel,v$ then $\tup{\nodel, v} \in \classK$.
\end{defn}

\subsection*{The proof of the Definability Theorem}
In this section we give a sketch of the proofs for the \emph{right to left} directions of the theorem. A detailed version can be found in~\cite[Section 2.6]{MLBOOK}.

\begin{proof}[Right to left of (i)] Suppose that \classK is closed under ultraproducts and bisimulations and \CclassK is closed under ultrapowers. The main ideas to prove this theorem are the following:
\begin{enumerate}
\item Propose a set $\Gamma = \text{`theory of \classK'}$ as a candidate set of formulas defining \classK. Every model of \classK trivially makes $\Gamma$ true. For $\Gamma$ to define \classK we still need to prove the other half, that is: If $\model,w \models \Gamma$ then $\tup{\model, w} \in \classK$.
\item Take any $\model,w \models \Gamma$, we will get to a contradiction by assuming that $\tup{\model,w} \in \CclassK$. We start by showing that there is a model $\tup{\nodel,v} \in \classK$ such that $\model,w \theq{} \nodel,v$. Here we will use that \classK is closed under ultraproducts.
\item As we did in the proof of the characterization theorem, we construct, for $\model,w$ and $\nodel,v$, \wsat extensions $\model^{*},w^{*}$ and $\nodel^{*},v^{*}$ which are elementary equivalent to their originators. That is, they have the same first-order theory and they are \wsat. Observe, again, that this implies $\model^{*},w^{*} \theq{} \nodel^{*},v^{*}$ and $\nodel^{*} \models \alpha(x)[v^{*}]$.

One \emph{important} difference with respect to the proof of the characterization proof is that here we use that \classK and \CclassK are closed under ultrapowers and conclude that $\tup{\model^{*},w^{*}} \in \classK$ and $\tup{\nodel^{*},v^{*}} \in \CclassK$.
\item Using Theorem~\ref{thm:bml:wsatbisim} seen in Chapter~\ref{chap:intro}, as $\model^{*},w^{*} \theq{} \nodel^{*},v^{*}$ (and they are saturated) we have that $\model^{*},w^{*} \bisim \nodel^{*},v^{*}$.
\item Finally, as $\model^{*},w^{*} \bisim \nodel^{*},v^{*}$ and \classK is closed under bisimulations then $\tup{\nodel^{*},v^{*}} \in \classK$. Absurd, in point 3 we had said that $\tup{\nodel^{*},v^{*}} \in \CclassK$.
\end{enumerate}
As seen before, one of the central tricks is the detour through \wsat models. The details are as follows: In the first point the set should be defined as
$$\Gamma = \setcomp{\varphi}{\text{for every model } \tup{\adel, u} \text{ in \classK; }{ \adel,u \models \varphi}}.$$
For the second point, let $\Sigma = \setcomp{\varphi}{\model,w \models \varphi}$ be the theory of $\tup{\model,w}$. If we find a model in \classK that models $\Sigma$ then it will be \bml equivalent to $\tup{\model,w}$.

The proof in~\cite{MLBOOK} hand-crafts an ultraproduct of models to make this step but we will take a route which keeps us away from the inner works of ultraproducts. Suppose that there is no such set in \classK making true all of $\Sigma$. By Theorem~\ref{thm:compactness} there exists a finite subset $\Sigma_{0} \subseteq \Sigma$ that is not satisfiable in \classK. Then $\lnot\bigwedge\Sigma_{0}$ would be true in \classK. In particular, $\model, w \not\models \bigwedge\Sigma_{0}$. This is absurd because $\Sigma_{0}$ is a subset of $w$'s theory.

Therefore, there exists a model $\tup{\nodel,v} \in \classK$ such that $\nodel,v \models \Sigma$ which implies that $\model,w \theq{} \nodel,v$.

The third and fourth points are justified as in the characterization theorem and the fifth point is self-explanatory.
\end{proof}

\begin{proof}[Right to left of (ii)]
Suppose that both \classK and \CclassK are closed under bisimulations and ultraproducts. Using the first part of this theorem we know that there exist two sets $\Gamma_{1},\Gamma_{2}$ respectively defining \classK and \CclassK. It is clear that their union should be unsatisfiable because no model can be in \classK and \CclassK at the same time.

Using the compactness theorem, as $\Gamma_{1}\cup\Gamma_{2}$ is unsatisfiable there must be a finite subset $\Gamma_{0} \subseteq \Gamma_{1}\cup\Gamma_{2}$ which is unsatisfiable. Let $\Gamma_{0} = \{\alpha_{1},\dots,\alpha_{n},\beta_{1},\dots,\beta_{m}\}$ where $\alpha_{i} \in \Gamma_{1}$ and $\beta_{j} \in \Gamma_{2}$. As $\Gamma_{0}$ is unsatisfiable we can say that $\models \alpha_{1}\land\dots\land\alpha_{n}\to\lnot(\beta_{1}\land\dots\land\beta_{m})$. We show that it is exactly $\varphi = \alpha_{1}\land\dots\land\alpha_{n}$ that defines \classK.

Trivially every model in \classK satisfies $\varphi$. For the converse, take $\model, w \models \alpha_{1}\land\dots\land\alpha_{n}$, then $\model, w \not\models \beta_{1}\land\dots\land\beta_{m}$ therefore $\tup{\model,w} \notin \CclassK$ which means that $\tup{\model,w} \in \classK$.
\end{proof}

To close this section we want to draw attention to one of the hypothesis in (ii): the need for \emph{both} classes to be closed under bisimulations. Observe that, as the bisimulation relation is symmetric we could've just asked for either \classK or \CclassK to be closed under bisimulations and that would've been enough. One can prove that \classK is closed under bisimulations if and only if \CclassK is.

On the other hand, in the proof of (ii) from right to left, we \emph{strongly} use that both classes are closed under bisimulations to get two sets that define each of the classes. What would happen now if we were talking about simulations? As simulations are not necessarily symmetrical we can't be sure that \classK is closed under simulations if and only if \CclassK is. This fault brings problems if we want to follow this same proof scheme.

In~\cite{KR97, KRTEMPO} there are alternative proofs for this result for model equivalence notions that aren't symmetrical. None of them are general enough to fit the framework that we will develop but both have proved of great inspiration for the results given in Section~\ref{sec:main:def}.

\clearpage

\chapter{The generalized framework}
\label{chap:framework}
In this chapter we set up a proper framework which will aid us to prove generalized results for modal logics which lay (in terms of expressivity) below first order logic. We start by stating in which sense our results pursue a generalization. We will focus on the following two axes.

\begin{paragraph}{\bf Arbitrary modal logic}
\mbox{}\\\noindent We want to obtain characterization and definability results which hold for an arbitrary modal logic. Due to the broad spectrum of different logics we still have to stop somewhere. When we say `arbitrary' we mean any modal logic with conjunction and disjunction (interpreted as usual) which is interpreted over extensions of Kripke models.

These logics may come with different model equivalent notions. We want to be able to derive results no matter what the simulation or bisimulation relation looks like. We will only put constraints on the `arity' of the relation, that is, it should link an element from the domain of one model to an element of the domain of other model. It will later become clear that this last generalization comes with a great price to pay: we know nothing about the structural properties involved in this notion.
\end{paragraph}
\begin{paragraph}{\bf Relativization to a particular class of models}
\mbox{}\\\noindent The results presented in Chapter~\ref{chap:knownbml} were stated with respect to the class of \emph{all models}. That is, \bml is the fragment of first order formulas which are bisimulation invariant in the class of \emph{all first models}. Think of the following motivational example.

The `Basic Temporal Logic' is a modal logic which is defined as follows: Its language has the full boolean connective set and two modalities $F$ and $P$ which are often called `future' and `past'. The classical perspective on this logic interprets it over Kripke models defined as a tuple $\tup{W,R,V}$ and its satisfaction definition is the following.
$$
\begin{array}{rcl}
\model, w \models F\varphi & \tiff & \mbox{there is a $v$ such that $wRv$} \mbox{ and } \model, v \models \varphi\\
\model, w \models P\varphi & \tiff & \mbox{there is a $v$ such that $vRw$} \mbox{ and } \model, v \models \varphi
\end{array}
$$
In the definition it is clear that the $F$ modality looks forward in the relation $R$ and the $P$ modality looks back on it, hence the names `future' and `past'. Observe that the $F$ modality can be thought as a normal `diamond' over the relation $R$ but that is not possible with the $P$ modality.

An alternative interpretation is as follows. Interpret the logic over Kripke models which are tuples $\tup{W,R_{1},R_{2},V}$ where $R_{1} = \finv{R_{2}}$. With this restriction we can give a different satisfaction definition for the modalities.
$$
\begin{array}{rcl}
\model, w \models F\varphi & \tiff & \mbox{there is a $v$ such that $wR_{1}v$} \mbox{ and } \model, v \models \varphi\\
\model, w \models P\varphi & \tiff & \mbox{there is a $v$ such that $wR_{2}v$} \mbox{ and } \model, v \models \varphi
\end{array}
$$
In this case, both modalities are simple `diamonds' (which have been given fancy names $F$ and $P$). Does a similar characterization theorem hold in this case? Which properties should the restricted model class have for the characterization to hold? These are the kind of questions that we will be adressing in the following chapters.

\bigskip
When talking about definability we can think of \emph{relative definability} as follows: Is the class \classK definable with a \bml formula \emph{given} that we only consider models within class \classK'? That is, is there a formula $\varphi$ such that for every model in \classK', $\model, w \models \varphi$ if and only if $\tup{\model,w} \in \classK$?

The results stated in Chapter~\ref{chap:knownbml} are valid for the special case where \classK' is the class of all models. In practice, depending on the domain of application, it is common to work with restricted classes of models such as finite models, tree models, acyclic models, etc. We want to know whether these restrictions give us extra information and turn classes that were previously undefinable into definable classes. A relativized version of the definability theorem should aid us in this quest.
\end{paragraph}


\section{Basic definitions}
\label{sec:basicdefs}
\begin{defn}[Languages and formulas] We note \cL and \cFO as the source and target languages respectively. The source language is an extension of the language
$$\cP = \tup{(p_{i})_{i\in\nat},\land,\lor,\top,\bot}$$
which has infinitely many propositional variables, conjunction and the \emph{true} and \emph{false} constants. The target language is a (countable) first-order language which may or may not contain equality.

$\form{\cA}$ is the set of formulas of the language $\cA$ and \form{\cFOu} is the subset of \form{\cFO} with at most one free variable (and that variable is $x$).
\end{defn}

During this thesis we will deal with source logics which are at most as expressive as first order logic. If \cL is less or equally expressive than \cFO we should be able to express in \cFO everything that is expressible in \cL. We have seen before that, for \bml, there exists a standard translation $ST_{x}$ from \bml to first order logic. In general we define a formula translation as follows.

\begin{defn}[Formula translation]\label{trans:formula}
A \emph{formula translation} is a function $$\Trf:\form{\cL} \to \form{\cFOu}$$ that translates formulas from the source language \cL to the first-order language \cFO.
This function is required to preserve conjunctions and disjunctions, that is, formally speaking: Let ${\varphi_{1}, \varphi_{2} \in \form{\cL}}$ and $\odot \in \{\land,\lor\}$ then for every first-order formula of the form $\Trf(\varphi_{1})\odot\Trf(\varphi_{2})$ there exists an \cL-formula $\psi$ such that ${\Trf(\psi) \theq{\cFO} \Trf(\varphi_{1})\odot\Trf(\varphi_{2})}$.
\end{defn}

As we saw before in the definition of \bml's standard translation, in general, formula translations are defined homomorphically with respect to the boolean connectives. 
\begin{eqnarray*}
\Trf(\varphi_{1}\land\varphi_{2}) &=& \Trf(\varphi_{1})\land\Trf(\varphi_{2})\\
\Trf(\varphi_{1}\lor\varphi_{2}) &=& \Trf(\varphi_{1})\lor\Trf(\varphi_{2})
\end{eqnarray*}

\begin{defn}[Models]\label{def:models}
We will be working with source logics that are interpreted over variations of Kripke models.\footnote{Even propositional logic can be thought of as a modal logic without modal operators and restricted to models with one single point.}
We define $\mods{\cL}$ to be the class of all models of the source logic and
$\mods{\cFO}$ to be the class of all models of the target first-order logic.

Sometimes we will use the notion of \emph{pointed models}. A \emph{pointed model} in the source logic \cL is a model-world pair. We can see a pointed model as an \cL-model where the evaluation point has been fixed. We define the class of pointed models for the source logic as
$$\pmods{\cL} = \setcomp{\tup{\model, w} }{ \model \in \mods{\cL} \text{ and } w \in |\model|}$$
Similarly, in the target logic \cFO, we use $\fmodel, g \models \varphi$ to note that a formula $\varphi$ is true in the model $\fmodel$ under the valuation (or assignment) $g$.\footnote{Observe that in this case $g$ is a valuation and not a point of the domain.} A \emph{pointed model} of the target logic \cFO is a model-assignment pair. We can see a pointed model as an \cFO-model where the assignment function has been fixed.

\begin{defn}[$x$-assignment]
Let \fmodel be an \cFO model. An \emph{x-assignment} for \fmodel is a function $$g:\{x\}\to|\fmodel|$$ which assigns an element for the variable $x$. It can be seen as a finite valuation specialized to the variable $x$.
\end{defn}

We will use the concept of $x$-assignment to define the class of first order pointed models. This notion is a technical detail needed to make things work in Definition~\ref{trans:model}. The problem and solution will become clear after that definition. We define the class of pointed models for the target logic as
$$\pmods{\cFO} = \setcomp{\tup{\model^{f}, g}}{\model^{f} \in \mods{\cFO} \text{ and } g \text{ is an $x$-assignment for } \fmodel}$$
\end{defn}
Observe that the formulas obtained through the translation defined in Definition~\ref{trans:formula} have \emph{at most} one free variable and that variable is $x$. Therefore, if we want to evaluate those formulas in a first order model \fmodel, an $x$-assignment $g$ is enough for $\fmodel, g \models \varphi(x)$ to be well defined.

\begin{nota}
Let $\tup{\fmodel,g}, \tup{\fnodel,h} \in \pmods{\cFO}$. We write $\fmodel,g \theq{\cFO} \fnodel,h$ to mean that for every first order formula $\alpha(x)$; $\fmodel,g \models \alpha(x)$ if and only if $\fnodel,h \models \alpha(x)$.
\end{nota}

There's one more thing to be taken into account, formulas from \cL and formulas from \cFO are not evaluated in the same models. The former are evaluated in Kripke models and the later are evaluated in first-order models. This is the reason why we are not yet able to compare $\varphi$ with $\Trf(\varphi)$.

We can think of models as `information bearers', they represent some information relative to the world in a way that is compatible with some logic. Therefore, the information is not in the model itself but somewhere else. We need to define some way to `look at' this information from different perspectives, one compatible with the source logic \cL and other compatible with the target logic \cFO. Following the same line we define a model translation that `converts' the information between the source and target logic.

\begin{defn}[Model translation]\label{trans:model} Given a class of models $\classK \subseteq \pmods{\cFO}$, a \emph{model translation} is a biyective function
$$\Trmk{\classK}:\pmods{\cL} \to \classK$$
\end{defn}

We write $\Trm$ instead of $\Trmk{\classK}$ when the class of models is clear from the context. As an abuse of notation we use $\Trm(\model)$ when we are not interested in the associated assignment and $\Tlm(\adel^{f},g)$ for the preimage of $\tup{\adel^{f},g}$. 

Returning to the need for $x$-assignments, note that if we allowed $g$ to be a standard assignment (i.e. $g:\fvar\to|\fmodel|$) in Definition~\ref{def:models} then for every pointed \cL-model $\tup{\model,w}$ we would have \emph{many} pointed \cFO-model $\tup{\fmodel,g_{i}}$ where $g_{i}(x) = w$ and the assignment for the rest of the variables changes arbitrarily. Therefore, this could carry problems at the moment of satisfying the \emph{suryectivity} requirement.

As an exercise, suppose that the class of pointed models is defined with standard assignments and try to define a model translation for \bml. You will observe that there is a cardinality problem.

\medskip
When we proved the results for \bml we did not use a model translation, at least not explicitly. On the other hand, this translation was implicitly present when we gave an informal way to `look at' models from both a \bml and a first order perspective. The model translation function will serve us in this task. We are now ready to set proper constraints on the translations.

\begin{defn}[Truth preserving pair of translations]\label{def:preservingtrans} A pair of translations $(\Trf, \Trm)$ is said to be \emph{truth-preserving} if
for all $\varphi \in \form{\cL}$ and all $\tup{\model,w} \in \pmods{\cL}$
$$\model,w \models \varphi \text{\ \ iff\ \ } \Trm(\model,w) \models \Trf(\varphi)$$
\end{defn}

 Let's fix $(\Trf, \Trmk{\classK})$ as our pair of \emph{truth-preserving} translations for the rest of the thesis. We will also want to translate formulas from \cL to \cFO and then go back to \cL-formulas. As we are not requiring \Trf to be injective this could lead to a problem. We make the following claim.

\begin{propo}\label{prop:trfpreimg}
For any $\alpha$, $\beta$ such that $\Trf(\alpha) = \Trf(\beta)$ we have ${\models_{\cL} \alpha \leftrightarrow \beta}$.
\end{propo}
\begin{proof}
Suppose that $\not\models_{\cL} \alpha \leftrightarrow \beta$, then we have a model \model and a point $w$ such that ${\model,w \models \alpha}$ and ${\model,w \not\models \beta}$. Then by definition of truth-preservation of the translations we get ${\Trm(\model, w) \models \Trf(\alpha)}$ and ${\Trm(\model, w) \not\models \Trf(\beta)}$. Absurd.
\end{proof}

We will use this proposition to make a simplification. First define the equivalence relation $\varphi \sim \psi \text{ iff } \models_{\cL} \varphi \leftrightarrow \psi$. Regarding \cL-formulas, we can always take the equivalence classes defined by the quotient set of \cL-formulas by $\sim$ and for each class choose a representative to work with.

To simplify the proofs in this thesis, and without loss of generality, we will assume that we are working with the set of formulas defined above. All of our proofs should also work with the original set of formulas but they would require excessive detours and justifications. In this setting we will be working up to formula equivalence and we will assume that our formula translation $\Trf$ is injective.

\begin{defn}\label{nota:thsub}
Let $\classK \subseteq \mods{\cL}$, \model be an \cL-model and $w \in |\model|$. We define the \emph{theory} of a pointed model, model and class of models as follows
\begin{eqnarray*}
\Th(\model,w) &=& \setcomp{\varphi }{ \model,w \models \varphi}\\
\Th(\model) &=& \setcomp{\varphi }{ \model \models \varphi}\\
\Th(\classK) &=& \setcomp{\varphi }{ \forall\model \in \classK \text{ it holds that } \model \models \varphi}
\end{eqnarray*}

Given two \cL-pointed models $\tup{\model, w}$ and $\tup{\nodel, v}$ we say that the \emph{pointed} models are modally equivalent (noted
$\model, w \theq{\cL} \nodel, v$) when $\Th(\model, w) = \Th(\nodel,v)$. We say that two models (\emph{not} pointed) are modally equivalent (noted
$\model \theq{\cL} \nodel$) when $\Th(\model) = \Th(\nodel)$.
We write $\model, w \thsub{\cL} \nodel, v$ when $\Th(\model, w) \subseteq \Th(\nodel,v)$ and
$\model \thsub{\cL} \nodel$ when $\Th(\model) \subseteq \Th(\nodel)$.
All these definitions can be similarly defined for the target logic \cFO and we will assume them defined.
\end{defn}

%
%
\medskip
The framework defined in this section will allow us to transfer results between the source and target logics. As an example we prove compactness for \cL under some special closure conditions (which will be addressed later).

\begin{lem}[\cL is compact]\label{lem:compact}
If \cL has a pair of truth-preserving translations $(\Trf, \Trmk{\classK})$ and \classK is closed under ultraproducts then \cL is compact.
\end{lem}
\begin{proof}

Let $\Gamma$ be a set of \cL-formulas and suppose that any finite set of $\Gamma$ is \cL-satisfiable. We
will show that $\Gamma$ is \cL-satisfiable.

Take any finite $\Delta_{f} \subseteq \Trf(\Gamma)$, we want to see that it is satisfiable in \classK. As our formula translation is injective we have a set $\Delta \subseteq \Gamma$ such that $\Delta = \finv{\Trf}(\Delta_{f})$. Observe that $\Delta$ is finite because $\Trf$ is injective. By hypothesis there exists $\tup{\model,w}$ such that $\model,w \models \Delta$ because $\Delta$ is a finite subset of $\Gamma$. Hence, by truth-preservation, $\Trm(\model,w) \models \Delta_{f}$ and $\Trm(\model,w) \in \classK$.

By Theorem \ref{thm:compactness} we conclude that there exists a model $\tup{\fnodel,g}$ in the class \classK such that ${\fnodel,g \models \Trf(\Gamma)}$. As the translations are truth preserving we get $\Tlm(\fnodel,g) \models \Gamma$.
\end{proof}

\section{General model equivalence}
In Chapter~\ref{chap:intro} of this thesis we introduced the general idea of model equivalence and, in particular the notions of simulation and bisimulation for some specific logics (namely \bmlsn and \bml). We want the framework we are developing to be able to handle several types of model equivalence relations.

Restricting ourselves to the definition of simulation and bisimulation we can see that, the latter can be seen as a special case of the first where there is a symmetrical atomic condition and a `back' clause. Looking at their common properties we can say that they both agree in the following points:
\begin{enumerate}[(i)]
\setlength{\itemsep}{0pt}
\item They relate a point in one model with a point in the other model. Thus, given \model and \nodel, if $Z$ is such relation then $Z \subseteq |\model|\times|\nodel|$.
\item They imply some kind of modal theory transfer. In the case of simulations, if $wZv$ then $\model,w \thsub{} \nodel,v$. On the other hand, bisimulations imply full modal equivalence: if $wZv$ then $\model,w \theq{} \nodel,v$.
\item For every (bi)simulation between \model and \nodel, the models' structure doesn't change. That is why the notion of (bi)simulation always links points from \model to points of \nodel. We will have to make a small change here to be able to handle (bi)simulations in dynamic logics.\footnote{Logics where the modal operators may change the model.}
\end{enumerate}

Following this analysis, given a source logic \cL, we will give a \emph{meta-definition} for model equivalence notions. That is, we will not define a relation but give the conditions that a valid model equivalence notion should satisfy. Any simulation or bisimulation relation satisfying the next definition fits into our framework.

As an abuse of language, we call it \emph{\cL-simulation}. We use this name because it reminds us of the properties that simulation relation defined for \bmlsn satisfies but it is \emph{not} the same. As it will be clear from the definition, we don't impose any structural constraint.

\begin{defn}[\cL-simulation]\label{def:sim}
Given two \cL models \model and \nodel we define an \emph{\cL-simulation} to be a \emph{non-empty} relation $Z \subseteq \pmods{\cL}\times\pmods{\cL}$ with the following constraint
$$\text{If } \tup{\model,w}Z\tup{\nodel,v} \text{ then } \model,w \thsub{\cL} \nodel,v$$
We write $\model, w \rsim_{\cL} \nodel, v$ to indicate that there exists a simulation between $w$ and $v$ and $\model \rsim_{\cL} \nodel$ to indicate that there exists a point $w \in |\model|$ such that $\model,w \rsim_{\cL} \nodel,v$ for some $v \in |\nodel|$. We write $w \rsim_{\cL} v$ when the models are clear from context.
\end{defn}

Note that the simulation definition for \bmlsn satisfies the above definition with minor changes. The only difference is that we have to take into account the `model' component of the simulation relation. It can be re-defined as follows.

\begin{exam}
A \bmlsn simulation is a non-empty binary relation between pointed models such that whenever $\tup{\model,w}Z\tup{\nodel,v}$ we have that: 

\medskip

\textbf{Atomic condition}: If $w \in V^{\model}(p)$ then $v \in V^{\nodel}(p)$ for all $p \in \prop$.

\textbf{Forth}: if $wR_{1}w'$, then there exists a point $v'$ in $\nodel$ such that $vR_{2}v'$ and $\tup{\model,w'}Z\tup{\nodel,v'}$.
\end{exam}
\noindent
When thinking about \bml's bisimulation, it may seem that this definition is missing something. We know that if $\model,w$ and $\nodel,v$ are related by a \emph{bisimulation} then $\model,w \equiv \nodel,v$ but the above definition only guarantees $\model,w \thsub{} \nodel,v$. Don't worry about that now, it will become clear in the next section that this condition is enough for what we need. Observe, also, that a bisimulation is a special case of Definition \ref{def:sim} where the relation is symmetric.

We want to stress that this definition of \cL-simulation does not cover all possible types of model equivalences and it isn't suitable for all types of modal languages. One example where this notion is not adequate is when the language doesn't have \emph{disjuction} nor negation. Let \model and \nodel be two models, the right notion of simulation for this language links \emph{sets} of points from \model to a point of \nodel. As we have defined our possible languages in Section \ref{sec:basicdefs} this will not be a problem because we always have disjunction in our source language. For more information about model theory on disjunction-free languages refer to~\cite{KR97}.



\section{Saturation}
We have seen that, in general, modal equivalence does not imply bisimilarity. It is also the case with \bmlsn's simulation that, in general $\model,w \rsim \nodel,v$ does not imply $\model,w \thsub{} \nodel,v$. This problem recurs with most model equivalence notions found in the literature.

We have stated that in the class of \wsat models, \bml equivalence implies bisimulation. For this framework we need to define a similar notion which fits the logics we will be working with. Let's formally define the general condition that we're pursuing so we can focus on it.

\begin{defn}[Hennessy-Milner Property]\label{def:hmp}
Let \classK be a class of \cL-pointed models, we say that \classK has the \emph{Hennessy-Milner property} if for every two \cL-models $\tup{\model,w}$ and $\tup{\nodel,v}$ in \classK, whenever $\model,w \thsub{\cL} \nodel,v$ we have $\model,w \rsim_{\cL} \nodel,v$.
\end{defn}

This definition should be interpreted as the converse of the \cL-simulation (Definition~\ref{def:sim}) requirement and will be the definition of Hennessy-Milner class used in our framework.

Is this definition general enough to cover the cases we have been talking about? We know that if we fix \cL as \bml and the simulation relation as \bml's bisimulation we have that $\model,w \theq{\cL} \nodel,v$ implies $\model,w \bisim_{\cL} \nodel,v$ but the definition above seems to impose a stronger constraint. We only have $\model,w \thsub{\cL} \nodel,v$ as hypothesis and we should conclude the same thesis. We make the following statement that explains why there is no problem with this.

\begin{propo}\label{prop:negsubeq}
If \cL has negation then $\model,w \thsub{\cL} \nodel,v$ if and only if $\model,w \theq{\cL} \nodel,v$.
\end{propo}

As \bml has negation, the seemingly weak hypothesis turns strong enough to prove the result in that particular case. The special case regarding saturation for \bml is nicely covered in~\cite{MLBOOK}.

Regarding $\omega$-saturation, the definition given in Definition~\ref{def:omegasat} is in terms of first order models. In that moment, as we were looking at \bml and first order models as if they were the same, that gave us no problems.

In this chapter we want to make an explicit differentiation between \mbox{\cL-models} and \mbox{\cFO-models}. To make our proofs simpler we choose the following definition for \wsat \mbox{\cL-models}.

\begin{defn}
We say that an \cL-model \model is \emph{\wsat} if and only if $\Trm(\model)$ is.
\end{defn}

For details on $\omega$-saturation, classical results can be found in~\cite{CK90}. Also, in Marco Hollenberg's thesis~\cite{mh98}, he extensively investigates Hennessy-Milner classes.

\clearpage

\chapter{Main Results}\label{chap:main}
\section{Adequate pair}
In Chapter~\ref{chap:framework}, while developing the framework, we explained the objectives we are pursuing when doing a generalization. In the following definition we will make explicit the requirements for the theorems in this chapter to hold for an arbitrary logic \cL and with respect to a class of models \classK.

\begin{defn}[Adequate pair]\label{adeq} A logic \cL and a class of models $\classK \subseteq \pmods{\cFO}$ is said to be an \emph{adequate pair} if they fulfill the following requirements
\begin{enumerate}
\setlength{\itemsep}{0pt}
\item{\label{adeq:closure}\classK is closed under ultraproducts (Definition \ref{def:upclosed}).}
\item{\label{adeq:trans}There exist truth-preserving translations \Trf, \Trmk{\classK} (Definition \ref{def:preservingtrans}).}
\item{\label{adeq:sim}There exists an \cL-simulation notion (Definition \ref{def:sim}).}
\item{\label{adeq:omegasat}The class of \wsat \cL-models should have the Hennessy-Milner property with respect to \cL-simulations (Definitions \ref{def:omegasat} and \ref{def:hmp}).}
\end{enumerate}
\end{defn}

We need to formally define the closure under ultraproducts and ultrapowers used above, as the ones given in Chapter~\ref{chap:knownbml} were specifically crafted for \bml.

\begin{defn}[Closure under ultraproducts]\label{def:upclosed}
A class $\classK \subseteq \mods{\cFO}$ is said to be \emph{closed under ultraproducts} if, let $\fmodel_{i}$ be a family of \cFO-models in \classK and let $U$ be an ultrafilter, the ultraproduct $\prod_{U}\fmodel_{i}$ is also in \classK. A more sophisticated definition is needeed for first-order pointed models.

A class $\classK \subseteq \pmods{\cFO}$ is said to be \emph{closed under ultraproducts} if, let $\tup{\fmodel_{i}, g_{i}}$ be a family of \cFO-pointed models and let $U$ be an ultrafilter. Let the $\prod_{U}\fmodel_{i}$ be the ultraproduct of the models then $\tup{\prod_{U}\fmodel_{i}, g^{*}_{i}} \in \classK$ for every $g^{*}_{i}$ defined as $g^{*}_{i}(x) = \lambda z.g_{i}(x)$ for all $x$.\footnote{For a formal definition of the lambda notation refer to~\cite{lambda}.}
\end{defn}

\begin{defn}[Closure under ultrapowers]
A class \classK of \cFO-models is said to be \emph{closed under ultrapowers} if it is closed under ultraproducts where every $\fmodel_{i}$ is the same model. A similar definition can be given for pointed models.
\end{defn}

Why are we requiring \classK to be closed under ultraproducts? We could have asked for more, such as \classK being definable by a first order formula, which implies closure under ultraproducts. We could've also tried to impose no restriction over \classK.

We decided to require \classK to be closed under ultraproducts because it is the weakest condition that lets us use the relativized version of the first order compactness theorem (stated and proved in the appendix as Theorem~\ref{thm:compactness}). In particular, all first-order definable classes and the class of `all models' will always fit in an adequate pair.

The second item in Definition~\ref{adeq} makes sure that \cL is less or equally expressive than first order and that there is some way to compare between the formulas and models of both logics.

In the same way, the third item only asks for the definition of a simulation notion which is essential to develop the model theory of \cL. All the results will be stated in terms of that \cL-simulation notion.

With enough practice, points one to three can be easily checked by just `looking at' \cL. It is only when we get to the last item that we face the strongest requirement. This points says that the class of \wsat \cL-models should have the Hennessy-Milner property.

In Chapter~\ref{chap:knownbml} we mentioned that the class of \wsat{} \bml models had the Hennessy-Milner property with respect to \bml bisimulations (although in that moment we didn't call it `Hennessy-Milner' yet). The proof of that result makes a link between the semantics of \bml and the structure of the \bml bisimulation. Therefore, it makes use of the \emph{structural} definition of the \bml bisimulation.

So far, given a logic \cL, we are looking at \cL-simulations as black boxes. All we know is that $w \rsim_{\cL} v$ implies $w \thsub{\cL} v$. We don't know which structural properties it imposes. This is the reason why we still need this item to be proved for the results to work.

We think that there's still much work to be done to weaken this last requirement and we will give our opinion on directions for further work in the conclusions.

\section{Characterization}\label{sec:charact}
One of the central notions in the characterization theorem for \bml was that of bisimulation invariance. Recall that bisimulations are defined between \bml models but the notion of bisimulation invariance is defined for first order formulas.
\begin{defn*}\label{thm:bml:bisiminv}
A first order formula $\alpha(x)$ is \emph{invariant for \bml bimulations} if for all \bml models $\model, \nodel$ and $w \in |\model|, v \in |\nodel|$ such that $\model,w \bisim \nodel,v$ the following holds: 
$$\model \models \alpha(x)[w] \text{ iff } \nodel \models \alpha(x)[v].$$
\end{defn*}
When working with \bml, this difference made no problem to us because we didn't really distinguish between \bml and first order models. It is time for us to give an invariance definition that fits our framework and there is an important decision to be made.

The property of `invariance' is thought for first order formulas and the notion of \cL-simulation is defined between \cL models. We have to options: The first one is to call a first order formula $\alpha(x)$ `invariant for \cL-simulations' if, for every two \cL models $\model,w$ and $\nodel,v$ such that $\model,w \rsim_{\cL} \nodel,v$ whenever $\alpha(x)$ holds in $\Trm(\model,w)$ it should hold in $\Trm(\nodel,v)$. In this case we are `mixing' the models through the translation.

The other option is to `lift' the \cL-simulation notion to \cFO models and define a simulation relation $\rsim_{\cFO}$. In this case we could just say that a first order formula $\alpha(x)$ is `invariant for \cL-simulations' if, for every two \cFO models $\fmodel,g$ and $\fnodel,h$ such that $\fmodel,g \rsim_{\cFO} \fnodel,h$ whenever $\alpha(x)$ holds in $\fmodel,g$ it should hold in $\fnodel,h$.

The advantage of the first option is that there's no need for new definitions; in contrast, the second one would require a formalization for the `lifting' to be defined along with the model translation. On the other hand, it would be nice to be able to check two \cFO models for `model equivalence' as we can do with \cL.

In this thesis we choose the first option because it is the most direct one in this setting. Observe that, in particular, the first option can be seen as a special case of the second one when the following `canonical lifting' is defined.
$$\fmodel,g \rsim_{\cFO} \fnodel,h \text{ \ iff \ } \Tlm(\fmodel,g) \rsim_{\cL} \Tlm(\fnodel,h)$$
Again, the problem with this definition is that it bears no structural information regarding model equivalence between first order models. It is just another detour.

\begin{defn}[\cL-simulation \classK-invariance]
Let $\tup{\cL,\classK}$ be an adequate pair. A formula $\alpha(x)$ of $\cFOu$ is \emph{\classK-invariant for \cL-simulations} if for all \cL-models $\model, \nodel$ and $w \in |\model|, v \in |\nodel|$: If $\model,w \rsim_{\cL} \nodel,v$ and $\Trm(\model,w) \models \alpha(x)$ then $\Trm(\nodel,v) \models \alpha(x)$.
\end{defn}

Before stating the characterization theorem let's see the importance and role of the class of models \classK in these definitions. What happens to formula equivalence when we change the class of models? As a motivating example we will work with a first-order formula $\varphi = \forall x.R(x,x)$ which holds in a model if and only if $R$ is a reflexive relation.

In the class of all models it is obvious that $\not\models \varphi$ ($\varphi$ is not valid) because we can come up with some models where $R$ is not reflexive. Given that ``reflexivity'' is not expressible in basic modal logic, we can conclude that $\varphi$ is not equivalent to the translation of any basic modal formula.

\begin{defn}[\classK-equivalence] Let $\classK \subseteq \mods{\cFO}$ and $\varphi,\psi \in \form{\cFO}$. We say that $\varphi$ and $\psi$ are \emph{\classK-equivalent} if and only if $\models_{\classK} \varphi \liff \psi$.
\end{defn}

Let's now restrict the class of models, let $\classK$ be the class of \emph{reflexive} models. Now $\models_{\classK} \varphi$ because it is valid in every model of the class. In this setting there \emph{is} a basic modal formula whose translation is \classK-equivalent to $\varphi$. Take $\psi = \top$ we have $\models_{\classK} ST_{x}(\psi) \leftrightarrow \varphi$ because $\models_{\classK} \top \leftrightarrow \varphi$. What happened here is that, restricting our class of models the number of valid formulas has grown and with them the number of ``formulas equivalent to a translation''.

Something similar occurs with \cL-simulation invariance. Again, we have seen that in the class of all models we can have two bisimilar models where one has a reflexive relation and the other doesn't. Therefore ``reflexivity'' is not invariant under bisimulations.

If we change the class of models to the class of reflexive models we see that now the property becomes invariant over bisimulations. This happens because it is trivially invariant \emph{all} over \classK. It is nice to observe that the concepts of invariance and equivalence are very closely related to each other when we change the class of models we are working with.

\begin{thm}[Characterization]\label{thm:charact}
Given an adequate pair $\tup{\cL,\classK}$ then 
\begin{center}
\begin{tabular}{l}
A formula $\alpha(x)$ of $\cFOu$\\
is \classK-equivalent to the translation of an $\cL$-formula iff\\
$\alpha(x)$ is \classK-invariant for $\cL$-simulations.
\end{tabular}
\end{center}
\end{thm}

\begin{proof}[Left to right]
Suppose $\alpha(x)$ is \classK-equivalent to the translation of an $\cL$-formula $\varphi$. We want to see that it is invariant over \cL-simulations. This is a consequence of the invariance of \cL-formulas over \cL-simulations. Suppose we have $\model,w \rsim \nodel,v$ and $\Trm(\model,w) \models \alpha(x)$ but $\Trm(\nodel,v) \not\models \alpha(x)$. As $\models_{\classK} \alpha(x) \leftrightarrow \Trf(\varphi)$ and the translations are truth-preserving it must hold that $\model,w \models \varphi$ and $\nodel,v \not\models \varphi$. But this is a contradiction because \cL-formulas are invariant under \cL-simulations and we have a simulation linking those points.
\end{proof}

\noindent \textit{Right to left.}
Suppose $\alpha(x)$ is \classK-invariant for $\cL$-simulations, we want to see that it is \classK-equivalent to the translation of an $\cL$-formula. Consider the following set of consecuences
$$\slc{\alpha} = \setcomp{\Trf(\varphi)}{\varphi \textrm{ is an \cL-formula and } \alpha(x) \models_{\classK} \Trf(\varphi)}.$$
We will prove that if $\slc{\alpha} \models_{\classK} \alpha(x)$ we are done.

\begin{propo}\label{prop:slclisto}
If $\slc{\alpha} \models_{\classK} \alpha(x)$ then $\alpha(x)$ is \classK-equivalent to the translation of an \cL-formula.
\end{propo}

\begin{proof}
Suppose $\slc{\alpha} \models_{\classK} \alpha(x)$, by relative compactness (Theorem \ref{thm:compactness}) there is a finite set $\Delta \subseteq \slc{\alpha}$ such that $\Delta \models_{\classK} \alpha(x)$, therefore $\models_{\classK} \bigwedge\Delta \to \alpha(x)$. Trivially (by definition) we have that $\models_{\classK} \alpha(x) \to \bigwedge\Delta$ so we can conclude $\models_{\classK} \alpha(x) \leftrightarrow \bigwedge\Delta$. As every $\beta \in \Delta$ is the translation of an \cL-formula and the translation preserves conjunction then $\bigwedge\Delta$ is also the translation of some modal formula.
\end{proof}

\begin{lem}\label{lem:slcvale}
$\slc{\alpha} \models_{\classK} \alpha(x)$.
\end{lem}

\noindent \textit{Proof.}
Suppose that $\Trm(\model,w) \models \slc{\alpha}$. We have to show that $\Trm(\model,w) \models \alpha(x)$. Define $\NTh^{w}(x)$ as
$$\NTh^{w}(x) = \setcomp{\neg\Trf(\varphi)}{\varphi \textrm{ is an \cL-formula and } \model,w \not\models \varphi}$$
Observe that, if \cL has negation then $\NTh^{w}(x)$ will be the translation of $w$'s modal theory and every model of $\NTh^{w}(x)$ will be modally equivalent to $w$. If \cL doesn't have negation we will only preserve formulas that are not true in $w$. This definition fits for both cases. Now define the set
$$\Sigma(x) = \{\alpha(x)\} \cup \NTh^{w}(x).$$
We will see that $\Sigma(x)$ has a model in \classK.

\begin{propo}
$\Sigma(x)$ has model in \classK.
\end{propo}
\begin{proof}
Let's suppose that there is no model in \classK for $\Sigma(x)$ and use the contrapositive of Theorem \ref{thm:compactness}. We can conclude that there must be a finite subset $\{\alpha(x),\lnot\delta_{1},\dots,\lnot\delta_{n}\} \subseteq \Sigma(x)$ with $\lnot\delta_{i} \in \NTh^{w}(x)$ which doesn't have model in \classK. Note that this set should include $\alpha(x)$, otherwise it would have a had model, namely $\Trm(\model,w)$.

Observe that, for every model $\adel^{f} \in \classK$, as $\adel^{f} \not\models \{\alpha(x),\lnot\delta_{1},\dots,\lnot\delta_{n}\}$ then $\adel^{f} \models \alpha(x) \rightarrow \neg(\lnot\delta_{1}\land\dots\land\lnot\delta_{n})$. This means that $\alpha(x) \rightarrow (\delta_{1} \lor \dots \lor \delta_{n})$ is valid in \classK, therefore $\alpha(x) \models_{\classK} \delta_{1} \lor \dots \lor \delta_{n}$. If $\delta_{1} \lor \dots \lor \delta_{n}$ is a \classK-consecuence of $\alpha(x)$ then, as the formula translation preserves disjunction, $\delta_{1} \lor \dots \lor \delta_{n} \in \slc{\alpha}$. But, as $\Trm(\model,w) \models \slc{\alpha}$ then $\Trm(\model,w) \models \delta_{1} \lor \dots \lor \delta_{n}$. This is absurd because $\Trm(\model,w) \not\models \delta_{i}$ for every $i$.
\end{proof}

As $\Sigma(x)$ is satisfiable in \classK we have a model \nodel and an element $v$ such that ${\Trm(\nodel,v) \models \Sigma(x)}$. We make the following proposition.

\begin{propo}\label{propo:modsinc}
$\nodel,v \thsub{\cL} \model,w$.
\end{propo}
\begin{proof}
Take the contrapositive. Suppose that $\model,w \not\models \varphi$ then $\neg\Trf(\varphi) \in \NTh^{w}(x)$ and because $\NTh^{w}(x) \subseteq \Sigma(x)$ we can state that $\Trm(\nodel,v) \models \lnot\Trf(\varphi)$ which implies that $\Trm(\nodel,v) \not\models \Trf(\varphi)$. By truth-preservation of the translations we get $\nodel,v \not\models \varphi$.
\end{proof}

We will need to link $\Trm(\model,w)$ and $\Trm(\nodel,v)$ in a way that lets us transfer the validity of $\alpha(x)$ from the second model to the first one. The next lemma will come handy.

\begin{lem}[Big Detour Lemma]\label{lem:detour}
Let $\alpha(x) \in \form{\cFOu}$ be \cL-bisimulation \classK-invariant, if $\nodel,v \thsub{\cL} \model,w$ and $\Trm(\nodel,v) \models \alpha(x)$ then $\Trm(\model,w) \models \alpha(x)$.
\end{lem}

\begin{proof}
We define some names to avoid cumbersome notation in this proof. We add a subscript $f$ to the first-order translations of \cL models, we add a superscript $+$ to first-order saturated models and a superscript $*$ to modal saturated models. 

Applying Theorem~\ref{thm:similarultrapowers} to $\model,w$ and $\nodel,v$ (with $\classM_{1} = \classM_{2} = \mods{\cL}$) we build up new models. The theorem explicitly states the relationship among them, we will use this result to prove this lemma. Hereafter we will use the same notation as in Theorem~\ref{thm:similarultrapowers}.

The following diagram helps to illustrate the actual situation along with the relationship between the various models. Think of it as a cube. The front face represents the models from the source language and the back face has the models from the first-order language.
\begin{figure}[h]
\centering
\def\pathcolor{black}
\tikzset{>=stealth',every on chain/.append style={join},every join/.style={->}}
\begin{tikzpicture}[ 
	back line/.style={densely dotted}, 
	cross line/.style={preaction={draw=white, -, 	line width=6pt}}
] 
\matrix (m) [
	matrix of math nodes, 
	row sep=3em, column sep=2.5em,
	text height=1.5ex, 
	text depth=0.25ex
	]{ 
	             & \nodel_{f},g_{v} & & \nodel_{f}^{+},g_{v}^{+} \\ 
	  \nodel,v     &            & &               & \nodel^{*},v^{*} \\ 
	             & \model_{f},g_{w} & & \model_{f}^{+},g_{w}^{+} \\ 
	  \model,w     &            & &               & \model^{*},w^{*} \\ 
	}; 
\path[<->]
(m-2-1) edge [cross line] node [above]{$\theq{\cL}$} (m-2-5);
\path[->]
(m-2-1)	edge [cross line] node [left]{$\thsub{\cL}$} (m-4-1);
\path[<->]
(m-4-5)	edge [cross line] node [above]{$\theq{\cL}$} (m-4-1);
\path[->]
(m-2-5)	edge [cross line, draw=\pathcolor, fill=\pathcolor] node [right]{$\rsim$/$\thsub{}$} (m-4-5);

\path[<->]
(m-3-2)	edge [back line, draw=\pathcolor, fill=\pathcolor] node[above left,font=\scriptsize]{\Trm} (m-4-1) 
(m-4-5) edge [back line, draw=\pathcolor, fill=\pathcolor] node[above right,font=\scriptsize]{\Trm} (m-3-4) 
(m-2-1)	edge [back line, draw=\pathcolor, fill=\pathcolor] node[above left,font=\scriptsize]{\Trm} (m-1-2) 
(m-1-4) edge [back line, draw=\pathcolor, fill=\pathcolor] node[above right,font=\scriptsize]{\Trm} (m-2-5); 

\path[<->] 
(m-1-2) edge [cross line, draw=\pathcolor, fill=\pathcolor] node [above]{$\theq{\cFO}$} (m-1-4)
(m-3-4) edge [cross line, draw=\pathcolor, fill=\pathcolor] node [above]{$\theq{\cFO}$} (m-3-2);
\end{tikzpicture} 
\caption{Directions for the detour.}
\end{figure}
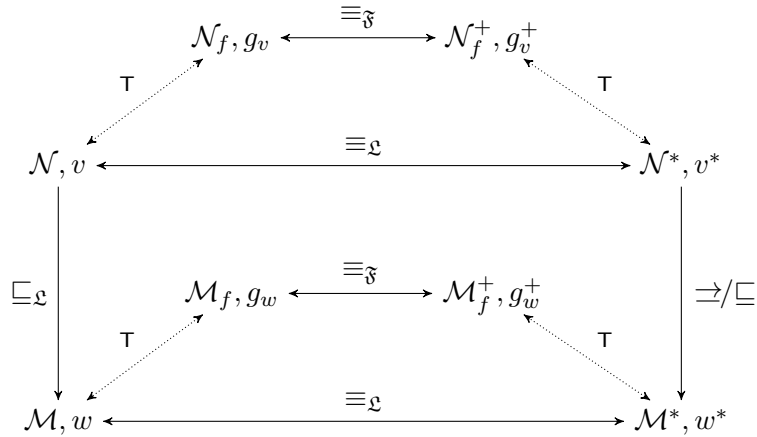

With this new notation the Big Detour Lemma can be restated as follows: Let $\alpha(x)$ be an \cL-bisimulation \classK-invariant \cFO-formula, if $\nodel,v \thsub{\cL} \model,w$ and $\nodel_{f},g_{v} \models \alpha(x)$ then $\model_{f},g_{w} \models \alpha(x)$.

Using a simple diagram chase argument we can see that, as $\nodel_{f},g_{v} \models \alpha(x)$ and $\nodel^{+}_{f},g^{+}_{v}$ is elementary equivalent to $\nodel_{f},g_{v}$, then $\nodel^{+}_{f},g^{+}_{v} \models \alpha(x)$. Because $\alpha(x)$ is invariant under \cL-simulations and $\nodel^{*},v^{*} \rsim_{\cL} \model^{*},w^{*}$ we know that $\model^{+}_{f},g^{+}_{w} \models \alpha(x)$. Again by elementary equivalence we conclude that $\model_{f},g_{w} \models \alpha(x)$ which is what we wanted to prove.
\end{proof}

Applying this lemma to $\model,w$ and $\nodel,v$ and having transfered the validity of $\alpha(x)$ from $\Trm(\nodel,v)$ to $\Trm(\model,w)$ we can conclude that $\slc{\alpha} \models \alpha(x)$. With this final affirmation we have just proved the right-to-left direction of the characterization result.

\clearpage
\section{Definability}\label{sec:main:def}
The study of class definability is not new. There exist, for example, several results for first order logic regarding the definability of classes of models. In that case, a class of models that is definable by means of a set of first order formulas is called \emph{elementary} and those that can be defined by means of a single formula are called \emph{basic elementary} classes.

To develop the theory of this section we will use a relativized version of the concept of first order definability. It is defined as follows.

\begin{defn}[\classC-elementary class] Let $\classC \subseteq \mods{\cFO}$.
\begin{enumerate}
\setlength{\itemsep}{0pt}
\item A class $\classK \subseteq \classC$ is called \emph{\classC-elementary} (noted $\classC\mbox{-}\elementary$) if there exists a set of first order formulas $\Gamma$ such that for all $\fmodel \in \classC$ it occurs that $\fmodel \models \Gamma \text{ iff } \fmodel \in \classK$.\footnote{A $\classC\mbox{-}\elementary$ class can be seen as the intersection of $\classC\mbox{-}\belementary$ classes. The $\Delta$ in the notation comes from the german word \emph{Durchschnitt} which means `cross-section' and makes reference to this fact.}
\item A class $\classK \subseteq \classC$ is called \emph{basic \classC-elementary} (noted $\classC\mbox{-}\belementary$) if there exists a first order formula $\varphi$ such that for all $\fmodel \in \classC$ it occurs that $\fmodel \models \varphi \text{ iff } \fmodel \in \classK$.
\end{enumerate}
\end{defn}

\begin{defn}[Elementary class] Let $\classK \subseteq \mods{\cFO}$.
\begin{enumerate}
\setlength{\itemsep}{0pt}
\item \classK is called \emph{elementary} (noted $\elementary$) if it is \classC-elementary for $\classC = \mods{\cFO}$.
\item \classK is called \emph{basic elementary} (noted $\belementary$) if it is basic \classC-elementary for $\classC = \mods{\cFO}$.
\end{enumerate}
\end{defn}

On the modal side, we will use \emph{pointed models} for a smoother proof. We need some further definitions before stating the main theorem of this section. The concept of `definability' in the source logic is given analogously to the one of the target logic.

\begin{defn}[Definability]
A class $\classM \subseteq \pmods{\cL}$ is said to be \emph{definable by a set of formulas} if there exists a set $\Gamma$ of \cL-formulas such that $\tup{\model,w} \in \classM$ if and only if $\model,w \models \Gamma$.
\end{defn}

\begin{defn}[Closure under simulations]
A class $\classM \subseteq \pmods{\cL}$ is said to be \emph{closed under simulations} if, whenever $\tup{\model,w} \in \classM$, and $\tup{\nodel,v}$ is an \cL-pointed model such that $\model,w \rsim_{\cL} \nodel,v$ then $\tup{\nodel,v} \in \classM$.
\end{defn}

As in first order and \bml, we distinguish between two types of classes. Those that can be defined by a set of formulas and those that can be defined by a single formula. Here we state the first theorem and then carry on with the second one.

\begin{thm}[Definability by a set]\label{thm:def:set}
Given an adequate pair $\tup{\cL, \classK}$ and a class of pointed models $\classM \subseteq \pmods{\cL}$, the following are equivalent
\begin{enumerate}[(i)]
\setlength{\itemsep}{0pt}
 \item \label{def:set:formulas}\classM is definable by a set of \cL-formulas.
 \item \label{def:set:closure}\classM is closed under \cL-simulations, $\Trm(\classM)$ is closed under ultraproducts and $\Trm(\CclassM)$ is closed under ultrapowers.
\end{enumerate}
\end{thm}

\begin{proof}[From \ref{def:set:formulas} to \ref{def:set:closure}]
Suppose that \classM is defined by the set $\Gamma$ of \cL-formulas.
\begin{enumerate}
	\item Suppose now that there is a model $\tup{\model,w} \in \classM$ such that $\model,w \rsim \nodel,v$ for some pointed model $\nodel,v$. As $\tup{\model,w} \in \classM$ it must occur that $\model,w \models \Gamma$. By simulation preservation we have $\nodel,v \models \Gamma$ therefore $\tup{\nodel,v} \in \classM$. Therefore \classM is closed under \cL-simulations.

	\item To see that $\Trm(\classM)$ is closed under ultraproducts take a family of models $\tup{\fmodel_{i},g_{i}} \in \Trm(\classM)$. Because every $\fmodel_{i},g_{i}$ is in $\Trm(\classM)$ we have that $\fmodel_{i},g_{i} \models \Trf(\Gamma)$ for all $i$. Let $\fmodel = \prod_D \fmodel_{i}$ be an ultraproduct of those models, by \cite[Theorem 4.1.9]{CK90} we have that $\fmodel,g^{*}_{i} \models \Trf(\Gamma)$ for $g^{*}_{i}(x) = \lambda z.g_{i}(x)$. Therefore $\tup{\fmodel,g^{*}_{i}} \in \Trm(\classM)$. Thus, the class is closed under ultraproducts.\footnote{This application is a corollary of The Fundamental Theorem of Ultraproducts. This same application can be seen in the proof of Theorem's 4.1.12 in the same book.}

	\item We still have to check that $\Trm(\CclassM)$ is closed under ultrapowers. Take $\tup{\fmodel,g} \in \Trm(\CclassM)$, by definition $\fmodel,g \not\models \Trf(\Gamma)$. Let $\fmodel_{*} = \prod_{D} \fmodel$ be an ultrapower of $\fmodel$, by \cite[Corollary 4.1.10]{CK90} the ultrapower is elementary equivalent to the original model. Hence, let $h(x) = \lambda z.g(x)$ be the canonical mapping, $\fmodel_{*},h \not\models \Trf(\Gamma)$. This means that $\tup{\fmodel_{*},h} \in \Trm(\CclassM)$ and therefore the class is closed under ultrapowers.\qedhere
\end{enumerate}
\end{proof}

\begin{proof}[From \ref{def:set:closure} to \ref{def:set:formulas}]
Suppose \classM is closed under \cL-simulations, $\Trm(\classM)$ is closed under ultraproducts and $\Trm(\CclassM)$ is closed under ultrapowers. Define the set $\Gamma = \Th(\classM)$.
Trivially $\classM \models \Gamma$, we still have to show that if $\model,w \models \Gamma$ then $\tup{\model,w} \in \classM$.
Define the following set
$$\NTh^{w}(x) = \setcomp{\neg\Trf(\varphi)}{\varphi \textrm{ is an \cL-formula and } \model,w \not\models \varphi}$$
Let's see that $\NTh^{w}(x)$ is finitely satisfiable in $\Trm(\classM)$. Suppose not, there is a finite subset $\Sigma_{0} \subseteq \NTh^{w}(x)$ such that $\Sigma_{0} = \{\lnot\sigma_{1},\dots,\lnot\sigma_{n}\}$ is not satisfiable in $\Trm(\classM)$. That means that the formula $\psi = \lnot(\lnot\sigma_{1} \land \dots \land \lnot\sigma_{n})$ is valid in $\Trm(\classM)$. Observe that $\psi$ is equivalent to $\psi' = \sigma_{1} \lor \dots \lor \sigma_{n}$. As the formula translation preserves disjunction and truth there exists an \cL-formula $\psi^{*}$ such that $\psi' \theq{\cFO} \Trf(\psi^{*})$. Hence $\Trf(\psi^{*})$ is valid in $\Trm(\classM)$ and therefore $\psi^{*} \in \Gamma$. This is absurd because it is obvious that $\model,w \not\models \psi^{*}$ and by hipothesis $\model,w \models \Gamma$.

Having proved that every subset of $\NTh^{w}(x)$ is satisfiable, by relative compactness, there is a model $\tup{\nodel,v} \in \classM$ such that $\Trm(\nodel, v) \models \NTh^{w}(x)$. We have already proved (in Proposition \ref{propo:modsinc}) that these models satisfy $\nodel, v \thsub{\cL} \model,w$.

Suppose that $\tup{\model,w} \in \CclassM$, using Theorem \ref{thm:similarultrapowers} (with $\classM_{1} = \classM$ and $\classM_{2} = \CclassM$) we can conclude that there exist models $\tup{\nodel^{*},v^{*}} \in \classM$ and $\tup{\model^{*},w^{*}} \in \CclassM$ such that $\nodel^{*},v^{*} \rsim_{\cL} \model^{*},w^{*}$. As \classM is closed under simulations then $\tup{\model,w} \in \classM$. Absurd, therefore $\tup{\model,w}$ must be in \classM.
\end{proof}

\begin{nota}
Let  $\tup{\fmodel,g}, \tup{\fnodel,h} \in \pmods{\cFO}$ we write $\fmodel,g \pisom \fnodel,h$ to mean that there exists a potential isomorphism $I$ between \fmodel and \fnodel such that $\tup{a} I \tup{b}$ where $a = g(x)$ and $b = h(x)$. That is, there is a potential isomorphism that links the elements assigned by $g$ and $h$.
\end{nota}

\begin{defn}[\classC-closure under potential isomorphisms] Let $\classC \subseteq \mods{\cFO}$. A class $\classK \subseteq \classC$ is \emph{\classC-closed  under potential isomorphisms} if for every $\fmodel \in \classK$ and $\fnodel \in \classC$ such that $\fmodel \pisom \fnodel$ then $\fnodel \in \classK$.

The definition for pointed models is similar. Let $\classC \subseteq \pmods{\cFO}$. A class $\classK \subseteq \classC$ is \emph{\classC-closed  under potential isomorphisms} if for every $\tup{\fmodel,g} \in \classK$ and $\tup{\fnodel,h} \in \classC$ such that $\fmodel,g \pisom \fnodel,h$ then $\tup{\fnodel,h} \in \classK$.
\end{defn}

\begin{lem}\label{lem:simthenpisom}
Let $\classM \subseteq \pmods{\cL}$. If \classM is closed under \cL-simulations and both $\Trmk{\classK}(\classM)$ and $\Trmk{\classK}(\CclassM)$ are closed under ultrapowers then $\Trmk{\classK}(\classM)$ and $\Trmk{\classK}(\CclassM)$ are \classK-closed under potential isomorphisms.
\end{lem}
\begin{proof}[Proof for $\Trm(\classM)$.]
Suppose that $\Trm(\classM)$ is not \classK-closed under potential isomorphisms. This means that there exist models $\tup{\fmodel,g} \in \Trm(\classM)$ and $\tup{\fnodel,h} \in \Trm(\CclassM)$ such that $\fmodel,g \pisom \fnodel,h$. Recall that $\classK \setminus \Trm(\classM) = \Trm(\CclassM)$. For a smoother proof, call their modal counterparts $\model,w$ and $\nodel,v$ respectively. Therefore $\tup{\model,w} \in \classM$ and $\tup{\nodel,v} \notin \classM$.

As $\fmodel,g \pisom \fnodel,h$ we know by~\cite[Proposition 2.4.4]{CK90} that $\fmodel,g \models \varphi(x)$ if and only if $\fnodel,h \models \varphi(x)$. In particular they have the same modal theory, $\model,w \theq{\cL} \nodel,v$. As this implies that $\model,w \thsub{\cL} \nodel,v$ we can use Theorem~\ref{thm:similarultrapowers} (instantiating with $\classK_{1} = \Trm(\classM)$, $\classK_{2} = \Trm(\CclassM)$ and \model, \nodel interchanged) and get models $\tup{\model^{*},w^{*}} \in \classM$ and $\tup{\nodel^{*},v^{*}} \in \CclassM$ such that $\model^{*},w^{*} \rsim_{\cL} \nodel^{*},v^{*}$.

Knowing that $\model^{*},w^{*} \rsim_{\cL} \nodel^{*},v^{*}$ and that \classM is closed under simulations we conclude that $\tup{\nodel^{*},v^{*}} \in \classM$. This is absurd because it contradicts $\tup{\nodel^{*},v^{*}} \in \CclassM$. Hence $\Trmk{\classK}(\classM)$ is \classK-closed under potential isomorphisms.
\end{proof}

\begin{proof}[Proof for $\Trm(\CclassM)$.]
To see that $\Trm(\CclassM)$ is \classK-closed under potential isomorphisms we argue by contradicction. Suppose not, then there exist $\tup{\fmodel,g} \in \Trm(\CclassM)$ and $\tup{\fnodel,h} \in \classK \setminus \Trm(\CclassM)$ such that $\fmodel,g \pisom \fnodel,h$. As $\tup{\fnodel,h} \in \classK\setminus\Trm(\CclassM)$ this means that $\tup{\fnodel,h} \in \Trm(\classM)$. We have just proved that $\Trm(\classM)$ is \classK-closed under potential isomorphism then, as $\fmodel,g \pisom \fnodel,h$, we conclude that $\tup{\fmodel,g} \in \Trm(\classM)$ which contradicts our hypothesis.\footnote{Here we use the symmetry of the potential isomorphism relation.} Absurd.
\end{proof}

\begin{thm}[Definability by a single formula]\label{thm:def:formula}
Given an adequate pair $\tup{\cL, \classK}$, and a class of models $\classM \subseteq \mods{\cL}$, the following are equivalent
\begin{enumerate}[(i).]
 \item \label{def:form:formulas}\classM is definable by a single \cL-formula.
 \item \label{def:form:closure}\classM is closed under \cL-simulations and both $\Trm(\classM)$ and $\Trm(\CclassM)$ are closed under ultraproducts.
\end{enumerate}
\end{thm}

\begin{proof}[From \ref{def:form:formulas} to \ref{def:form:closure}]
Suppose \classM is definable by a single \cL-formula $\varphi$.
\begin{enumerate}
 \item Let's see that $\Trm(\classM)$ and $\Trm(\CclassM)$ are closed under ultraproducts. Recall that \classM is definable by a single \cL-formula $\varphi$. Take the class of first order models defined by $\Trf(\varphi)$ and call it $\classM^{e}$. Observe that $\classM^{e}$ can be expressed as the disjunct union $\classM^{e} = \Trm(\classM) \cup \classM'$ between the translation of \classM and some other models that do not fall in \classK. Therefore $\Trm(\CclassM) = \overline{\classM^{e}} \cap \classK = \mods{\lnot\Trf(\varphi)} \cap \classK$. The following diagram helps illustrate the different classes. The box represents the class of all \cFO models, \classK is the class with an irregular border and $\classM^{e}$ is the oval.
 \begin{center}
 \includegraphics{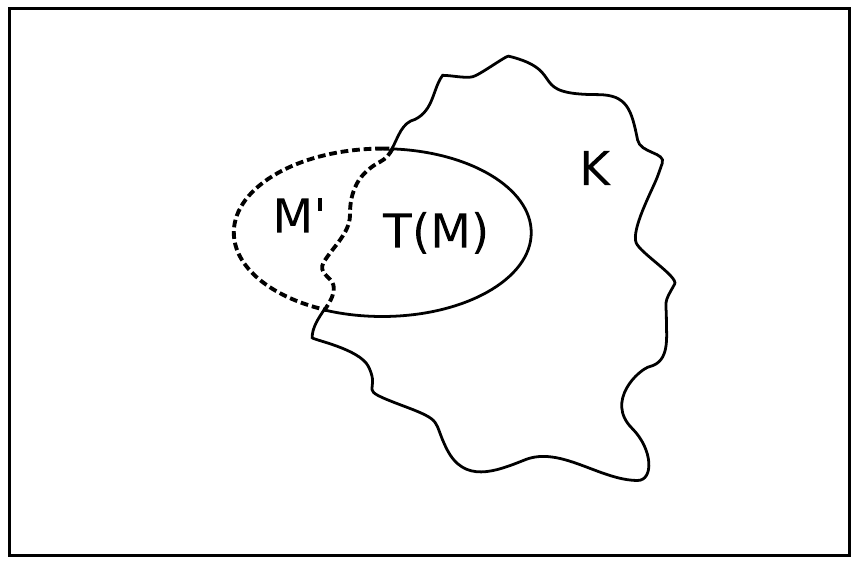}
 \end{center}
 Using Theorem \ref{thm:def:set}, as \classM is defined by the singleton set $T = \{\varphi\}$, $\Trm(\classM)$ is closed under ultraproducts. To see that $\Trm(\CclassM)$ is closed under ultraproducts proceed as follows.
\begin{enumerate}[(a)]
\item As \classK is closed under ultraproducts then, any ultraproduct from \classK must reside in \classK. In particular, any ultraproduct from $\Trm(\CclassM)$ must reside in \classK.
\item As $\overline{\classM^{e}}$ is defined by $\lnot\Trf(\varphi)$, it is closed under ultraproducts \cite[Corollary 6.1.16]{CK90}. This means that any ultraproduct from $\overline{\classM^{e}}$ must reside in $\overline{\classM^{e}}$, in particular, any ultraproduct from $\Trm(\CclassM)$ must reside in $\overline{\classM^{e}}$.
\end{enumerate}
From (a) and (b) we can conclude that any ultraproduct from $\Trm(\CclassM)$ must be in $\classK \cap \overline{\classM^{e}} = \Trm(\CclassM)$.
 \item Using Theorem \ref{thm:def:set}, as \classM is defined by the singleton set $T = \{\varphi\}$ we can be sure that \classM is closed under \cL-simulations. \qedhere
\end{enumerate}
\end{proof}

\begin{proof}[From \ref{def:form:closure} to \ref{def:form:formulas}]
Suppose that \classM is closed under \cL-simulations and both $\Trm(\classM)$ and $\Trm(\CclassM)$ are closed under ultraproducts. Using Theorem \ref{thm:def:set} we have a set of formulas $\Gamma$ defining \classM. By Lemma \ref{lem:simthenpisom} $\Trm(\classM)$ and $\Trm(\CclassM)$ are \classK-closed under potential isomorphisms. Now we use the relativized version of first order's definability result, Theorem \ref{thm:def:relative}, and conclude that there is a first order formula $\alpha(x)$ such that for every $\tup{\fmodel,g} \in \classK$; $\fmodel,g \models \alpha(x)$ if and only if $\tup{\fmodel,g} \in \Trm(\classM)$.

As \classM is closed under \cL-simulations then $\alpha$ is \classK-invariant for \cL-simulations. Using the Characterization Theorem (Theorem \ref{thm:charact}) we can conclude that $\alpha(x)$ is \classK-equivalent to the translation of a modal formula $\varphi$.
Therefore there exists $\varphi$ that defines \classM.
\end{proof}

\clearpage

\chapter{Applications}\label{chap:app}
In this chapter we will use the results that we have developed in the previous chapters and derive the characterization and definability theorems for particular cases of modal logics. 

\section{Memory Logics}
Memory logics are a novel family of modal logics introduced in~\cite{AFFM08}. They allow to model dynamic behavior through explicit memory operators that \textit{change} the evaluating structure. This proposal introduces a framework for studying the notion of state in a more general way, without bounding the analysis to any fixed domain (like knowledge change, time flow, linguistics contexts, etc.). Most of the work that has been done in this direction implicitly adds some specific native behavior in the ``dynamic component''. The approach presented in this paper wants to study some of the dynamic capabilities of the above mentioned approaches from a more abstract point of view, and analyze the different aspects of this family in terms of logic properties. 

This family of logics present several ``memory operators'' that can be considered modularly. We first present the syntax, signature and models for a broad set of operators and then analyze different possible combinations which form interesting logics.

It is important to note that there are no Characterization and Definability results known for this family of logics at this moment. Therefore, the results obtained through the use of the framework that we've developed will be original. We present the results for the unimodal case but it can be easily generalized for the multi-modal case.

\begin{defn}[Signatures]\label{signatures}
Let $\prop=\{p_1, p_2, \dots\}$ (the propositional symbols)
be a countable infinite set of symbols
and $\rel=\{r\}$ (the relational symbols) be disjoint. The source
signature is defined to be $\ssig=\tup{\prop,\rel}$.

Let $\fprop=\{P_1, P_2, \dots\}$ (the propositional predicates) and
$\frel=\{R\}$ (the relational predicates)
The target first-order signature is defined to be
${\tsig=\tup{\fprop \cup \frel,\emptyset,\emptyset}}$ \emph{with} equality.
\end{defn}

\begin{defn}[Syntax]\label{ml:syntax}
The syntax of the Memory Logics family over a given signature $\tup{\prop, \rel}$ is defined as an extension of the propositional calculus with the following operators:
$$
\varphi
::=  \dots
    \mid \known
    \mid \remember \varphi
    \mid \erase \varphi
    \mid \forget \varphi
    \mid \diam{r} \varphi
    \mid \ddiam{r} \varphi
$$

\medskip
\noindent
where $r \in \rel$. We define the dual of $\ddiam{r}$ in the usual way: for all $r \in \rel$, $\bbox{r}\varphi$ can be defined as $\lnot \ddiam{r} \lnot \varphi$. We usually call these operators `known', `remember', `erase', `forget' and `double diamond'. Every logic of this family will be \emph{required} to have at least the $\remember$ and $\known$ operators and can have any combination of the other operators.
\end{defn}
Observe that, defined this way, in the family of memory logics one has two types of diamonds: the `single' diamond of \bml and the `double diamond'. The semantic of the first is defined as usual and the later will be defined shortly. 

\bigskip
The family of memory logics that we will be working with are all evaluated in an extension of Kripke models with a set that we call the `memory' of the model. It is defined as follows, along with its first-order equivalent.
\begin{defn}[Models] A \emph{model} for the source language over the signature \ssig is a tuple $\model = \tup{W, R, V, S}$ satisfying
\begin{enumerate}[(i)]
\noitemsep
\item $W$ is a nonempty set,
\item $R \subseteq W\times W$ is a binary relation over $W$,
\item $V:\prop \to \pw{W}$ is a valuation function, and
\item $S \subseteq W$ is the memory of the model.
\end{enumerate}
An \emph{\cFO-model} for the target language is a tuple 
$\tmodel = \tup{W^{t}, R^{t}, \tprops{\fprop},K}$ where
\begin{enumerate}[(i)]
\noitemsep
\item $W^{t}$ is a nonempty set,
\item $K \subseteq W$,
\item $R^{t} \subseteq W^{t}\times W^{t}$ is a binary relation over $W^{t}$, and
\item $\tprops{\fprop}$ are unary relations over $W^{t}$.
\end{enumerate}
\end{defn}

\begin{nota}
In the rest of this section the following notation will be useful. Let $\model = \tup{W,R,$ $V,S}$ be a model, $w \in W$, and $S' \subseteq W$ then we define
$$
\begin{array}{lcl}
\model[+w] & = & \tup{W,R,V,S \cup \{w\}}\\
\model[-w] & = & \tup{W,R,V,S \setminus \{w\}}\\
\model[+S'] & = & \tup{W,R,V,S \cup S'}\\
\model[-S'] & = & \tup{W,R,V,S \setminus S'}\\
\model[*] & = & \tup{W,R,V,\emptyset}.
\end{array}
$$
We usually write $\model[w]$ instead of $\model[+w]$.
\end{nota}

\begin{defn}[Semantics]\label{def:memory-semantics}
Given a model $\model = \tup{W, R, V, S}$ and $w \in W$, we extend the \emph{propositional} part of the semantics presented in Definition~\ref{def:ml-semantics} with the following rules:
$$
\begin{array}{rcl}
\model, w \models \known & \tiff & w \in S \label{known} \\
\model, w \models \remember \varphi & \tiff & \model[w], w \models \varphi \label{remember}\\
\model, w \models \forget \varphi & \tiff & \model[-w], w \models \varphi \label{forget}\\
\model, w \models \erase\varphi & \tiff & \model[*], w \models \varphi \\
\model, w \models \diam{r} \varphi & \tiff & \exists w' \in W,w R w' \textrm{ and } \model, w' \models \varphi.\\
\model, w \models \ddiam{r} \varphi & \tiff & \exists w' \in W,w R w' \textrm{ and } \model[w], w' \models \varphi.
\end{array}
$$
\end{defn}
Observe that the double diamond acts as a normal diamond \emph{but} it always remembers the current state before moving. Hence, it can be thought as if the formula were leaving a trace while being evaluated in the model.

\begin{defn}[Formula translation]
We will not give an explicit translation for these logics, as we will not use it explicitly.
We know there exists a translation to $\cFOu_{=}$ because, in~\cite{AFFM08}, there is an explicit translation from memory logics to \hl and there is a translation from $\hl$ to $\cFOu_{=}$ given in~\cite{BBW06}. This translation preserves both conjunction and disjunction as needed. Let's call it $\Trf: \form{\ssig} \to \form{\tsig}$.
\end{defn}

\begin{defn}[Model translation] Let \classK be the class of all models for the signature \tsig. Let $\model = \tup{W, R, V, S}$ and $\tmodel = \tup{W^{t}, R^{t}, \tprops{\fprop},K}$.
Define the \emph{model translation} $\Trmk{\classK}(\model,w) = \tup{\model^{t},g^{t}}$ to be the function induced by the following equations
\begin{center}
$
\begin{array}{rcl}
W^{t} & = & W \\
P^{t}_i & = & V(p_i) \\
K & = & S \\
R^{t} & = & R \\
g^{t}(x) & = & w
\end{array}
$
\end{center}
\end{defn}

The simulation notion for a logic of this family allows a very modular definition. Let $\sim$ be a binary relation between memory pointed models. So $\sim$ relates tuples
$\tup{\model, m}$ with $\tup{\nodel,n}$.

A simulation for a memory logic \cL can be defined imposing restrictions to $\sim$ depending on the operators that \cL has. In the following table we summarize the restrictions associated with each operator. We write $S^{\model}$ to refer to the memory of the model \model. $R^{1}_{r}$ is used to denote a relation in \model and $R^{2}_{r}$ is used to denote a relation in \nodel.
\begin{figure}[h]
\centering
\begin{small}
\renewcommand{\arraystretch}{1.2}
\begin{tabular}{|c|c|p{9.5cm}|}
\hline
\textit{always} & (nontriv) & $\sim$ is not empty.\\
\hline
\textit{always} & (agree) & If $\tup{\model, m} \sim \tup{\nodel, n}$, then $m$ and $n$ make the same propositional variables true.\\
\hline
{\normalsize$\known$} & (kagree) & If $\tup{\model, m} \sim \tup{\nodel, n}$, then $m \in S^{\model}$ if and only if $n \in S^{\nodel}$.\\
\hline
{\normalsize$\remember$} & (remember) & If $\tup{\model, m} \sim \tup{\nodel, n}$, then  $\tup{\model[m], m} \sim \tup{\nodel[n], n}$.\\
\hline
{\normalsize$\forget$} & (forget) & If $\tup{\model, m} \sim \tup{\nodel, n}$, then  $\tup{\model[-m], m} \sim \tup{\nodel[-n], n}$.\\
\hline
{\normalsize$\erase$} & (erase) & If $\tup{\model, m} \sim \tup{\nodel, n}$, then  $\tup{\model[*], m} \sim \tup{\nodel[*], n}$.\\
\hline
{\normalsize$\diam{r}$} & (forth) & If $\tup{\model, m} \sim \tup{\nodel, n}$ and
$R_r^1(m,m')$, then there exists ${n' \in |\nodel|}$
such that $R_r^2(n,n')$  and $\tup{\model, m'} \sim \tup{\nodel, n'}$.\\
& (back) & If $\tup{\model, m} \sim \tup{\nodel, n}$  and
$R_r^2(n,n')$, then there exists ${m' \in |\model|}$
such that $R_r^1(m,m')$  and $\tup{\model , m'} \sim \tup{\nodel, n'}$.\\
\hline
{\normalsize$\ddiam{r}$} & (mforth) & If $\tup{\model, m} \sim \tup{\nodel, n}$ and
$R_r^1(m,m')$, then there exists ${n' \in |\nodel|}$
such that $R_r^2(n,n')$  and $\tup{\model[m], m'} \sim \tup{\nodel[n], n'}$.\\
& (mback) & If $\tup{\model, m} \sim \tup{\nodel, n}$  and
$R_r^2(n,n')$, then there exists ${m' \in |\model|}$
such that $R_r^1(m,m')$ and $\tup{\model[m], m'} \sim \tup{\nodel[n], n'}$.\\
\hline
\end{tabular}
\renewcommand{\arraystretch}{1}
\end{small}
\caption{Operator restrictions for a modular memory simulation definition.}
\label{fig:memmodular}
\end{figure}

\begin{defn}[Memory simulation]
From now on, given a memory logic \cL, we will refer as `the simulation for \cL' to the simulation defined by the sum of the necessary conditions of Figure~\ref{fig:memmodular} for the operators in \cL.
\end{defn}

Observe that, as every memory logic has negation, the simulation notion for memory logics will be symmetrical. Therefore we will use the bisimulation symbol $\bisim$ to note memory simulations.

\begin{exam}\label{exam:ml}
To give some examples of possible memory logics we cite the following ones which are named in~\cite{phdmera}.
\begin{center}
$
\begin{array}{rcl}
\cMLRK & = & \{\remember, \known, \diam{r}\} \\
\cMLRKM & = & \{\remember, \known, \ddiam{r}\} \\
\cMLRKF & = & \{\remember, \known, \forget, \diam{r}\} \\
\cMLRKMF & = & \{\remember, \known, \forget, \ddiam{r}\} \\
\cMLRKE & = & \{\remember, \known, \erase, \diam{r}\} \\
\cMLRKME & = & \{\remember, \known, \erase, \ddiam{r}\} \\
\cMLRKA & = & \{\remember, \known, \forget, \erase, \diam{r}\}
\end{array}
$
\end{center}
It is clear from the satisfaction definition of each operator that these logics have different capabilities. A detailed insight on the expressive power of these logics can be found in the aforementioned PhD. thesis.
\end{exam}

\begin{thm}
Let \cL be a memory logic, let $\tup{\model,w}$ and $\tup{\nodel,v}$ be two memory models. If $\tup{\model,w} \bisim_{\cL} \tup{\nodel,v}$ then $\tup{\model,w} \theq{\cL} \tup{\nodel,v}$.
\end{thm}
\begin{proof}
Part of the proof can be found in~\cite{phdmera}, it can be easily extended to he full set of operators. We will not present this proof here as it exceeds the focus of this thesis.
\end{proof}

Before starting with the proof of the main theorem of this section we will prove some lemmas that will be useful. The model may change during the evaluation of a formula. For our special case, it will be enough to prove that adding a state to the memory preserves $\omega$-saturation.
\begin{lem}\label{lem:addwsat}
If \model is \wsat then $\model[+A]$ is \wsat too for all finite $A \subseteq |\model|$.
\end{lem}
\begin{proof}
The proof of this lemma can be found in~\cite[Lemma 5.2.2]{phdmera}.
\end{proof}

\begin{lem}\label{lem:wsatsuc}
Let $\model = \tup{W,R,\dots}$ be a \wsat Kripke model whose translation preserves the structure of the domain and the relations, that is, $\Trm(\model) = \tup{W,R,\dots}$ where $R \subseteq W\times W$.

Let $\Sigma$ be a set of modal formulas and ${w \in W}$. If every finite subset $\Delta \subseteq \Sigma$ satisfies $\model,v_{\Delta} \models \Delta$ where $v_{\Delta}$ is an $R$-successor of $w$ then there exists $v$, an $R$-successor of $w$, such that $\model,v \models \Sigma$.
\end{lem}
\begin{proof}
Recall that the definition of $\omega$-saturation lets us extend the first order language with a constant $\underline{a}$ for each element $a \in W$. Define $\Sigma^{*} = \{R\underline{w}x\} \cup \Trf(\Sigma)$.

If we show that $\Sigma^{*}$ is satisfiable in some pointed model $\Trm(\model,v)$ it is clear that $\Sigma$ will be satisfiable in a successor of $w$. This is because the domain and relations of \model and $\Trm(\model)$ are the same and if $\Trm(\model) \models R\underline{w}x[a]$ this means that $a$ is a successor of $w$. The rest of the proof will focus on proving that $\Sigma^{*}$ is satisfiable in the pointed model $\Trm(\model,v)$.

\medskip
Take a finite subset $\Sigma_{0} \subseteq \Sigma^{*}$. Observe that this set should satisfy the following inclusion $\Sigma_{0} \subseteq \{R\underline{w}x,\sigma_{1},\dots,\sigma_{n}\}$ with $\sigma_{i} \in \Trf(\Sigma)$. Therefore, if we show that this bigger set is satisfiable, it will also be the case with $\Sigma_{0}$.

By hypothesis, every finite subset of $\Sigma$ is satisfiable in a successor of $w$. Take the finite subset $\Delta$ such that $\Trf(\Delta) = \{\sigma_{1},\dots,\sigma_{n}\}$. This $\Delta$ is satisfiable in some successor $v_{\Delta}$ which means that $R(w,v_{\Delta})$. We can conclude that $\Trm(\model,v_{\Delta}) \models R\underline{w}x$ and $\Trm(\model,v_{\Delta}) \models \Trf(\Delta)$.

We have taken an arbitrary finite subset of $\Sigma^{*}$ and shown that it is satisfiable. By $\omega$-saturation we can conclude that the set $\Sigma^{*}$ is also satisfiable. 
\end{proof}

To be able to derive the characterization and definability results using the framework developed in the previous chapters we need to prove that, for every memory logic \cL, the class of \wsat models has the Hennessy-Milner property with respect to \cL-simulations. Each logic will have its own definition of simulation with the proper restrictions listed above.

As we want to consider all the possible logics from the family of memory logics we will need to prove that, given two models $\tup{\model,w}$ and $\tup{\nodel,v}$ such that $\tup{\model,w} \theq{\cL} \tup{\nodel,v}$ we can construct an \cL-simulation between them. We will do this by considering every possible operator and show that we can construct a simulation that satisfies the constraints associated for that operator.

\begin{thm}\label{app:mem:wsat}
Let \cL be a memory logic, the class of \wsat models has the Hennessy-Milner property with respect to \cL-simulations.
\end{thm}
\begin{proof}
Given two \wsat models \model, \nodel it suffices to give an \cL-simulation between them. We propose the binary relation $\sim$ defined as $$\tup{\model, w} \sim \tup{\nodel, v}\ \ \textrm{iff}\ \ \model, w \equiv_{\cL} \nodel, v$$
Suppose that $\tup{\model, w} \sim \tup{\nodel, v}$. We first show that this relation satisfies the \emph{(nontriv)} and \emph{(agree)} restrictions which apply for every combination of operators, then we will undertake the proof for each special operator.
\begin{paragraph}{\bf Basic restrictions.}
We can see that if we are given two equivalent worlds in two different models then, by definition, the relation will have at least one element and therefore \emph{(nontriv)} will be satisfied. Also, the definition of the relation implies that $w$ and $v$ make true the same propositional variables and therefore \emph{(agree)} is satisfied.
\end{paragraph}
\begin{paragraph}{\bf Restrictions for $\known$.}
We need to show that $w$ is known in \model if and only if $v$ is known in \nodel. The proof goes through easily using the satisfaction definition of the known operator $$w \in S^{\model} \iff \model, w \models \known \iff \nodel, v \models \known \iff v \in S^{\nodel}.$$
The first and last implications are because of the semantics of $\known$, the implication in the middle is because of \cL-equivalence between $w$ and $v$. This proves that \emph{(kagree)} is satisfied.
\end{paragraph}
\begin{paragraph}{\bf Restrictions for $\remember$.}
As we suppose that $\tup{\model, w} \sim \tup{\nodel, v}$ then for every $\varphi$ we have $$\model,w \models \varphi\ \ \iff\ \ \nodel,v \models \varphi$$
so, given a formula $\psi$ we can instantiate the equivalence and get $$\model,w \models \remember\psi\ \ \iff\ \ \nodel,v \models \remember\psi$$ which by satisfaction definition holds precisely when $$\model[+w],w \models \psi\ \ \iff\ \ \nodel[+v],v \models \psi$$ that means that those two states are equivalent and we can conclude (by def. of $\sim$) that $$\tup{\model[w], w} \sim \tup{\nodel[v], v}$$
This proves that \emph{(remember)} is satisfied.
\end{paragraph}
\begin{paragraph}{\bf Restrictions for $\forget$.}
As in the last paragraph, for every $\varphi$ we have $$\model,w \models \varphi\ \ \iff\ \ \nodel,v \models \varphi$$
so, given a formula $\psi$ we can instantiate the equivalence and get $$\model,w \models \forget\psi\ \ \iff\ \ \nodel,v \models \forget\psi$$ which by satisfaction definition holds precisely when $$\model[-w],w \models \psi\ \ \iff\ \ \nodel[-v],v \models \psi$$ that means that those two states are equivalent and we can conclude (by def. of $\sim$) that $$\tup{\model[-w], w} \sim \tup{\nodel[-v], v}$$
This proves that \emph{(forget)} is satisfied.
\end{paragraph}
\begin{paragraph}{\bf Restrictions for $\erase$.}
We proceed as usual, for every $\varphi$ we have $$\model,w \models \varphi\ \ \iff\ \ \nodel,v \models \varphi$$
so, given a formula $\psi$ we can instantiate the equivalence and get $$\model,w \models \erase\psi\ \ \iff\ \ \nodel,v \models \erase\psi$$ which by satisfaction definition holds precisely when $$\model[*],w \models \psi\ \ \iff\ \ \nodel[*],v \models \psi$$ that means that those two states are equivalent and we can conclude (by def. of $\sim$) that $$\tup{\model[*], w} \sim \tup{\nodel[*], v}$$
This proves that \emph{(erase)} is satisfied.
\end{paragraph}
\begin{paragraph}{\bf Restrictions for $\diam{r}$.}
As we have $\tup{\model, w} \sim \tup{\nodel, v}$ then $\model, w \equiv \nodel, v$. Suppose that $w'$ is a successor of $w$. Let $\Sigma$ be the set of all the formulas true at $\model, w^\prime$. For every finite subset $\Delta \subseteq \Sigma$ we have $\model, w^\prime \models \bigwedge\Delta$ and therefore \mbox{$\model, w \models \Diamond\bigwedge\Delta$}. By \cL-equivalence we have \mbox{$\nodel, v \models \Diamond\bigwedge\Delta$} which means that for every $\Delta$ we have a $v$-succesor which satisfies it. By Lemma~\ref{lem:wsatsuc} we can conclude that there exists $v^\prime$ a $v$-succesor so that $\nodel, v^\prime \models \Sigma$.

As $\model, w^\prime$ and $\nodel, v^\prime$ make the same formulas true then they are \cL-equivalent and by definition they will be related by the simulation. We conclude that $\tup{\model, w'} \sim \tup{\nodel, v'}$. This proves that \emph{(forth)} is satisfied. The proof for \emph{(back)} is similar but switching the models.

An alternative proof of this lemma, which uses a notion called m-saturation, can be found in~\cite{MLBOOK}.
\end{paragraph}
\begin{paragraph}{\bf Restrictions for $\ddiam{r}$.}
As we have $\tup{\model, w} \sim \tup{\nodel, v}$ then for every $\varphi$ we have $$\model,w \models \varphi\ \ \iff\ \ \nodel,v \models \varphi$$
therefore if, given an arbitrary $\psi$ we instantiate $\varphi = \remember\psi$ we get\footnote{We can use `remember' here because we required that every memory logic should have it. Without this requirement we would've been able to use only $\ddiam{r}$.}
$$\model, m \models \remember\psi\ \iff\ \nodel, n \models \remember\psi$$
which, by satisfaction definition holds exactly when
\begin{equation}\label{mlrkm:equiv}
\model[m], m \models \psi\ \iff\ \nodel[n], n \models \psi
\end{equation}
Observe that equation~\ref{mlrkm:equiv} implies that $\model[m], m \theq{\cL} \nodel[n], n$. Using Lemma~\ref{lem:addwsat} we also know that $\tup{\model[m], m}$ and $\tup{\nodel[n], n}$ are both \wsat.

Suppose that $w'$ is a successor of $w$. Let $\Sigma$ be the set of all the formulas true at $\model[w], w'$. For every finite subset $\Delta \subseteq \Sigma$ we have $\model[w], w' \models \bigwedge\Delta$ and therefore \mbox{$\model[w], w \models \ddiam{r}\bigwedge\Delta$}. By \cL-equivalence we have \mbox{$\nodel[v], v \models \ddiam{r}\bigwedge\Delta$} which means that for every $\Delta$ we have a $v$-succesor which satisfies it. By Lemma~\ref{lem:wsatsuc} we can conclude that there exists $v'$ a $v$-succesor so that $\nodel[v], v' \models \Sigma$.

As $\model[w], w'$ and $\nodel[n], v'$ make the same formulas true then they are \cL-equivalent and by definition they will be related by the simulation. This proves that \emph{(mforth)} is satisfied because ${\tup{\model[w], w'} \sim \tup{\nodel[v], v'}}$. The proof for \emph{(mback)} is similar but switching the models.
\end{paragraph}

After analyzing all the possible operators we have shown that, for every case, the \cL-simulation relation will satisfied the required constraints. This proves that given two \wsat \cL-equivalent models we are able to construct an \cL-simulation between them. Therefore, the \wsat class of models has the Hennessy-Milner property with respect to \cL-simulations.
\end{proof}

\begin{cor}
The definitions given above, along with Theorem~\ref{app:mem:wsat}, prove that the pair $\tup{\cL,\classK}$ is an adequate pair for every memory logic \cL. Therefore, the Characterization and Definability theorems~(\ref{thm:charact}, \ref{thm:def:set}~and~\ref{thm:def:formula}) hold for this family of logics. In particular, these theorems hold for the logics in Example~\ref{exam:ml}.
\end{cor}
\section{Hybrid Logics}
Hybrid Logics augment modal logics with machinery for describing and 
reasoning about identity, which is a crucial in many settings.
The notion of identity comes with the introduction of `nominals'
and operators to reason about them. In spirit, nominals work
mostly as propositional variables but they have the particularity
of being true in \emph{at most} one point. We start by defining the
signatures for these logics.

\begin{defn}[Signatures]
Let $\prop=\{p_1, p_2, \dots\}$ (the propositional symbols),
$\nom=\{i_{1}, i_{2},\dots\}$ (the nominal symbols)
and $\rel=\{r_1, r_2, \dots\}$ (the relational symbols) be
disjoint, countable infinite sets of symbols. The source
signature is defined to be $\ssig=\tup{\prop,\rel,\nom}$.

Let $\fprop=\{P_1, P_2, \dots\}$ (the propositional predicates),
$\frel=\{R_{1}, R_{2},\dots \}$ (the relational predicates) and
$\fconst=\{c_{1}, c_{2},\dots \}$ (the constants).
The target first-order signature is defined to be
${\tsig=\tup{\fprop \cup \frel,\fconst,\emptyset}}$ \emph{with} equality.
\end{defn}

In this thesis we will (re)prove the characterization and definability theorem for a small family of hybrid logics. We will only consider the cases which extend \bml with nominals and possibly the $@$ operator. There exist other hybrid important logics such as, for example, the ones which include the downarrow binder $\downarrow$. Results for these logics are nicely developed in~\cite[Chapter 12]{BBW06}.

\begin{defn}[Syntax]\label{hyb:syntax}
The syntax of the Hybrid Logic family over a given signature $\tup{\prop, \rel, \nom}$ is defined as an extension of the propositional calculus with the following operators:
$$
\varphi
::=  \dots
    \mid i
    \mid @_{i} \varphi
    \mid \diam{r} \varphi
$$

\medskip
\noindent
where $i \in \nom$ and $r \in \rel$. We define the dual of $\diam{r}$ in the usual way.
\end{defn}

\begin{defn}[Models] A \emph{hybrid model} for the source language is a tuple
$$\model = \tup{W, \rels{\rel}, V, G}$$ which satisfies
\begin{enumerate}[(i)]
\noitemsep
\item $W$ is a nonempty set,
\item $R_{r} \subseteq W\times W$ are binary relations over $W$,
\item $V:\prop \to \pw{W}$ is a valuation function, and
\item $G:\nom \to W$ is an assignment for the nominals.\footnote{In the literature one can find an equivalent definition where $V:\prop \cup \nom\to\pw{W}$ and $G$ doesn't exist. In this case we should add $V(i) = |1|$ for all $i \in \nom$ as a restriction. It is easy to see that this two definitions are equivalent.}
\end{enumerate}
An \emph{\cFO-model} for the target language is a tuple 
$$\tmodel = \tup{W^{t}, \trels{\frel}, \tprops{\fprop},\tconsts{\fconst}}$$ which satisfies
\begin{enumerate}[(i)]
\noitemsep
\item $W^{t}$ is a nonempty set,
\item $R^{t}_{r} \subseteq W^{t}\times W^{t}$ are binary relations over $W^{t}$,
\item $\tprops{\fprop}$ are unary relations over $W^{t}$, and
\item $c^{t}_{i}$ are constants.
\end{enumerate}
\end{defn}

\begin{defn}[Semantics]\label{def:hybrid-semantics}
Given a model $\model = \tup{W, R, V, G}$ and $w \in W$, we extend the semantics presented in Definition~\ref{def:ml-semantics} (\bml semantics) with the following rules:
$$
\begin{array}{rcl}
\model, w \models i & \tiff & w = i \\
\model, w \models @_{i}\varphi & \tiff & \model, G(i) \models \varphi
\end{array}
$$
\end{defn}
Observe that the satisfaction definition for the nominals acts as an identity checker and the $@$-operator lets us `jump' to an identified world.

\begin{defn}[Formula translation]
A formula translation that meets our requirements is given in~\cite{MLBOOK,BBW06}. We will not give the explicit definition because we will not need to use it.
\end{defn}

\begin{defn}[Model translation]
Let \classK be the class of all first order models over the target signature \tsig. Let $\model = \tup{W, \rels{\rel}, V, G}$ and $$\tmodel = \tup{W^{t}, \trels{\frel}, \tprops{\fprop},\tconsts{\fconst}}.$$ Define the model translation $\Trmk{\classK}(\model,w) = \tup{\model^{t},g^{t}}$ to be the function induced by the following equations
\begin{center}
$
\begin{array}{rcl}
W^{t} & = & W \\
P^{t}_i & = & V(p_i)\ \textrm{for}\ p_{i} \in \prop \\
c^{t}_i & = & G(i)\ \textrm{for}\ i \in \nom \\
R^{t}_{i} & = & R_i \\
g^{t}(x) & = & w
\end{array}
$
\end{center}
\end{defn}

As with memory logics, the small family of hybrid logics that we will analyze also allows a modular simulation definition. Let $\sim$ be a binary relation between hybrid pointed models. A simulation for a hybrid logic \cL can be defined imposing restrictions to $\sim$ depending on the operators that \cL has. It is important to stress that every hybrid logic \emph{should} have nominals. In the following table we summarize the restrictions associated with each operator. We write $G_{1}$ for the nominal assignment of \model and $G_{2}$ for the nominal assignment of \nodel.
\begin{figure}[h]
\centering
\begin{small}
\renewcommand{\arraystretch}{1.2}
\begin{tabular}{|c|c|p{8cm}|}
\hline
\textit{nominals} & (nagree) & If $\tup{\model, m} \sim \tup{\nodel, n}$, then $G_{1}(i) = m$ if and only if $G_{2}(i) = n$ for all $i\in\nom$.\\
\hline
{\normalsize @} & (nom) & If $G_{1}(i) = w$ and $G_{2}(i) = v$ for some $i\in\nom$ then $\tup{\model,w} \sim \tup{\nodel,v}$.\\
\hline
\end{tabular}
\renewcommand{\arraystretch}{1}
\end{small}
\caption{Operator restrictions for a modular hybrid simulation definition.}
\label{fig:hybmodular}
\end{figure}

\begin{defn}[Hybrid simulation]
From now on, given a hybrid logic \cL, we will refer as `the simulation for \cL' to the simulation defined by the sum of the necessary conditions of Figure~\ref{fig:hybmodular} for the operators in \cL and the \emph{(nontriv)} and \emph{(agree)}.
\end{defn}

\begin{thm}
Let \cL be a hybrid logic defined as in Definition~\ref{hyb:syntax}, let $\tup{\model,w}$ and $\tup{\nodel,v}$ be two hybrid models. If $\tup{\model,w} \bisim_{\cL} \tup{\nodel,v}$ then $\tup{\model,w} \theq{\cL} \tup{\nodel,v}$.
\end{thm}
\begin{proof}
The proof of this theorem can be found in~\cite{BBW06}.
\end{proof}

In the following theorem we will prove that the \wsat class of models has the Hennessy-Milner property with respect to the simulations for the following hybrid logics.
\begin{center}
$
\begin{array}{rcl}
\hl & = & \{\text{nominals},\diam{r}\} \\
\hlA & = & \{\text{nominals}, \diam{r}, @\}
\end{array}
$
\end{center}
We will achieve this by showning that, given two equivalent pointed hybrid models, we can construct a simulation between them.

\begin{thm}\label{app:hl:wsat}
Let \cL be a hybrid logic as in Definition~\ref{hyb:syntax}, the class of \wsat models has the Hennessy-Milner property with respect to \cL-simulations.
\end{thm}
\begin{proof}
Given two \wsat models \model, \nodel it suffices to give an \cL-simulation between them. We propose the binary relation $\sim$ defined as $$\tup{\model, w} \sim \tup{\nodel, v}\ \ \textrm{iff}\ \ \model, w \equiv_{\cL} \nodel, v$$
Suppose that $\tup{\model, w} \sim \tup{\nodel, v}$. The proof for the \emph{(nontriv)}, \emph{(agree)}, \emph{(forth)} and \emph{(back)} restrictions are the same as for memory logics (see Theorem~\ref{app:mem:wsat}). We prove the restrictions for nominals and the $@$ operator.
\begin{paragraph}{\bf Restrictions for nominals.}
This proof goes through using the satisfaction definition for the nominals. Remember that the nominals can only be true in \emph{one} world.
$$G^{\model}(i) = w \iff \model, w \models i \iff \nodel, v \models i \iff G^{\nodel}(i) = v$$
the first and last implications are because of the semantics of nominal satisfaction and the implication in the middle is because of the \cL-equivalence between $\tup{\model,w}$ and $\tup{\nodel,v}$.
\end{paragraph}
\begin{paragraph}{\bf Restrictions for $@$.} 
Suppose that $G_{1}(i) = w$ and $G_{2}(i) = v$. As the relation is non-empty we can always get two equivalent worlds $a \in |\model|$ and $b \in |\nodel|$. Then we have
$$\model, a \models \varphi\ \textrm{iff}\ \nodel, b \models \varphi$$
for all $\varphi$. Given an arbitrary formula $\psi$ we can instantiate $\varphi = @_{i}\psi$ thus obtaining
$$\model, a \models @_{i}\psi\ \textrm{iff}\ \nodel, b \models @_{i}\psi$$
which by semantic definition means that
$$\model, G_{1}(i) \models \psi\ \textrm{iff}\ \nodel, G_{2}(i) \models \psi.$$
By hypothesis we can replace $G_{1}(i)$ and $G_{2}(i)$ and get
$$\model, w \models \psi\ \textrm{iff}\ \nodel, v \models \psi$$
therefore $\model,w \theq{\cL} \nodel,v$ and by definition $\tup{\model,w} \sim \tup{\nodel,v}$.
\end{paragraph}

After analyzing all the possible operators we have shown that, for every case, the \cL-simulation relation will satisfy the required constraints. This proves that given two \wsat \cL-equivalent models we are able to construct an \cL-simulation between them. Therefore, the \wsat class of models has the Hennessy-Milner property with respect to \cL-simulations.
\end{proof}

\begin{cor}
The definitions given above, along with Theorem~\ref{app:hl:wsat}, prove that the pair $\tup{\cL,\classK}$ is an adequate pair for every hybrid logic \cL. Therefore, the Characterization and Definability theorems~(\ref{thm:charact}, \ref{thm:def:set}~and~\ref{thm:def:formula}) hold for this family of logics.
\end{cor}

\clearpage

\chapter{Conclusions and further work}\label{chap:conclusions}
When developing a notion of simulation for a given logic \cL we need to be sure that we end up with the adequate notion. This means that it should \emph{exactly} characterize model equivalence. If we prove that
\begin{equation}\label{con:mitad}
\text{If } \tup{\model,w} \sim_{\cL} \tup{\nodel,v} \text{ then } \model,w \thsub{\cL} \nodel,v
\end{equation}
we have the half of the work done but the notion could still be wrong. Suppose that, for example, we say that the right notion of simulation for \bmlsn is the bisimulation notion of \bml. It is obvious that we will be able to prove (\ref{con:mitad}) but we are not working with the right notion of simulation: is is \emph{too strong} for \bmlsn.

In the process of finding the right simulation notion, candidates are often checked `against' finite or finitely branching models. In those cases, one expects to be able to prove the converse of (\ref{con:mitad}). As we have seen, these classes of models are special cases of \wsat models. In the development of this thesis we arrive to the conclusion that if we can prove the converse of (\ref{con:mitad}) for any \wsat model then we can, with little work, derive the Characterization and Definability theorems. This observation stresses the important relationship of \wsat models and the right simulation notion for a given logic.

\medskip
When we defined the notion of adequate pair in Definition~\ref{adeq} we explained the strength of the Hennessy-Milner requirement. One would expect that a true generalization of the Characterization and Definability theorem doesn't require the proof of a lemma. Instead, it should give a series of easily checkable properties that a logic should satisfy.

In sake of trying to give a result that could handle a broad spectrum of simulation notions we faced a big problem: we had no information regarding the structural properties of a simulation. In the applications chapter we saw that this information was \emph{essential} to prove that \wsat models had the Hennesy-Milner property.

We think that an important line of work lies in the effort of trying to prove the Hennessy-Milner property without much information about the simulation notion. In future work we plan to integrate the results of this thesis with an approach similar to the one presented in~\cite{dc:coind}, where coinductive model semantics are given.

Recall that in the beginning of Chapter~\ref{chap:framework} we presented two equivalent definitions of the `Basic Temporal Logic'. The classical one had custom made modalities $F$ and $P$ and the alternative view considered them as normal diamonds over a restricted class of models. The work done by Areces and Gor\'in in~\cite{dc:coind} generalizes this idea for (almost) any modality which can be defined with the pattern of the diamond operator ($\forall\exists$).

From our perspective, the most important point of their work is that, by restricting the class of models, we get a \emph{unique} notion of model equivalence for every logic that fits in their framework. The right simulation notion turns to be the same as \bml's bisimulation.

As far as we know, to the moment, there was no direct way to prove Characterization and Definability results using the framework developed in~\cite{dc:coind}. The problem laid in the restriction applied to the class of models, there is no classical proof which takes this kind of restrictions into account. With the results developed in this thesis it should be easy to prove a more general result using their framework.

\medskip
Good as it is, the framework developed in~\cite{dc:coind} has its limits. Not every modality can be expressed with the pattern of a diamond. For example, there exists an extension of basic temporal logic which adds the Since and Until operators~\cite{BBW06}. This operators don't respect the pattern of the definition of a normal diamond and therefore don't fit in the framework.

We think that one of the most exciting ways to continue this work is to try to expand the framework developed by Areces and Gor\'in to support more types of operators. This would allow us to give a `canonical' simulation notion for a broader set of logics and therefore be able to prove the Hennessy-Milner property for them.

This line of work definitely looks as a promising path to give an automatic derivation of the Characterization and Definability theorems for a greater set of modal logics.

\setcounter{chapter}{1}
\setcounter{section}{0}
\renewcommand{\thechapter}{\Alph{chapter}}
\newtheorem{apthm}{Theorem}[chapter]
\newtheorem{apcoro}{Corollary}[chapter]
\newtheorem{apdefn}[apthm]{Definition}

\theoremstyle{plain}
\newtheorem{apthmcite}{Theorem}[chapter]

\chapter*{Appendix A: Auxiliary results}
\addcontentsline{toc}{chapter}{Appendix A: Auxiliary results}
\chaptermark{Appendix}
\markboth{Appendix}{Appendix}

\begin{apthm}[First order compactness relative to a class of models]\label{thm:compactness} Let \classC be a class of first-order models which is closed under ultraproducts and let $\Sigma$ be a set of first order formulas. If every finite set $\Delta \subseteq \Sigma$ has a model in $\classC$ then there is a model in $\classC$ for $\Sigma$.
\end{apthm}
\begin{proof}
Let $\fmodel_{i}$ be a model for each finite subset $\Delta_i \subseteq \Sigma$, algebraic proofs of the compactness theorem \cite[Theorem 4.3]{KULT08} show that the ultraproduct of the models $\model = \prod_{U} \fmodel_{i}$ satisfies $\model  \models \Sigma$.\footnote{With a suitable ultrafilter $U$.} As each $\fmodel_{i}$ is in $\classC$ and $\classC$ is closed under ultraproducts we conclude that $\model \in \classC$.
\end{proof}

\begin{apthm}[First order definability relative to a class of models]\label{thm:def:relative} Let \classC be a class of first-order models which is closed under ultraproducts and let $\classK \subseteq \classC$.
\begin{enumerate}[(i)]
\item {\classK is a \classC-elementary class (noted $\classC\mbox{-}\elementary$) if and only if \classK is closed under ultraproducts, \classK is \classC-closed under potential isomorphisms and $\CclassK\cap\classC$ is closed under ultrapowers.}
\item {\classK is a basic \classC-elementary class (noted $\classC\mbox{-}\belementary$) if and only if both \classK and $\CclassK\cap\classC$ are closed under ultraproducts and \classC-closed under potential isomorphisms.}
\end{enumerate}
\end{apthm}

\begin{proof} Left to right directions are left to the reader. Let's prove right-to-left directions.
\begin{enumerate}[(i)]
\item {\label{thm:def:relative:i}
Let \classC be a class of first-order models which is closed under ultraproducts, let $\classK \subseteq \classC$ be such that \classK is closed under ultraproducts, \classK is \classC-closed under potential isomorphisms and $\CclassK\cap\classC$ is closed under ultrapowers.

Let $\Gamma = \setcomp{\varphi}{\models_{\classK} \varphi}$, we show that $\Gamma$ defines $\classK$. For the easy part, take a model $\fmodel \in \classK$. By definition of $\Gamma$ we have that $\fmodel \models \Gamma$.

For the hard part, let $\fmodel \in \classC$ be such that $\fmodel \models \Gamma$. Define the first order theory of the model $\fmodel$ as
$$\Sigma = \setcomp{\varphi}{\varphi \text{ is a sentence and } \fmodel \models \varphi}.$$
Let's see that there is a model for $\Sigma$ which lays in \classK. Suppose not, by Theorem \ref{thm:compactness} there is a finite subset $\Sigma_{0} = \{\sigma_{1},\dots,\sigma_{n}\}$ of $\Sigma$ which is unsatisfiable in \classK. Hence, $\models_{\classK} \lnot(\sigma_{1}\land\dots\land\sigma_{n})$ which means that $\lnot(\sigma_{1}\land\dots\land\sigma_{n}) \in \Gamma$. As $\fmodel \models \Gamma$ we arrive to an absurd. We have proved that there exists $\fnodel \in \classK$ such that $\fnodel \models \Sigma$.

By \cite[Theorem 6.1.15]{CK90}, $\fmodel \theq{\cFO} \fnodel$ if and only if there exist ultrapowers $\fmodel_{*}$ and $\fnodel_{*}$ such that $\fmodel_{*} \pisom \fnodel_{*}$. Because \classK is closed under ultraproducts, in particular it is closed under ultrapowers, therefore, $\fnodel_{*} \in \classK$. As both classes are closed under ultrapowers, $\fmodel$ and $\fmodel_{*}$ belong to the same class. Last but not least, as \classK is \classC-closed under potential isomorphisms and $\fmodel_{*} \pisom \fnodel_{*}$ then $\fmodel_{*} \in \classK$. Finally we conclude that $\fmodel \in \classK$.
}
\item {
Let \classC be a class of first-order models which is closed under ultraproducts, let $\classK \subseteq \classC$ be such that both \classK and $\CclassK\cap\classC$ are closed under ultraproducts and \classC-closed under potential isomorphisms.

By (\ref{thm:def:relative:i}) we know there exist two sets $\Gamma,\Gamma_{c}$ respectively defining $\classK$ and $\CclassK\cap\classC$. Observe that the union $\Gamma \cup \Gamma_{c}$ is not satisfiable in \classC. By Theorem \ref{thm:compactness} there exists a finite subset $\Sigma_{0} \subseteq \Gamma \cup \Gamma_{c}$ which is unsatisfiable in \classC. Call $\Sigma_{0} = \{\alpha_{1},\dots,\alpha_{n},\beta_{1},\dots,\beta_{m}\}$ with $\alpha_{i} \in \Gamma$ and $\beta_{j} \in \Gamma_{c}$. As $\Sigma_{0}$ is unsatisfiable in \classC this means that $\models_{\classC} \alpha_{1}\land\dots\land\alpha_{n} \to \lnot(\beta_{1}\land\dots\land\beta_{m})$. Let's see that it is exactly $\varphi = \alpha_{1}\land\dots\land\alpha_{n}$ that defines \classK.

Let $\fmodel \in \classC$. If $\fmodel \in \classK$ then trivially $\fmodel \models \varphi$. Suppose $\fmodel \models \varphi$ then $\fmodel \not\models \beta_{1}\land\dots\land\beta_{m}$ therefore $\fmodel \not\models \Gamma_{c}$ hence $\fmodel \notin \CclassK\cap\classC$. We conclude that $\fmodel \in \classK$.\qedhere
}
\end{enumerate}
\end{proof}



\begin{apthm}\label{thm:similarultrapowers}
Let $\tup{\cL,\classK}$ be an adequate pair and let $\classM_{1}, \classM_{2} \subseteq \mods{\cL}$ be two classes such that $\Trm(\classM_{1})$ and $\Trm(\classM_{2})$ are closed under ultrapowers. Let $\model \in \classM_{1}$ and $\nodel \in \classM_{2}$ be two \cL-models such that for some $w \in |\model|$, $v \in |\nodel|$ they satisfy $\nodel,v \thsub{\cL} \model,w$ then
there exist models $\model^{*} \in \classM_{1}$ and $\nodel^{*} \in \classM_{2}$ and elements $w^{*} \in |\model^{*}|$, $v^{*} \in |\nodel^{*}|$ such that
\begin{enumerate}
\item $\Trm(\model,w) \theq{\cFO} \Trm(\model^{*},w^{*})$ and $\Trm(\nodel,v) \theq{\cFO} \Trm(\nodel^{*},v^{*})$\\ Their translations are pairwise elementary equivalent.
\item $\model,w \theq{\cL} \model^{*},w^{*}$ and $\nodel,v \theq{\cL} \nodel^{*},v^{*}$\\ They are pairwise equivalent.
\item $\nodel^{*},v^{*} \rsim_{\cL} \model^{*},w^{*}$\\ There is a simulation from $\nodel^{*},v^{*}$ to $\model^{*},w^{*}$.
\end{enumerate}
\end{apthm}
\begin{proof}
We define some names for the models which we will be working on before starting with the proof. Call $\model_{f}, g_{w} = \Trm(\model, w)$ and $\nodel_{f}, g_{v} = \Trm(\nodel, v)$. Take $\model^{+}_{f}, \nodel^{+}_{f}$ to be \wsat ultrapowers of $\model_{f}$ and $\nodel_{f}$ (their existance is proved in Theorem~\ref{thm:satultra}). As the classes are closed under ultrapowers the saturated models are in the same class as their originators.

By~\cite[Corollary 4.1.13]{CK90} we have an elementary embedding $d:|\model_{f}|\to|\model^{+}_{f}|$. Let $g^{+}_{w}$ be an assignment for $\model^{+}_{f}$ such that $g^{+}_{w}(x) = d(g_w(x))$. Take the modal preimage of $\model^{+}_{f}, g^{+}_{w}$ and call it $\model^{*}, w^{*} = \Tlm(\model^{+}_{f}, g^{+}_{w})$. We make the same process and assign similar names to models and points deriving from \nodel.
\begin{enumerate}
\item As a corollary of~\cite[Corollary 4.1.13]{CK90}, as there is an elementary embedding, we have that $\model_{f},g_{w} \theq{\cFO} \model^{+}_{f},g_{w}^{+}$. The same proof works with $\nodel_{f}$ and $\nodel^{+}_{f}$.
\item Following the last point, we can conclude, through the translations' truth-preservation, that $\model,w \theq{\cL} \model^{*},w^{*}$. The same proof works with $\nodel,v$ and $\nodel^{*},v^{*}$. Corollary: $\nodel^{*},v^{*} \thsub{\cL} \model^{*},w^{*}$.
\item As both $\model^{+}_{f}$ and $\nodel^{+}_{f}$ are \wsat, by definition of adequate pair, that implies that they have the Hennesy-Milner property. Therefore, because we've just proved that $\nodel^{*},v^{*} \thsub{\cL} \model^{*},w^{*}$ we can conclude that $\nodel^{*},v^{*} \rsim_{\cL} \model^{*},w^{*}$. \qedhere
\end{enumerate}
\end{proof}

\chapter*{Appendix B: Filters and ultraproducts}
\addcontentsline{toc}{chapter}{Appendix B: Filters and ultraproducts}
\chaptermark{Appendix}
\markboth{Appendix}{Appendix}

\setcounter{chapter}{2}
\setcounter{apthm}{0}

The ultraproduct construction is a uniform method of building models of first 
order theories which has applications in many areas of mathematics. It is attractive 
because it is algebraic in nature, but preserves all properties expressible in first order 
logic. In this section we will make a brief summary of the tools presented in~\cite{KULT08}
with some additional notes that may guide the reader. Unless explicitly stated the proofs can
be checked in the publication that we've just mentioned.

\begin{apdefn}[Filter, proper filter, ultrafilter]
Let $I$ be a non-empty set.\\
A \emph{filter} $U$ over $I$ is a set of subsets of $I$ such that:
\begin{enumerate}[(i)]
\setlength{\itemsep}{0pt}
\item $I \in U$.
\item $U$ is closed under supersets; if $X \in U$ and $X \subseteq Y$ then $Y \in U$.
\item $U$ is closed under finite intersections; if $X \in U$ and $Y \in U$ then $X\cap Y \in U$.
\end{enumerate}
A \emph{proper filter} over $I$ is a filter $U$ over $I$ such that:
\begin{enumerate}
\item[(iv)] $\emptyset \notin U$.
\end{enumerate}
An \emph{ultrafilter} over $I$ is a proper filter $U$ over $I$ such that:
\begin{enumerate}
\item[(v)] For each $X \subseteq I$ exactly one of $X,I\backslash X$ belongs to $U$.
\end{enumerate}
\end{apdefn}

\begin{apthm}[Ultrafilter Theorem, Tarski]
Every proper filter over the set $I$ can be extended to an ultrafilter over $I$.
\end{apthm}

We first define the ultraproduct operation on \emph{sets}. Let $U$ be an ultrafilter over $I$, 
and for each $i \in I$ let $A_{i}$ be a non-empty set. The ultraproduct 
$\prod_{U} A_{i}$ is obtained by first taking the cartesian product
$C = \prod_{i\in I} A_{i}$. Observe that $C$ is the set of all functions $f$ such that
for each $i \in I$, $f(i) \in A_{i}$. We continue by identifying elements which 
are equal for $U$-almost all $i \in I$. The formal definition is as follows.

\begin{apdefn}[$U$-equivalence]
Let $U$ be an ultrafilter over $I$. Two elements $f,g$ of the cartesian product
$\prod_{i\in I} A_{i}$ are said to be $U$-equivalent, noted $f =_{U} g$, if the set
$\setcomp{i }{ f(i) = g(i)}$ belongs to $U$. The $U$-equivalence class of $f$ is the set
$f_{U} = \setcomp{g }{ f =_{U} g}$.
\end{apdefn}

\begin{apdefn}[Ultraproduct of sets]
Let $U$ be an ultrafilter over $I$, the \emph{ultraproduct of sets} $\prod_{U} A_{i}$ is defined as the set of $U$-equivalence classes
$$\prod_{U} A_{i} = \setcomp{f_{U} }{ f \in \prod_{i\in I} A_{i}}.$$
An \emph{ultrapower of sets} of $A$ modulo $U$ is defined as the ultraproduct $\prod_{U} A = \prod_{U} A_{i}$ where $A_{i} = A$ for each $i \in I$.
The \emph{natural} or \emph{cannonical embedding} is the mapping $d:A\to\prod_{U} A_{i}$ such that $d(a)$ is the $U$-equivalence class of the constant function with value $a$. That is $d(a) = (\lambda x.a)_{U}$.
\end{apdefn}

We now introduce the ultraproduct operation on first order structures. For each
$i \in I$, let $\fmodel_{i}$ be a first order model with universe set $A_{i}$.
Briefly, the ultraproduct of models $\prod_{U} \fmodel_{i}$ is the unique first order model with universe 
$\prod_{U} A_{i}$ such that each \emph{basic} formula 
holds in the ultraproduct if and only if it holds in 
$\prod_{U} \fmodel_{i}$ for $U$-almost all $i$.
Here is the formal definition.

\begin{apdefn}[Ultraproduct of models]\label{apx:upmod}
Let $U$ be an ultrafilter over $I$, and let $\fmodel_{i}$ be a familiy of $\mathcal{L}$-structures with universe set $A_{i}$.
The \emph{ultraproduct of models} $\prod_{U} \fmodel_{i}$ is the unique model $\fmodel$ such that:\footnote{The definition found in~\cite{KULT08} has a mistake in point (ii). The subscript of $\mathcal{A}$ is missing in the set definition after the `iff', it should be $\mathcal{A}_{i}$.}
\begin{enumerate}[(i)]
\item The universe of $\fmodel$ is $M = \prod_{U} A_{i}$.
\item For each atomic formula $\varphi(x_{1},\dots,x_{k})$ which has at most one symbol from the vocabulary $\mathcal{L}$, and each $f_{1},\dots,f_{k} \in \prod_{i\in I} A_{i}$,
$$\fmodel \models \varphi(f_{1U},\dots,f_{kU}) \text{ iff } \setcomp{i }{ \fmodel_{i} \models \varphi(f_{1}(i),\dots,f_{k}(i))} \in U.$$
\end{enumerate}
\end{apdefn}
Using the properties of ultrafilters one can check that such structure is unique and thus well-defined. Similarly, the \emph{ultrapower of models} of the model $\fmodel$ modulo $U$ is defined as the ultraproduct $\prod_{U} \fmodel = \prod_{U} \fmodel_{i}$ where $\fmodel_{i} = \fmodel$ for each $i \in I$.

Finally we present the theorem of \L o\'s which makes ultraproducts useful in model theory. It shows that each formula 
holds in the ultraproduct if and only if it holds in $\prod_{U} \fmodel_{i}$ for $U$-almost all $i$. Observe that in this case there is no restriction to basic formulas as before in Definition \ref{apx:upmod}.

\begin{apthm}[Fundamental theorem of Ultraproducts, \L o\'{s}]
Let $U$ be an ultrafilter over $I$, and let $\fmodel_{i}$ be a family of  $\mathcal{L}$-structures for each $i \in I$. For each formula $\varphi(x_{1},\dots,x_{k})$, and each $f_{1},\dots,f_{k} \in \prod_{i\in I} A_{i}$, we have
$$\prod_{U} \fmodel_{i} \models \varphi(f_{1U},\dots,f_{kU}) \text{ iff } \setcomp{i }{ \fmodel_{i} \models \varphi(f_{1}(i),\dots,f_{k}(i))} \in U.$$
\end{apthm}

\begin{apcoro}
For each set of first-order sentences $\Gamma$, and family of models $\fmodel_{i}$. If $\fmodel_{i} \models \Gamma$ for all $i \in I$ then $\prod_{U} \fmodel_{i} \models \Gamma$.
\end{apcoro}

\begin{apcoro}[{\cite[Corollary A.21]{MLBOOK}}]\label{coro:ultraelem}
Let $\prod_{U} \fmodel$ be an ultrapower of $\fmodel$, the diagonal mapping $d(a) = (\lambda x.a)_{U}$ is an elementary embedding. That is, for any first order formula $\varphi(x_{1},\dots,x_{k})$ and $a_{1},\dots,a_{k} \in \fmodel$
$$\fmodel \models \varphi(a_{1},\dots,a_{k}) \text{ if and only if } \prod_{U} \fmodel \models \varphi(d(a_{1}),\dots,d(a_{k})).$$
\end{apcoro}

Using this results we can state an important theorem, the existence of elementary equivalent \wsat ultrapowers.

\begin{apthm}\label{thm:satultra}
Let \cFO be countable and $\fmodel$ be an \cFO model then there exists an \mbox{\wsat} ultrapower $\prod_{U} \fmodel$ such that $\fmodel \theq{\cFO} \prod_{U} \fmodel$.
\end{apthm}
\begin{proof}
Follows from a direct combination of~\cite[Theorem 5.6]{KULT08} and Corollary~\ref{coro:ultraelem}.
\end{proof}

\clearpage

\backmatter


\addcontentsline{toc}{chapter}{Bibliography}
\bibliographystyle{alpha}
{\small
\bibliography{thesis}
}

\end{document}